\documentclass{aa}

\usepackage{graphicx}
\usepackage{txfonts}
\usepackage{placeins}
\usepackage{float}
\usepackage{subcaption}
\usepackage{lscape}
\begin{document}

   \title{An imaging line survey of OMC-1 to OMC-3}

   \subtitle{Averaged spectra of template regions}

   \author{N. Brinkmann
          \inst{1}
          \and
          F. Wyrowski \inst{1}
          \and
          J. Kauffmann \inst{2}
          \and
          D. Colombo  \inst{1}
          \and
          K.M. Menten \inst{1}
          \and
          X. D. Tang \inst{1,3,4}
          \and
          R. G\"usten \inst{1}
          }
   \institute{Max-Planck-Institut f\"ur Radioastronomie, Auf dem Hügel 69,
              D-53121 Bonn, Germany\\
              \email{brinkmann@mpifr-bonn.mpg.de}
         \and
            Haystack Observatory, Massachusetts Institute of Technology,
            Westford, MA 01886
         \and
             Xinjiang Astronomical Observatory, Chinese Academy of Sciences, 
             830011 Urumqi, PR China
         \and
            Key Laboratory of Radio Astronomy, Chinese Academy of Sciences,
            830011 Urumqi, PR China
             }

   \date{Received:~~~~~~~~~~~~; Accepted:~~~~~~~~~~~~}

% \abstract{}{}{}{}{} 
% 5 {} token are mandatory
 
  \abstract
  % context heading (optional)
  % {} leave it empty if necessary  
   {Recently, sensitive wide-bandwidth receivers in the millimetre regime 
   have enabled us to combine large spatial and spectral coverage
   for observations of molecular clouds. The resulting capability to map the distributions of lines
   from many molecules simultaneously yields unbiased coverage of the various 
   environments within star-forming regions.}
  % aims heading (mandatory)
   {Our aim is to identify the dominant molecular cooling lines and characteristic emission features 
   in the 1.3 mm 
   window of distinct regions in the northern part of the Orion A molecular cloud. By defining and
   analysing template regions, we also intend to help with the interpretation of
   observations from more distant sources which cannot be easily spatially resolved.}
  % methods heading (mandatory)
   {We analyse an imaging line survey covering the area of OMC-1 to OMC-3 from 200.2 to 281.8 GHz
   obtained with
   the PI230 receiver at the APEX telescope.
   Masks are used to define regions with distinct properties (e.g. column density or temperature ranges)
   from which we obtain averaged spectra. Lines of 29 molecular species (55 isotopologues) 
   are fitted for each region to obtain the respective total intensity.}
  % results heading (mandatory)
   {We find that strong sources like Orion KL have a clear impact on the emission on larger scales.
   Although not  spatially extended, their line emission contributes substantially to spectra
   averaged over large regions. Conversely, the emission signatures of dense, cold regions 
   like OMC-2 and OMC-3 (e.g. enhanced N$_{2}$H$^{+}$ emission and low HCN/HNC ratio) seem to be 
   difficult to pick up on larger scales, where they are eclipsed by signatures of stronger sources. 
   In all regions, HCO$^{+}$ appears to contribute between 3\% and
   6\% to the total intensity, the most stable value for all bright species. N$_{2}$H$^{+}$ shows the 
   strongest correlation with column density, but not with typical high-density tracers like HCN, 
   HCO$^{+}$, H$_{2}$CO, or HNC. Common line ratios 
   associated with UV illumination, CN/HNC and CN/HCO$^{+}$, show ambiguous results on larger scales,
   suggesting that the identification of UV illuminated material may be more challenging. The HCN/HNC 
   ratio may be related to temperature over varying scales.}
  % conclusions heading (optional), leave it empty if necessary 
   {}


   \maketitle
%
%
%________________________________________________________________

\section{Introduction}

The northern part of the Orion A molecular cloud is one of the most prominent 
regions
of current low- to intermediate-mass star formation, whose close proximity of just $\sim400 \ 
\textrm{pc}$ \citep{Menten2007,Kounkel2017} 
enables us to spatially resolve its physically and chemically different
regions. Continuum maps show substructures which divide Orion A into 
morphologically different regions: 
the bright OMC-1 in the south, with a group of filaments radiating away from its 
central region (\citealt{ODell2008} and references therein, e.g. 
\citealt{Wiseman1998}), 
and the less prominent  OMC-2/3 in the north.

OMC-1 hosts star formation and is heavily influenced
by intense UV radiation from the young massive Trapezium stars. 
In addition to the Trapezium, Orion BN/KL (hereafter KL) and Orion South are
sites of recent of star formation \citep{ODell2008}. Their positions are indicated
in Fig. \ref{fig:int_maps}.
OMC-1 also 
includes an archetypical photon-dominated region (PDR), the Orion Bar,
outflows, and the eponymous hot core (e.g. \citealt{Masson1984}) containing a rich chemistry of 
complex molecules, but also more quiescent gas around it. 
OMC-2 and OMC-3, located northwards of OMC-1, appear to be a continuation of the 
gas in OMC-1,
apart from a shift in velocity (e.g. \citealt{ODell2008,Peterson2008}). 
A large number of pre-stellar Class 0 and Class I 
objects following
the filamentary structure have been discovered through submillimetre 
observations (e.g. \citealt{Chini1997,Johnstone1999,Lis1998}), in addition to
a number of brown dwarfs \citep{Peterson2008dwarfs}. In contrast to OMC-1, 
there are no massive O/B stars, resulting in very different conditions with 
outflows
driven by young embedded stars (\citealt{Peterson2008} and references therein, 
e.g. \citealt{Yu1997}), but without strong UV radiation.

Investigating the characteristics of these distinct regions helps us
to understand how low- to high-mass star formation influences the surrounding 
material and vice versa.
Previous line surveys could be biased in that they have often been focused on a
single or  few positions 
(e.g. \citealt{Sutton1985,Blake1987,Tercero2010,Tercero2011,Esplugues2013,Johnstone2003}), or spatially extended 
regions were mapped in selected molecular lines only (e.g. \citealt{Ungerechts1997}), missing 
out on information a complete frequency coverage offers.

New opportunities arise with sensitive, wide-bandwidth receivers which combine 
an extensive frequency coverage with 
a mapping speed sufficient for large fields of view. This enables us now to obtain 
a comprehensive and
unbiased picture of a molecular cloud. It facilitates the deduction of a
variety of physical and chemical conditions using several molecular
tracers. These possibilities start being utilised in lower frequency regimes
(\citealt{Pety2016} and companion papers, e.g. \citealt{Gratier2017,Bron2018} focusing on 
Orion B or the LEGO project started with \citealt{Kauffmann2017}). Our data set described in 
Section \ref{obs} deals with higher
energies, and is thus sensitive to higher critical densities and constitutes a 
useful complement to these observations.

This will help us to better understand the conditions (e.g. column density, temperature, 
strength of UV illumination) under which different molecules are excited, 
and in particular identify
those molecules predominantly present in very specific environments (e.g. 
$\textrm{N}_{2}\textrm{H}^{+}$ in dense cores; see  also \citealt{Pety2016}).
This information is crucial for understanding emission of more distant molecular clouds, 
including extragalactic sources.

Although CO emission is dominant in the 1.3 mm window, its share of 
the total intensity (and thus 
cooling) changes for distinct regions (see also \citealt{Goldsmith2001}). Comparing 
the influence of different molecular species to
the cooling in Orion A under various conditions will help us to develop templates. These can be used
to `reproduce' spectra of other clouds when regions with different physical and chemical conditions are
not resolved.

%%%%%%%%%%%%%%%%%%%%%%%%%%%%%%%%%%%%%%%%%%%%%%%%%%%%

\section{Observations and data reduction}\label{obs}
\subsection{Observations}
All observations were carried out with the Atacama Pathfinder Experiment (APEX) $12 \ \mathrm{m}$ 
submillimetre telescope 
\citep{Guesten2006} using the PI230 receiver operating in the $1.3 \ 
\mathrm{mm}$
atmospheric window. 
We covered the area of OMC-1 to OMC-3 from $200.2 
~\mathrm{to}~ 281.8 ~ \mathrm{GHz}$ with data 
collected over several observing periods from October 2015 to November 2018.

The PI230 receiver offers $32 ~ \mathrm{GHz}$ bandwidth per tuning, apportioned into two sidebands
and two polarisations,  with the two polarisation mixers co-aligned on sky. Eight Fourier 
Transform Spectrometer (FFTS4G)  
backends each provide 65536 channels for $4 ~ \mathrm{GHz}$. 
An overlap of $0.2 ~ \mathrm{GHz}$ between two backends results in $7.8 ~ 
\mathrm{GHz}$ coverage for each sideband and polarisation per tuning. The sideband rejection
is $\approx 20~\mathrm{dB}$. The width of the telescope beam changes from around $22\arcsec$ to
$31\arcsec$ over the $1.3 \ \mathrm{mm}$ window.

\begin{figure}
 \centering
 \includegraphics[width=8cm]{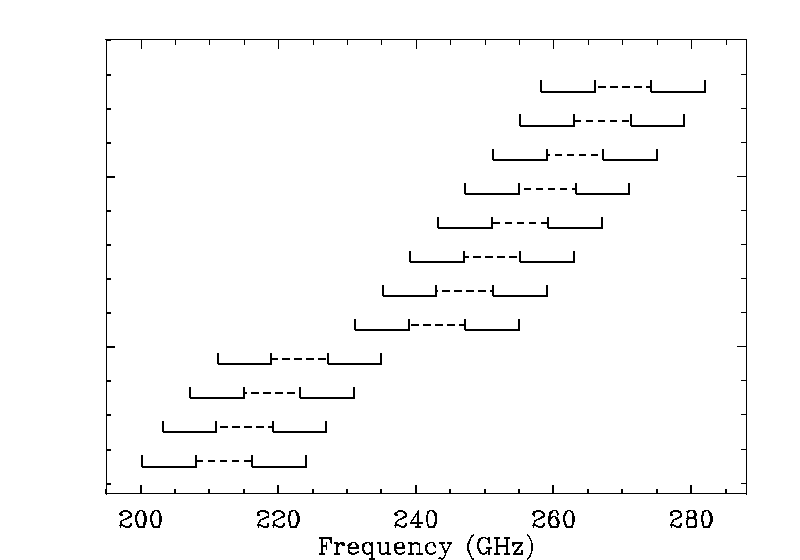}
 \caption{Twelve overlapping frequency set-ups were used to cover the $1.3 \ 
\mathrm{mm}$ window, each providing $2 \times 7.8 ~ \mathrm{GHz}$ bandwidth.}
 \label{setup}
\end{figure}

The overall spatial coverage of around $400\arcsec \times 1900\arcsec$, 
corresponding to roughly $0.8 \times 3.8 ~ \mathrm{pc}$ at the distance 
of the Orion Nebula Cluster, was achieved with 
on-the-fly maps scanned in the x- and 
y-directions in steps of $8\arcsec$ with a dump time of $0.3 ~ \mathrm{seconds}$. 
The reference position was $\alpha_{2000} = 5^{\mathrm{h}} 31^{\mathrm{m}} 
14.5^{\mathrm{s}}$, 
$\delta_{2000} = -5^{\circ} 52^{\arcmin} 29.0^{\arcsec}$.
The frequency coverage was obtained with 12 overlapping frequency set-ups as shown in Fig. \ref{setup}.
The mean and median system temperatures in main-beam brightness temperature scale ($T_{\rm mb}$)  determined over all observations 
are $T_{\mathrm{sys}}^{\mathrm{mean}} \approx 266 ~ \mathrm{K}$
and $T_{\mathrm{sys}}^{\mathrm{median}} \approx 239 ~ \mathrm{K}$, respectively.
Mean and median receiver temperatures were $T_{\mathrm{rec}}^{\mathrm{mean}} \approx 81 ~ \mathrm{K}$
and $T_{\mathrm{rec}}^{\mathrm{median}} \approx 78 ~ \mathrm{K}$, respectively.

%%%%%%%%%%%%%%%%%%%%%%%%%%%%%%%%%%%%%%%%%%

\subsection{Calibration and data reduction}\label{obs_calis}
All data reduction was done with the GILDAS\footnote{http://www.iram.fr/IRAMFR/GILDAS}
software package. 
This included analysis of spectra and maps of the lines' intensity distribution. In the first 
processing step the data were calibrated and 
correction factors for the beam efficiencies applied separately for 
each observing period.
These correction factors were obtained from Jupiter continuum 
pointings at different frequencies. The observed
peak temperature of the planet was compared to the expected value calculated by the GILDAS 
ASTRO module, yielding the main beam efficiency. These main beam efficiencies 
were then fitted by the Ruze formula, facilitating a frequency dependent correction for every observing
period, summarised in Table \ref{table_efficiencies}. The exact value of the surface accuracy of the dish
has only a marginal impact on the  data observed in the $1.3 ~ \mathrm{mm}$ window, however. The main 
beam efficiency varies by 
$\lesssim 5 ~\%$ over the observed frequency range.
The spectra are also converted from corrected antenna temperature, $T^*_{\rm A}$, 
to $T_{\mathrm{mb}}$ units in this step, which 
is thus the temperature scale used in this paper.

\begin{table}
\small
\caption{ Main beam efficiencies ($\eta^{0}_{\mathrm{mb}}$) and surface accuracy ($\sigma$) used to correct 
the data from
different observing periods}        
\label{table_efficiencies}     
\centering 
\begin{tabular}{l@{\hskip 1cm} c@{\hskip 1cm} c}
\hline\hline  
Observing period &  $\eta^{0}_{\mathrm{mb}}$ & $\sigma$ [$\mu \mathrm{m}$] \\ [3pt]
\hline                  
\\
October 2015 - December 2015 & 0.62 & 30 
\\ 
July 2016 - November 2016    & 0.73 & 30
\\
April 2017                   & 0.74 & 30 
\\
July 2017                    & 0.66 & 30
\\
May 2018 - July 2018         & 0.73 & 15 
\\
November 2018                & 0.75 & 13 
\\
\hline    
\end{tabular}
\end{table}

A substantial amount of data was collected during the commissioning phase of the PI230 receiver, 
entailing larger uncertainties. However, parts of the maps have been observed repeatedly in different observing 
periods  such that efficiency corrected spectra could be compared and the earlier data
validated, suggesting our calibration uncertainty to be around $30\%$ overall.

In the next processing step we removed spikes, artefacts,  and spurious lines  that only appeared during 
the early commissioning of the receiver/backend system and affect parts of the data. Furthermore, spectra
with 
$T_{\mathrm{sys}} \geq 600 ~ \mathrm{K}$ were rejected. 
While single channel spikes are
readily identified by simple scripts, the artefacts needed to be manually identified 
and blanked as they showed continuous behaviour towards their edges, making it more challenging 
to discriminate between them and unaffected channels. A larger number of spectra were checked 
for each backend and each frequency setting for the different observing days.

To identify and remove spurious lines, we did a first assignment of all observed 
lines to molecular transitions from the JPL\footnote{http://spec.jpl.nasa.gov/} \citep{Pickett1998} and
CDMS\footnote{http://www.astro.uni-koeln.de/cdms/} \citep{Endres2016} molecular 
spectroscopy databases using the WEEDS \citep{Maret2011} extension for GILDAS. 
If a line could not be identified, we checked if it was present in all relevant backends and frequency 
settings. 
The respective channels in the originating backends were blanked if the line 
only appeared in some instances.

%%%%%%%%%%%%%%%%%%%%%%%%%%%%%%%%%%%%%%%%%%%%%

\subsection{Final data set}
In the following processing step, a first-order baseline was subtracted from each spectrum. The boundaries 
of the line windows were based on the Orion KL spectrum, where we expect the broadest lines and strongest
overlap between them.
To simplify the analysis and speed up both imaging and the extraction of spectra,
we assembled all data for each backend and each frequency set-up and gridded it using the 
same template cube with a smoothed beam size of $32\arcsec$. 
This results in 96 data cubes (8 backends $\times$ 12 frequency set-ups), each containing 
around 3000 spectra and each of which with 
65536 channels  (corresponding to a resolution of $0.07 -\ 0.09 ~ \mathrm{km s}^{-1}$ or
$61~\mathrm{kHz}$). 
These spectra are intensity corrected, (mostly) artefact-free, baseline 
subtracted, and arranged on the same grid, and form the basis for all further analyses.

\begin{figure*}[ht]
\centering
        \includegraphics[width=\textwidth]{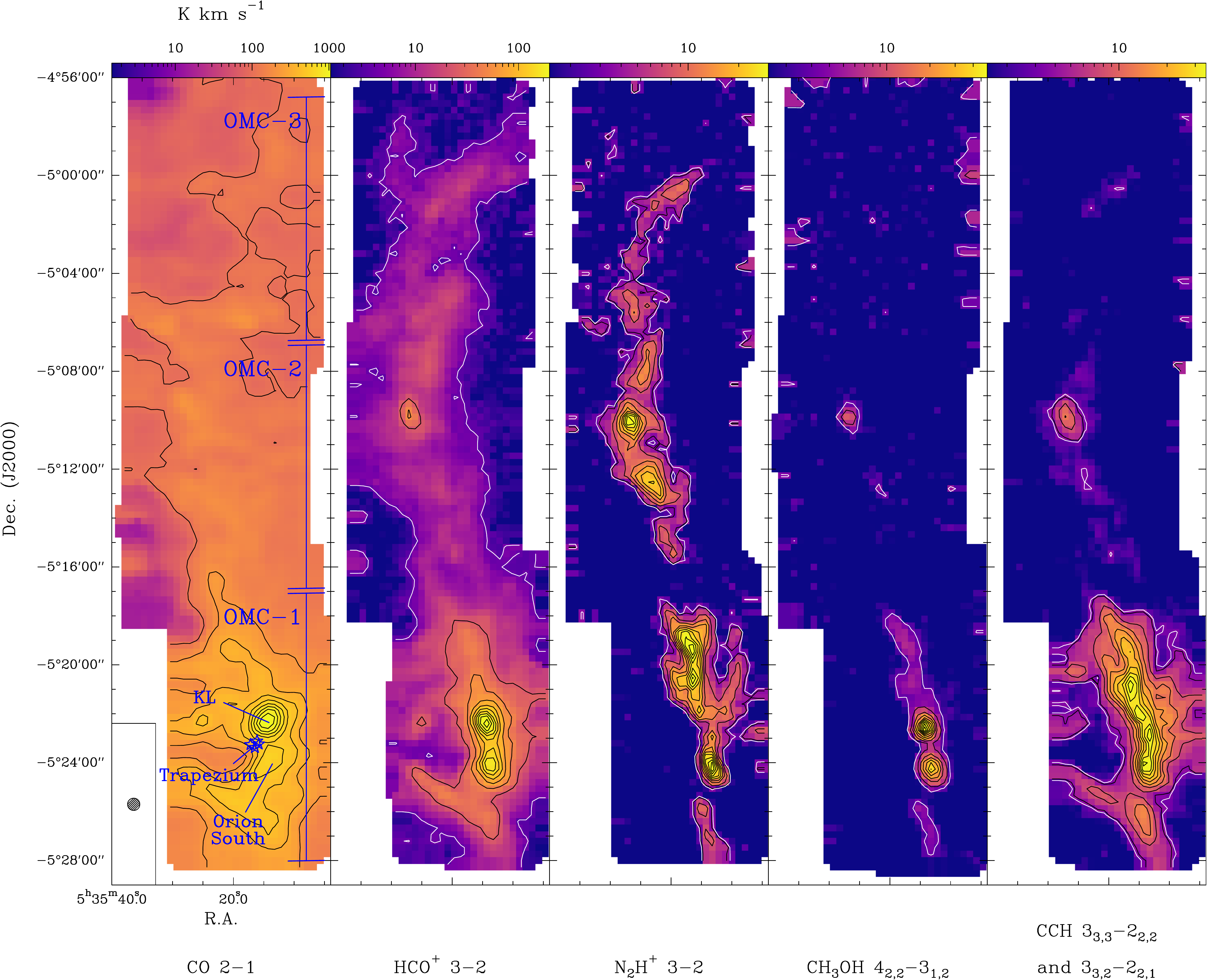}
        \caption{ Selected integrated intensity [5,15] $\mathrm{km~s}^{-1}$ maps. Black contours
        run from 10\% to 90\% of the maximum value in steps of 10\%, while the white contour shows the
        $3\,\sigma$ level.
        The maps highlight some of the molecules 
        typically used as tracers: CO for the bulk of molecular gas, $\mathrm{HCO}^{+}$
        as an indicator of high density, $\mathrm{N}_{2}\mathrm{H}^{+}$ for cold dense gas, 
        $\mathrm{CH}_{3}\mathrm{OH}$ for shocked material, and $\mathrm{C}_{2}\mathrm{H}$ associated
        with UV irradiation. The velocity range used for the maps does not encompass the whole
        line width (which differs strongly, especially comparing OMC-1 to OMC-3), but covers all
        central velocities.}
        \label{fig:int_maps}
\end{figure*}

The 1.3 mm window gives us access to a variety of molecular species and transitions. In this paper we 
want to concentrate on 29 species (55 isotopologues), which are listed in 
Table \ref{table_species}.
We can work with our data set both in terms of imaging and spectral analysis.
An example of the differing spatial extensions of typical molecular tracers is given in the integrated 
intensity maps of Fig. \ref{fig:int_maps}. The maps are resampled to a resolution
of $0.4 ~ \mathrm{km~s}^{-1}$, resulting in a typical rms noise of around 0.5 K.
However,  not all species contained in our data set can be mapped as  for an individual image pixel their 
intensity may be too low. Those molecules (like $\mathrm{CF}^{+}$) are detectable when we average over
larger areas. These averaged spectra will be further described in Sections \ref{regions} and 
\ref{total_intensities}.

\begin{table}
\tiny
\caption{All considered species and isotopologues.}        
\label{table_species}     
\centering 
\begin{tabular}{l@{\hskip 1cm} l}
\hline\hline  
species &  isotopologues 
\\
\hline                       
\\
CO & $\mathrm{CO}~ , ^{13}\mathrm{CO} ~, \mathrm{C}^{18}\mathrm{O} ~, \mathrm{C}^{17}\mathrm{O} ~,
^{13}\mathrm{C}^{18}\mathrm{O}$
\\ 
$\mathrm{c-C}_{3}\mathrm{H}_{2}$ & $\mathrm{c-C}_{3}\mathrm{H}_{2}$ 
\\
$\mathrm{C}_{2}\mathrm{H}$ & $\mathrm{C}_{2}\mathrm{H} ~ , \mathrm{C}_{2}\mathrm{D}$ 
\\
$\mathrm{CF}^{+}$ & $\mathrm{CF}^{+}$ 
\\
$\mathrm{CH}_{3}\mathrm{CCH}$ & $\mathrm{CH}_{3}\mathrm{CCH}$ 
\\
$\mathrm{CH}_{3}\mathrm{CN}$ & $\mathrm{CH}_{3}\mathrm{CN}$ 
\\
$\mathrm{CH}_{3}\mathrm{OH}$ & $\mathrm{CH}_{3}\mathrm{OH}$ 
\\
CN & CN, $^{13}\mathrm{CN} $ 
\\
CS & CS, $^{13}\mathrm{CS} ~, \mathrm{C}^{34}\mathrm{S} ~, \mathrm{C}^{33}\mathrm{S}$ 
\\
HCCCN & HCCCN 
\\
HCN & HCN, DCN, $\mathrm{H}^{13}\mathrm{CN}~, \mathrm{HC}^{15}\mathrm{N}$ 
\\
HCO & HCO 
\\
$\mathrm{HCO}^{+}$ & $\mathrm{HCO}^{+} ~, \mathrm{DCO}^{+} ~, \mathrm{H}^{13}\mathrm{CO}^{+} ~ ,
\mathrm{HC}^{18}\mathrm{O}^{+} ~ , \mathrm{HC}^{17}\mathrm{O}^{+}$ 
\\
$\mathrm{HCS}^{+}$ & $\mathrm{HCS}^{+}$ 
\\
HDO & HDO 
\\
$\mathrm{H}_{2}\mathrm{CCO}$ & $\mathrm{H}_{2}\mathrm{CCO}$ 
\\
$\mathrm{H}_{2}\mathrm{CO}$ & $\mathrm{H}_{2}\mathrm{CO} ~ , \mathrm{HDCO} ~,
\mathrm{H}_{2}^{13}\mathrm{CO}$ 
\\
$\mathrm{H}_{2}\mathrm{CS}$ & $\mathrm{H}_{2}\mathrm{CS} ~ , \mathrm{HDCS} ~,
\mathrm{H}_{2}\mathrm{C}^{34}\mathrm{S}$ 
\\
$\mathrm{H}_{2}\mathrm{S}$ & $\mathrm{H}_{2}\mathrm{S}$ 
\\
HNC& HNC, DNC, $\mathrm{HN}^{13}\mathrm{C}$ 
\\
HNCO & HNCO 
\\
$\mathrm{N}_{2}\mathrm{H}^{+}$ & $\mathrm{N}_{2}\mathrm{H}^{+} ~, \mathrm{N}_{2}\mathrm{D}^{+}$ 
\\
NO & NO 
\\
NS & NS 
\\
OCS & OCS 
\\
SiO & SiO, $^{29}\mathrm{SiO} ~, ^{30}\mathrm{SiO}$ 
\\
SO & SO, $^{34}\mathrm{SO}$ 
\\
$\mathrm{SO}^{+}$ & $\mathrm{SO}^{+}$ 
\\
$\mathrm{SO}_{2}$ & $\mathrm{SO}_{2}$ 
\\ \\
\hline    
\end{tabular}
\tablefoot{Summary of the species and isotopologues used for our analysis. A List of all
transitions is found in Sect. \ref{all_transitions}}
\end{table}

%%%%%%%%%%%%%%%%%%%%%%%%%%%%%%%%%%%%%%%%%%%%%%
\section{Analysis}\label{analysis}

The final data set offers a starting point for varied analyses. In this paper, we want 
to concentrate on the cooling by different molecular species depending on their environment, 
focusing on an observational point of view.

\subsection{Ancillary data}\label{ancillary}

\begin{figure}
\centering
    \includegraphics[width=\hsize]{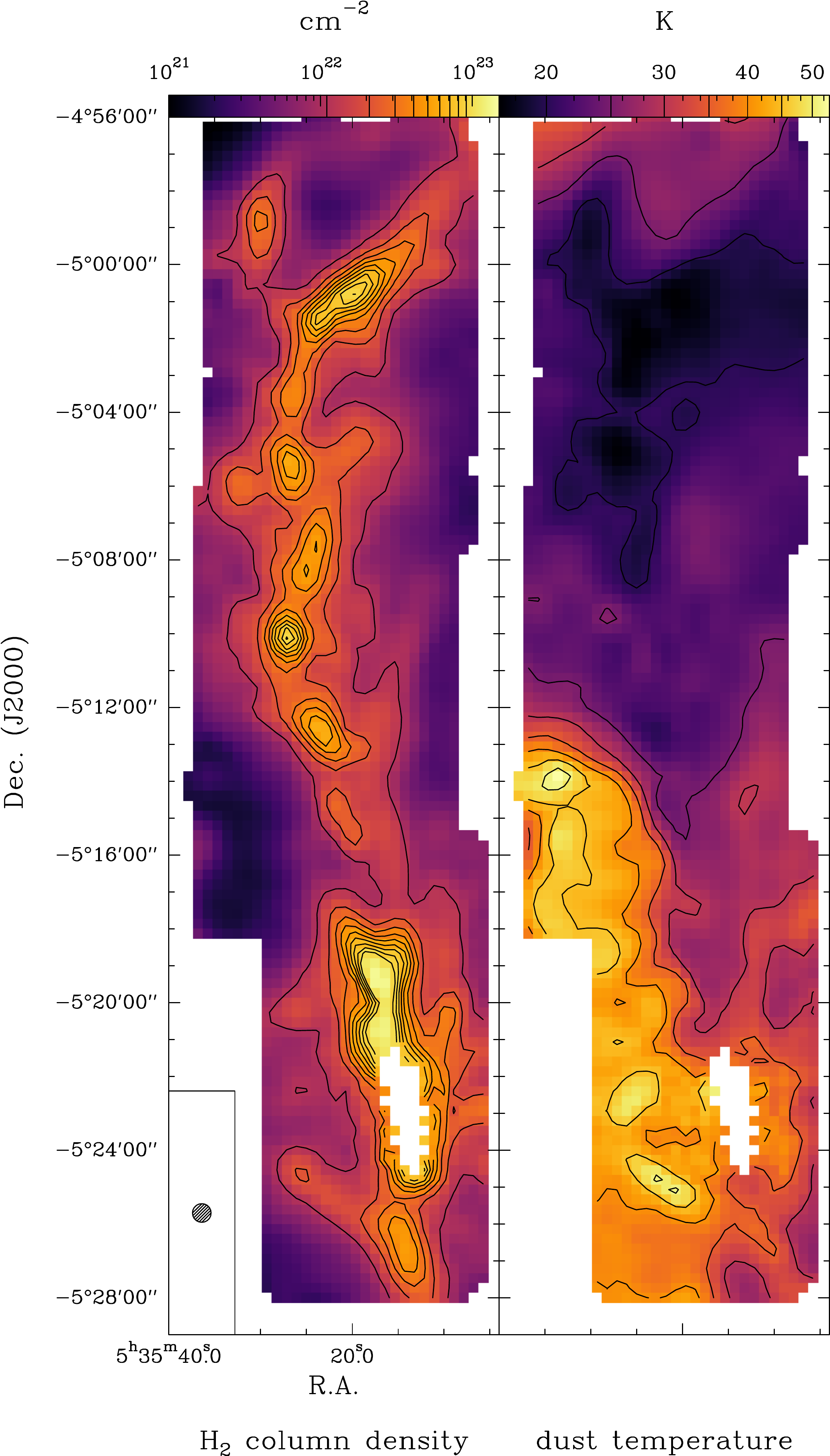}
    \caption{Dust derived column density and temperature map, modified (reprojected and cropped, 
    the column density unit converted) from \citet{Guzman2015}. The area around KL and Orion South is saturated and blanked.}
    \label{fig:density_temp}
\end{figure}

Attributing column density, temperature, or the strength of UV irradiation to different 
parts of Orion A is done with the help of ancillary data. Dust column density\footnote{Converted from 
$\mathrm{g~cm}^{-2}$ to $\mathrm{cm}^{-2}$ under the simplifying assumption that all mass comes
from $\mathrm{H}_{2}$, $1\mathrm{g} \approx 2.134 \times 10^{23}$ molecules.} and
temperature based on \textit{Herschel} data are   from \citet{Guzman2015}, while 
$\mathrm{C}65\alpha$ emission data (here used to define the dense PDR region) are from  \citet{Wyrowski1997}.
Dust and gas temperature are expected to be coupled for densities $\sim 3 \times 10^{4} ~
\mathrm{cm}^{-3}$ \citep{Galli2002}, and \citet{Guzman2015} found that for the examined
MALT90 clumps \citep{jackson2013,Foster2013,Foster2011} under $22 ~ \mathrm{K}$ ammonia and dust 
temperatures agree within $\pm 3 ~ \mathrm{K}$,
while the uncertainties become larger with increasing temperature. We will mainly use the temperature
map to differentiate between colder ($<25 ~ \mathrm{K}$) and warmer ($ \geq 25 ~ \mathrm{K}$) regions.
Maps of gas kinetic temperatures based on NH$_{3}$ (Orion A) and H$_{2}$CO (OMC-1 only) observations 
can be found in \citet{Friesen2017} and \citet{Tang2018}, respectively. As the utilised 
molecular lines require higher column densities to be excited, these maps do not cover all of the ambient 
material within our maps. Hence we decided on the use of dust-derived temperatures.
Reproductions of the column density and dust temperature maps (reprojected using our template data cube) 
are shown in Fig. \ref{fig:density_temp}.

\subsection{Regions and masks}\label{regions}

Masks were used to define regions meeting certain conditions (e.g. column density or temperature 
ranges). These masks were mostly based on dust column density and temperature maps, 
but in some cases also selected around chosen coordinates. 
To this end, the maps were reprojected using the same template cube as
for the gridding of our observational data, such that combinations of masks (e.g. high column density
concurrent 
with low temperature) were possible. See Table \ref{table_regions} for a summary of the selected 
regions and their
basic properties. For a better visual idea of the spatial extension of these regions, 
images of the masks are
included in the Appendix (Fig. \ref{fig:masks}).

\begin{figure*}[ht]
\centering
        \includegraphics[width=0.88\textwidth]{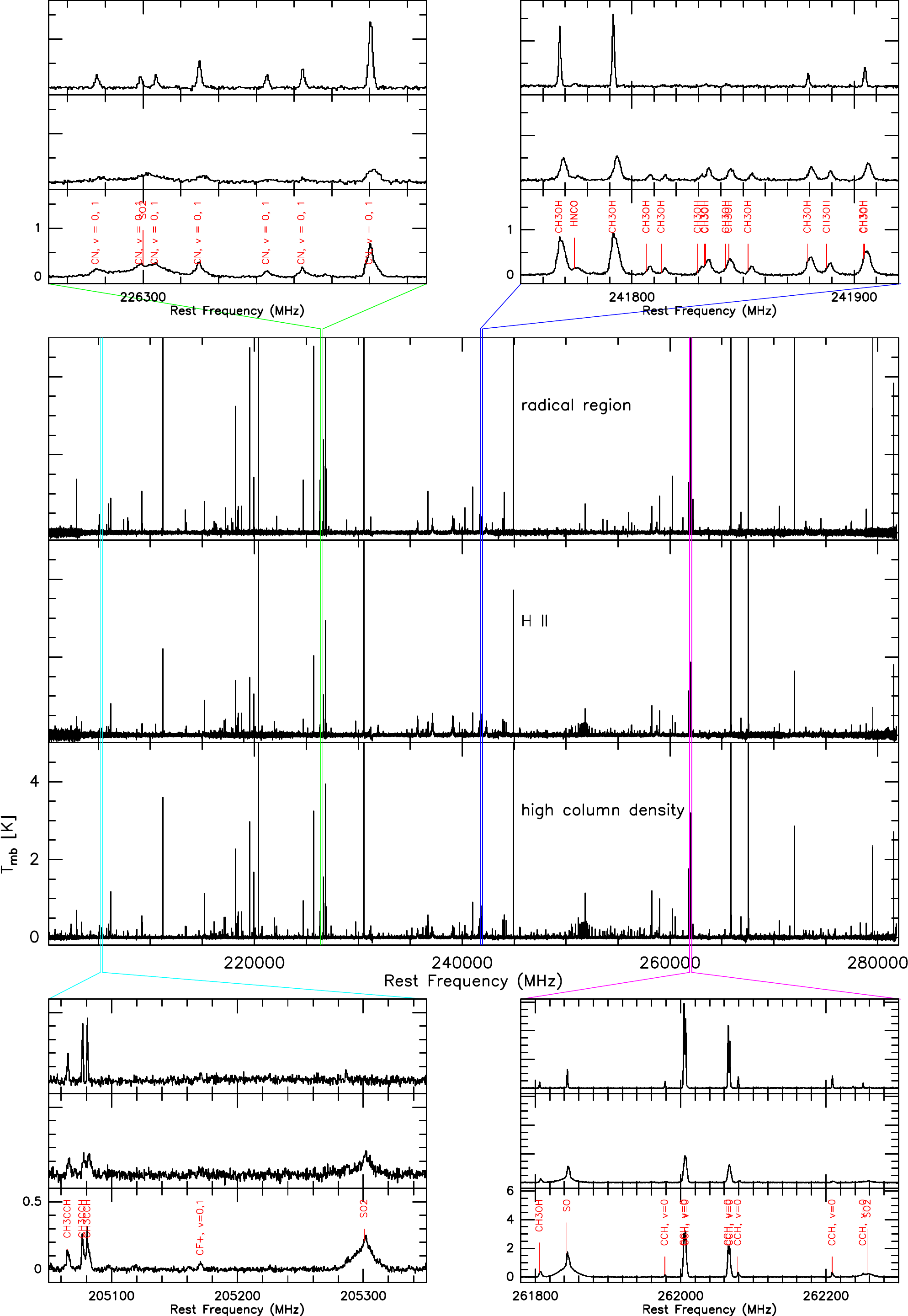}
        \caption{ Spectra obtained from different masks as described in Sections \ref{regions}
         and \ref{total_intensities}. 
        The restriction of the y-axis to 5 K for the middle panel is done to make 
        the weaker lines visible (the CO line has a peak temperature of roughly 57 K), while the
        zoomed-in boxes illustrate some of the molecular variety.}
        \label{fig:whole_spectrum}
\end{figure*}

Our choice of masks aims to assess  the effects of temperature and column density, for example. For the reasons described in Section \ref{total_intensities},
KL and Orion South are excluded in some masks, despite matching the column density or temperature criteria.
We have masks with only one condition (`high column density' vs. `low column density',
`high temperature without KL and Orion South' vs. `low temperature'), but also more restrictive masks, which
also define other parameters (`high column density, low temperature' vs. `low column density, 
low temperature' vs.
`low column density, high temperature' vs. `high column density, high temperature without 
KL and Orion South'). The
masks `high column density' and `high column density without KL and Orion South' intend 
to show how the spatially quite compact
emission of KL and Orion South influences the larger scale emission. Additionally, UV irradiation is 
considered in the masks `\ion{H}{ii}' and `dense PDR'. 
Choosing this `dense PDR' traced by radio recombination lines, instead of a more `diffuse PDR' traced 
by \ion{C}{ii} (\citealt{Pabst2019}), has the advantage of the edge-on view in Orion. This means we will only probe the 
actual PDR and are not affected by projection effects.
Earlier papers (\citealt{Turner1977}) pointed out that the `radical region' may be a chemically 
distinct environment within 
OMC-1, although subsequent observations were not conclusive (\citealt{Greaves1992}). 
We decided to include this region in our analysis, hoping that our unbiased data set might advance its 
characterisation. Column density and temperature values listed in Table \ref{table_regions} for the `radical 
region' are thus not its defining features.
Lastly, using no mask at all (`all averaged') and employing
the results from the other regions, we can evaluate which features or regions dominate on the largest
considered scale and which might disappear, indicating that those latter regions may be
hard to identify in other, spatially unresolved sources.

The exact values we chose for the column density and temperature thresholds
($N = 3.6 \times 10^{22}~\mathrm{cm}^{-2}~, ~ T= 25 ~ \mathrm{K}$) are somewhat arbitrary. We selected 
them such that they divide Orion A broadly into the denser parts of the filament ($N \geq 3.6 \times
10^{22}~\mathrm{cm}^{-2}$) and the ambient material ($N < 3.6 \times
10^{22}~\mathrm{cm}^{-2}$), and distinguish between the historically established regions OMC-1  
($T > 25 ~ \mathrm{K}$) and OMC-2/3 ($T < 25 ~ \mathrm{K}$). Based on these basic distinctions, 
combinations of masks are added and complemented by the `\ion{H}{ii}', `dense PDR' and `radical region' masks.

Our masks are thus not devised to divide Orion A into strictly disjointed regions, but to explore the emission of 
template regions meeting physical or chemical conditions. A listing of all overlaps between the selected regions is
given in Table \ref{table_overlaps}.

\begin{landscape}
\begin{table}
\small
\caption{Selected regions and their fundamental properties.}        
\label{table_regions}     
\centering 
\begin{tabular}{l l c c c c c l}     
\hline\hline
region & explicit mask & approx. size & column density range & median column density & temperature range 
& median temperature &
label$^{(\dag)}$
\\
 % table heading
& & [arcmin$^{2}$] & $\lbrack \mathrm{cm}^{-2} \rbrack$ & $\lbrack \mathrm{cm}^{-2} \rbrack$ & $\lbrack 
\mathrm{K} \rbrack$ & 
$\lbrack \mathrm{K} \rbrack$
\\ [3pt]
\hline                        
\\
all averaged & no mask & 210.0 & $9.3 \times 10^{20} ~ - ~ >1.4 \times 10^{23}$  $^{(*)}$ 
& $1.0 \times 10^{22}$ $^{(*)}$ & 17 - 53 $^{(*)}$ & 25 $^{(*)}$ & 1 
\\  \\
high column density & $N \geq 3.6 \times 10^{22}~\mathrm{cm}^{-2}$ & 23.3 & $3.6 \times 10^{22} ~ - ~ >1.4 \times 10^{23}$
$^{(*)}$ & $4.9 \times 10^{22}$ $^{(*)}$ & 17 - 51 $^{(*)}$ & 27 $^{(*)}$ & 2
\\ \\
high column density without & $N \geq 3.6 \times 10^{22}~\mathrm{cm}^{-2}$ & 21.5 & $3.6 \times 10^{22} ~ - ~ 1.4 \times
10^{23}$ & $4.9 \times 10^{22}$ &  17 - 51 & 27 & 3 
\\
KL and Orion South & &  
\\ \\
high column density, & $N \geq 3.6 \times 10^{22}~~\mathrm{cm}^{-2}$ & 9.2 & $3.6 
\times 10^{22} ~ - ~ 9.3 \times 10^{22}$ & $4.7 \times 10^{22}$ & 17 - < 25  & 19 & 4
\\
low temperature & $~ T < 25 ~ \mathrm{K}$ & & 
\\ \\
high column density, high & $N \geq 3.6 \times 10^{22}~\mathrm{cm}^{-2}$ & 11.5 & $3.6 
\times 10^{22} ~ - ~ 1.4 \times 10^{23}$ & $5.5 \times 10^{22}$  & 26 - 51 & 32 & 5
\\
temperature, without KL  & $~ 25 < T < 55 ~ \mathrm{K}$ & &  
\\
and Orion South & & 
\\ \\
low column density & $ N < 3.6 \times 10^{22}~\mathrm{cm}^{-2}$ & 186.7 & $9.3 \times 10^{20} ~ - ~ < 3.6 \times 10^{22}$ 
& $9.0 \times 10^{21}$ & 17 - 53 & 25 & 6
\\ \\
low temperature & $T < 25 ~ \mathrm{K} $ & 102.1 & $2.1 \times 10^{21} ~ - ~ 9.3 \times 10^{22}$ 
& $1.2 \times 10^{22}$ & 17 - < 25 & 21 & 7
\\ \\ 
low column density,  & $N < 3.6 \times 10^{22}~~\mathrm{cm}^{-2}$ & 92.9 & $2.1\times 
10^{21} ~ - ~ < 3.6 \times 10^{22}$& $1.1 \times 10^{22}$ & 17 - < 25 & 22 & 8 
\\
low temperature & $~ T < 25 ~ \mathrm{K}$ & &
\\ \\
low column density, & $N < 3.6 \times 10^{22}~~\mathrm{cm}^{-2}$ & 93.7 & $9.3\times 
10^{20} ~ - ~ < 3.6\times 10^{22}$ & $7.5 \times 10^{21}$ & 25 - 53 & 34 & 9
\\
high temperature & $~ 25 < T < 55 ~ \mathrm{K}$
\\ \\
high temperature,  & $25 < T < 55 ~ \mathrm{K}$ & 105.2 & $9.3 \times 10^{20} ~ - ~ 1.4 \times 10^{23}$ 
& $8.5 \times 10^{21}$ & 25 - 53 & 33 & 10
\\
without KL and Orion South & &   
\\ \\
\ion{H}{ii} & high $\mathrm{H31}\alpha$ emission $^{(**)}$ & 6.4 & $7.9 \times 10^{21}- ~ 6.7 \times 
10^{22}$ $^{(*)}$& $1.7 \times 10^{22}$ $^{(*)}$& 36 - 51$^{(*)}$ & 42 & 11
\\ \\
radical region $^{(***)}$ & $130\arcsec \times 130\arcsec$ around & 5.1 & $1.8 \times 10^{22} ~ - ~ 1.4 \times 
10^{23}$ & $5.8 \times 10^{22}$ & 27 - 38 & 30 & 12 
\\ 
& $\alpha_{2000} = 5^{\mathrm{h}} 35^{\mathrm{m}} 16.51^{\mathrm{s}}$ 
 & &  
\\
& $\delta_{2000} = -5^{\circ} 19^{\arcmin} 33.0^{\arcsec}$ & & 
\\
\\
dense PDR & $N < 3.6 \times 10^{22}~~\mathrm{cm}^{-2}$ & 4.7 & $8.8 \times 10^{21} ~ - ~ 3.5 \times 10^{22}$ 
& $1.8 \times 10^{22}$ & 37 -  51 & 44 & 13 
\\
& high $\mathrm{C}65\alpha$ emission  $^{(****)}$&  
\\ \\
\hline    
\end{tabular}
\tablefoot{(\dag) To be used as an identifier in later plots. (*) The area around Orion KL and Orion 
South is saturated, thus the given maximum column density,
maximum temperature, and median values are actually lower limits. (**) Mask based on an integrated [-20, 15] 
$\mathrm{km ~ s}^{-1}$ intensity image of the $\mathrm{H31}\alpha$ emission at $210501.788 ~ \mathrm{MHz}$, 
where pixels with an integrated intensity over $ \approx 3.1 ~ \mathrm{K ~ km ~ s}^{-1}$ were selected.
(***) Coordinates based on \citet{Greaves1992}. (****) Mask based on an integrated intensity image of the 
$\mathrm{C65}\alpha$ emission at $23415.9609 ~ \mathrm{MHz}$ from \citet{Wyrowski1997}, where pixels with 
an integrated intensity over $ 4 ~ \mathrm{K ~ km ~ s}^{-1}$ were selected.}
\end{table}
\end{landscape}

%%%%%%%%%%%%%%%%%%%%%%%%%%%%%%%%%%%%%%%%%%%%%%%%

\subsection{Obtaining total intensities}\label{total_intensities}

The total intensity of different molecular species for the different regions was determined in three steps.
Firstly, the spectrum of the region was extracted from our data cubes using the masks described in section
\ref{regions}. 
All spectra within a mask were combined into a single spectrum, the `regional spectrum' or average.
The idea is that local effects should average 
out in this step when the area is large, while common characteristic
features will add up.
Smoothing along the frequency axis to a resolution of around $305 ~ \mathrm{kHz}$ 
was done to improve the signal-to-noise ratio.
Three of these averaged spectra are  shown as examples in Fig. \ref{fig:whole_spectrum}.

In the second step, the spectral lines listed in  Table \ref{all_transitions} were considered in Gaussian 
fits. Line-of-sight velocities do not vary significantly within the mapped area, from $\sim 9 ~ \mathrm{km~s}^{-1}$ 
for most of OMC-1 
(especially when excluding KL and Orion South with slightly lower velocities) and increasing
to $\sim 12 ~ \mathrm{km~s}^{-1}$ towards OMC-3. Different velocity components thus overlap in a spectrum. 
The list of lines is based on the spectrum obtained with the high column density mask (which includes Orion KL and Orion South),
where we expect the greatest number of lines. 
Omitted are complex molecules with the exception of
CH$_{3}$OH, CH$_{3}$CN, and CH$_{3}$CCH. The lines of  CH$_{3}$OCH$_{3}$ or CH$_{3}$OCHO, for example,  are numerous but
weak and their contribution to the overall emission as typical hot core molecules should be negligible on 
larger scales. 
The main goal is to capture most of the line emission, which in some cases means that not each individual transition
is fitted. The automated fitting is more robust when strongly overlapping lines are not fitted separately, 
but covered in a single Gaussian fit. This favouring of robustness was abandoned when the overlapping lines 
belonged to different species. 
In that case, the lines were fitted separately. Overlap between lines becomes an issue whenever 
KL and Orion South are involved, hence the decision to only include them in two larger regions, as
automated fitting is not feasible otherwise.

In the third and final step all lines above $5\sigma$ with reasonable widths and velocity were considered to be real 
detections and the area of their fit added to the 
total intensity of their respective species.

Noise levels vary within a single spectrum depending on frequency, and between different spectra
due to the varied spatial extension of the regions. This influences the possibility to detect very weak 
lines and will be addressed further in Fig. \ref{fig:total_intensities}.

%%%%%%%%%%%%%%%%%%%%%%%%%%%%%%%%%%%%%%%%%%%%%%%%

\section{Results}\label{results}

The averaged total intensity for each species and region is listed in Table \ref{table_all_total_intensities},
together with the detection limit for each region.
Owing to space constraints, the formal fit errors are listed separately  in   Table 
\ref{table_all_total_intensities_errors}, but they usually amount to $2-6\%$
for all species and regions. Due to the overall low noise levels and the high number of
free parameters, the fit errors for a single line are very small. The actual uncertainties are thus 
dominated by the calibration uncertainties discussed in Section \ref{obs_calis}.

The high column density region has the highest averaged total intensity with $\sim 1360 ~ \mathrm{K ~ km ~ s}^{-1}$. 
Removing just the region around KL and Orion South has a significant impact on the overall emission 
and reduces the averaged total intensity by $48\%$, although the removed area represents only $8\%$ of 
the high column density region. The regions `high column density, high temperature, without KL and Orion South' and 
`\ion{H}{ii}' have comparable averaged total intensities ($\sim 954 ~ \mathrm{K ~ km ~ s}^{-1}$ and
$\sim 925 ~ \mathrm{K ~ km ~ s}^{-1}$), which are a factor of $\sim3$ above average
($\sim 319 ~ \mathrm{K ~ km ~ s}^{-1}$). The averaged total CO intensity is also similar for these two
regions, their differences lying in the species with less emission. The `dense PDR' and `radical' regions both 
emit a factor of $\sim2$ above average, but their share of CO emission differs by around  $20\%$, 
indicating that their emission profiles (meaning the breakdown of the total emission into contributions from
different species) are distinct from each other. The `high temperature without KL
and Orion South' region is most similar to the `averaged' region not only in terms of total intensity, 
but also regarding the most prominent species. The remaining regions, including `high column density, low
temperature', emit below average.

\begin{figure*}
    \centering
\begin{subfigure}[t]{\textwidth}
\centering
        \includegraphics[width=11cm, angle=90]{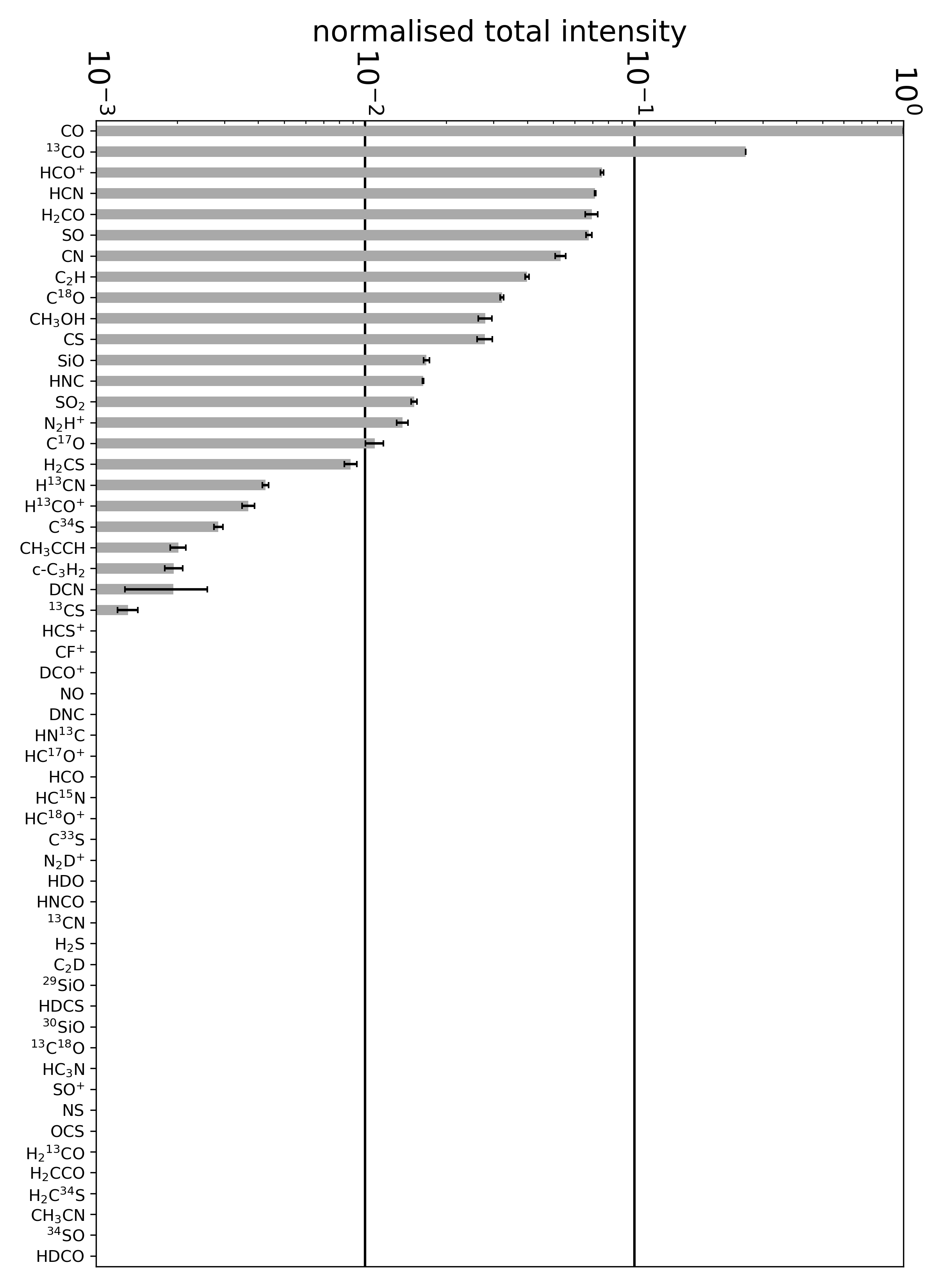}
        \caption{ all averaged normalised total intensities}
        \label{fig:ti_averaged}
    \end{subfigure}
    ~
    \begin{subfigure}[t]{\textwidth}
    \centering
        \includegraphics[width=0.55\textwidth]{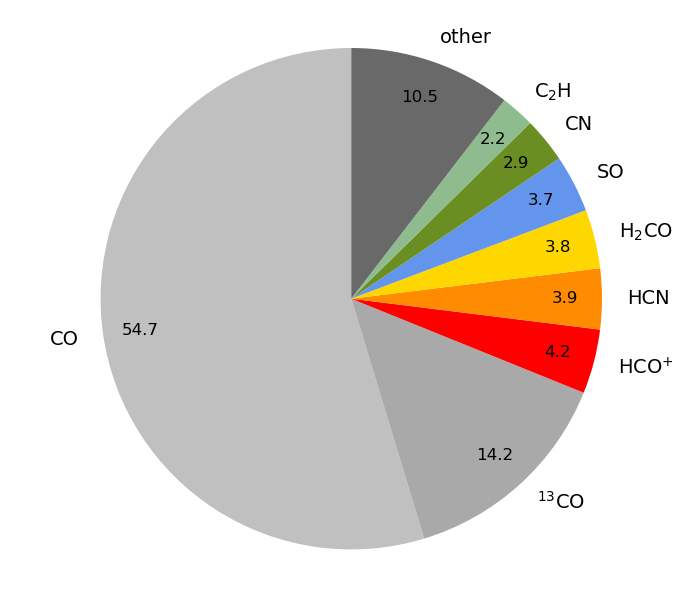}
        \caption{ all averaged pie chart}
        \label{fig:pie_averaged}
    \end{subfigure}

    \caption{Normalised total intensity and pie chart for our averaged data. The order of species in the
    normalised plot is also used for every other region (see Fig. \ref{fig:total_intensities}). The pie chart
    gives  the percentage of the total intensity originating from different
    species. Shares under $2\%$ are summed  under `other'.}\label{all_averaged_charts}
\end{figure*}

For easier comparison, hereafter the data is   presented in three complementing ways, each highlighting
different aspects.
Figure \ref{all_averaged_charts} shows the visualisation for the different species with 
the example of our average data (where no mask was used). 
The normalised intensity of all species in descending order is shown in Fig. \ref{fig:ti_averaged} with a 
logarithmic plot, down to $0.1\%$ of the CO intensity.
This order of species is used in the corresponding total intensity plots for all other regions (see 
Fig. \ref{fig:total_intensities}). For the comparison of regions, this representation of the data 
helps to make shifts in the influence of species more apparent.
The pie charts (Fig. \ref{fig:pie_averaged} and Fig. \ref{pie_charts}) visualise the percentage of the 
total intensity originating from different species, with shares under $2\%$ summed  under `other'. 
Plots in this form are more conducive from an observational point of view,
as they highlight the most dominant and accessible species. 

The third visualisation (Fig.\ref{fig_share_heatmap}) concentrates more on a comparison between species. 
The plot again shows 
the percentaged share of each molecule, while the colour bar 
helps to highlight in which region each species has its largest or smallest share and which species are
overall stronger or weaker emitters.

Of the 55 species considered in our analysis, 15 are  seen to contribute over $2\%$ each of the total intensity 
for at least one region. Members of these 15 species account for around $88\%$ to $94\%$ of 
line emission in the $1.3 \ \mathrm{mm}$ window in all cases (or $5.6\%$ to $53\%$ when excluding CO and its 
isotopologues). 
In addition to CO and $^{13}$CO, only HCO$^{+}$ and
H$_{2}$CO are prominent in every region, each contributing between $3\%$ and $6\%$ in all cases.

In terms of averaged total intensity, there is a factor of $\sim 8$ between the lowest (`low column density, 
low temperature') and highest (`high column density') regions. 
The high column density region has the lowest share of CO and the highest diversity of species noticeably involved
in cooling. SO$_{2}$, with a share of
$11.4\%$ being the most important coolant after CO in the high column density region, is no longer relevant when KL 
and Orion South are removed (its share drops to around $1\%$; see also Fig. \ref{fig:pie_hd},
\ref{fig:pie_hd_wo_kl} and Fig. \ref{fig:ti_hd}, \ref{fig:ti_hd_wo_KL}). SO and CH$_{3}$OH show similar trends. 
In the high column density region, their total intensities exceed that of the typically strong emitters HCN, 
HCO$^{+}$, and H$_{2}$CO, while they lose importance without KL and Orion South.

The regions `high column density, low temperature' and `high column density, high temperature, without KL and Orion South'
are parts of `high column density without KL and Orion South' (see also Fig. \ref{fig:masks}), coinciding with
OMC-2/OMC-3 and OMC-1, respectively. Both in terms of averaged total intensity and as suggested by the pie
charts and detailed normalised intensities (Fig. \ref{fig:pie_hd_wo_kl}, \ref{fig:pie_lt_hd},
\ref{fig:pie_ht_hd_wo_KL} and Fig. \ref{fig:ti_hd_wo_KL}, \ref{fig:ti_lt_hd}, \ref{fig:ti_ht_hd_wo_KL}),
OMC-1 dominates the emission. Higher N$_{2}$H$^{+}$ emission, indicative of OMC-2/OMC-3, is not seen in
the `high column density without KL and Orion South' region, which is instead very similar to OMC-1 alone.

The regions `low column density', `low temperature', `low column density, low temperature' and `low column density, high 
temperature' are similar in the sense of high CO shares ($60\%$ to $67\%$ for CO, 
$15\%$ to $21\%$ for $^{13}$CO) and only two to four species over
$2\%$ (Fig. \ref{fig:pie_low_density}, \ref{fig:pie_lt}, \ref{fig:pie_lt_ld}, and \ref{fig:pie_ht_ld}). 
Their averaged total intensities are similar and below average. Their distinctions lie in the lower
intensity species (Fig. \ref{fig:ti_ld}, \ref{fig:ti_lt}, \ref{fig:ti_lt_ld}, and \ref{fig:ti_ht_ld}).

The regions `high temperature, without KL and Orion South' and `dense PDR' are similar to each other and---with the
the exception of SO emission and slightly different CO shares---to the `averaged' region (Fig.
\ref{fig:pie_ht_wo_KL}, \ref{fig:pie_bar}, \ref{fig:pie_averaged}). 
The averaged total intensity is  noticeably higher in the `dense PDR' region than in the two others. The 
differences between regions lie again in the fainter species (Fig. \ref{fig:ti_ht_wo_KL}, \ref{fig:ti_bar},
\ref{fig:ti_averaged}). 

The `\ion{H}{ii}' region is the only other region, in addition to  `high column density',  with relevant SO$_{2}$
emission and also has  higher shares of CH$_{3}$OH and SO, which is
the dominant coolant after CO and $^{13}$CO.
The `radical region' has the highest shares of C$_{2}$H and CN and is the only other region besides 
`high column density, low temperature' (dense part of OMC-2/OMC-3) with notable shares of N$_{2}$H$^{+}$.

\begin{table*}
\tiny
\caption{Averaged total intensities $\int T_{\mathrm{mb}}dv$ $\lbrack \mathrm{K~km ~ s}^{-1} \rbrack$ for 
all regions and species. The detection limit is based on the median line width for each region and assumes a
$5\sigma_\mathrm{median}$ feature. Real detections below this limit may occur, for example when  
$\sigma_\mathrm{local}<\sigma_\mathrm{median}$ or when the line is narrow.}   
\label{table_all_total_intensities}     
\centering 
\begin{tabular}{l | c c c c c c c c c c c c c}
\hline\hline   % table heading
\\
species 
& \rotatebox{90}{all averaged} 
&\rotatebox{90}{high column density} 
& \rotatebox{90}{high column density without}
\rotatebox{90}{KL and Orion South} 
& \rotatebox{90}{high column density,} \rotatebox{90}{low temperature} 
& \rotatebox{90}{high column density,} \rotatebox{90}{high temperature,}
\rotatebox{90}{without KL and Orion South}
& \rotatebox{90}{low column density} 
& \rotatebox{90}{low temperature} 
&\rotatebox{90}{low column density,} \rotatebox{90}{low temperature}
& \rotatebox{90}{low column density,} \rotatebox{90}{high temperature}
& \rotatebox{90}{high temperature}\rotatebox{90}{without KL and Orion South}
& \rotatebox{90}{\ion{H}{ii}} & \rotatebox{90}{radical region} & \rotatebox{90}{dense PDR}
 
\\
\hline                       
\\
CO               & 174.21& 393.21 &  303.48 & 137.13 & 405.67 & 146.64 & 112.19 & 110.06 & 187.64 &  209.34 & 431.08 & 279.67 & 391.86 \\ 
$^{13}$CO        & 45.19 & 90.51  &  81.37  & 46.54  & 103.78 & 39.91  & 38.91  & 38.10  & 41.71  &  48.31  & 69.57  & 81.02  & 71.17  \\
C$^{18}$O        & 5.61  & 12.82  &  11.22  & 9.00   & 12.78  & 4.83   & 6.19   & 5.86   & 3.79   &  4.58   & 6.41   & 10.19  & 4.74   \\
C$^{17}$O        & 1.89  & 4.13   &  3.82   & 3.19   & 3.98   & 1.62   & 2.16   & 2.03   & 1.19   &  1.50   & 2.12   & 3.46   & 1.73   \\
$^{13}$C$^{18}$O & -     & 0.15   &  0.14   & 0.14   & 0.15   & -      & 0.11   & -      & -      &  -      & -      & 0.16   & -      \\
c-C$_{3}$H$_{2}$ & 0.34  & 3.27   &  2.71   & 0.36   & 4.09   & -      & -      & -      & 0.38   &  1.02   & 1.41   & 3.10   & 3.83   \\
C$_{2}$H         & 6.96  & 30.33  &  25.48  & 7.01   & 37.11  & 3.96   & 2.11   & 1.65   & 6.19   &  9.72   & 18.66  & 35.05  & 13.82  \\
C$_{2}$D         & -     & 0.77   &  0.70   & 0.36   & 0.86   & -      & -      & -      & -      &  -      & -      & 0.78   & -      \\
CF$^{+}$         & 0.13  & 0.25   &  0.21   & -      & 0.21   & 0.12   & 0.11   & -      & -      &  -      & -      & -      & -      \\
CH$_{3}$CCH      & 0.35  & 10.06  &  5.14   & 0.52   & 8.18   & -      & -      & -      & -      &  0.36   & 3.74   & 5.74   & -      \\
CH$_{3}$CN       & -     & 30.02  &  1.20   & -      & 1.25   & -      & -      & -      & -      &  1.36   & 1.73   & -      & -      \\
CH$_{3}$OH       & 4.87  & 87.74  &  17.96  & 4.05   & 25.79  & 0.73   & 0.58   & 0.43   & 0.85   &  1.90   & 40.63  & 17.96  & 0.92   \\
CN               & 9.26  & 36.02  &  26.50  & 9.33   & 38.77  & 5.62   & 3.29   & 2.05   & 4.76   &  8.38   & 29.60  & 35.05  & 17.98  \\
$^{13}$CN        & -     & 0.37   &  0.32   & -      & 0.48   & -      & -      & -      & -      &  -      & -      & 0.32   & -      \\
CS               & 4.86  & 29.57  &  18.69  & 3.27   & 27.50  & 2.03   & 1.09   & 0.96   & 3.05   &  5.88   & 18.63  & 16.48  & 8.96   \\
$^{13}$CS        & 0.23  & 2.47   &  1.33   & -      & 2.01   & -      & -      & -      & -      &  -      & 1.80   & 0.97   & 0.58   \\
C$^{34}$S        & 0.50  & 3.32   &  2.17   & -      & 3.23   & -      & -      & -      & 0.27   &  0.56   & 2.25   & 1.75   & 0.93   \\
C$^{33}$S        & -     & 0.84   &  0.44   & -      & 0.74   & -      & -      & -      & -      &  -      & 0.58   & 0.38   & 0.28   \\
HC$_{3}$N        & -     & 17.21  &  2.28   & -      & 3.23   & -      & -      & -      & -      &  -      & 6.57   & 0.84   & 0.27   \\
HCN              & 12.46 & 70.94  &  36.67  & 9.37   & 49.07  & 4.38   & 3.02   & 2.32   & 6.45   &  11.62  & 50.43  & 26.35  & 21.14  \\
DCN              & 0.34  & 2.51   &  1.51   & 0.35   & 1.96   & -      & -      & -      & -      &  0.33   & 1.69   & 1.09   & 0.34   \\
H$^{13}$CN       & 0.74  & 7.43   &  2.24   & 0.25   & 2.73   & -      & -      & -      & 0.19   &  0.53   & 2.47   & 1.61   & 0.70   \\
HC$^{15}$N       & -     & 2.42   &  0.56   & -      & 0.77   & -      & -      & -      & -      &  -      & 0.97   & 0.30   & -      \\
HCO              & -     & 0.13   &  0.15   & -      & 0.54   & -      & -      & -      & -      &  -      & -      & 0.25   & 0.43   \\
HCO$^{+}$        & 13.22 & 61.37  &  39.97  & 14.92  & 52.21  & 7.29   & 6.39   & 5.59   & 9.42   &  14.78  & 50.30  & 29.17  & 27.65  \\
DCO$^{+}$        & 0.12  & 1.06   &  0.81   & 0.97   & 0.64   & 0.07   & 0.22   & 0.15   & -      &  -      & -      & 0.48   & -      \\
H$^{13}$CO$^{+}$ & 0.64  & 3.31   &  2.55   & 1.27   & 3.09   & 0.15   & 0.36   & 0.27   & 0.30   &  0.62   & 2.11   & 2.30   & 0.44   \\
HC$^{18}$O$^{+}$ & -     & 0.30   &  0.21   & -      & 0.28   & -      & -      & -      & -      &  -      & 0.24   & 0.23   & -      \\
HC$^{17}$O$^{+}$ & -     & -      &  -      & -      & -      & -      & -      & -      & -      &  -      & -      & -      & -      \\
HCS$^{+}$        & 0.13  & 1.88   &  1.44   & -      & 2.15   & -      & -      & -      & -      &  0.37   & 1.21   & 1.60   & -      \\
HDO              & -     & 0.59   &  -      & -      & -      & -      & -      & -      & -      &  -      & -      & -      & -      \\
H$_{2}$CCO       & -     & 0.81   &  -      & -      & -      & -      & -      & -      & -      &  -      & -      & 0.16   & -      \\
H$_{2}$CO        & 12.10 & 64.62  &  41.28  & 17.95  & 53.55  & 6.15   & 5.75   & 4.42   & 7.75   &  12.51  & 40.88  & 36.38  & 21.44  \\
HDCO             & -     & 2.09   &  0.35   & 0.18   & 0.95   & -      & -      & -      & -      &  -      & 0.37   & 0.61   & -      \\
H$_{2}^{13}$CO & -     & 2.34   &  0.39   & 0.15   & 1.04   & -      & -      & -      & -      &  -      & 0.71   & 0.29   & -      \\
H$_{2}$CS        & 1.54  & 18.10  &  10.15  & -      & 16.35  & -      & -      & -      & 0.32   &  2.15   & 10.07  & 10.97  & 0.27   \\
HDCS             & -     & -      &  -      & -      & 0.30   & -      & -      & -      & -      &  -      & -      & 0.26   & -      \\
H$_{2}$C$^{34}$S & -     & -      &  -      & -      & -      & -      & -      & -      & -      &  -      & -      & -      & -      \\
H$_{2}$S         & -     & 2.30   &  -      & -      & 0.49   & -      & -      & -      & -      &  -      & -      & -      & -      \\
HNC              & 2.86  & 14.37  &  10.25  & 5.53   & 12.94  & 1.56   & 1.70   & 1.31   & 1.74   &  2.97   & 7.51   & 10.66  & 3.71   \\
DNC              & 0.08  & 0.63   &  0.50   & 0.53   & 0.56   & -      & 0.11   & -      & -      &  -      & -      & 0.52   & -      \\
HN$^{13}$C       & -     & 0.58   &  0.47   & 0.22   & 0.64   & -      & 0.06   & -      & -      &  -      & 0.27   & 0.58   & -      \\
HNCO             & -     & 2.44   &  -      & -      & 0.49   & -      & -      & -      & -      &  -      & -      & -      & -      \\
N$_{2}$H$^{+}$   & 2.40  & 11.14  &  10.67  & 9.75   & 11.40  & 1.28   & 2.37   & 1.62   & 0.84   &  2.05   & 2.67   & 13.39  & -      \\
N$_{2}$D$^{+}$   & -     & 0.19   &  0.20   & 0.44   & -      & -      & -      & -      & -      &  -      & -      & -      & -      \\
NO               & 0.10  & 2.23   &  1.01   & 0.44   & 1.24   & 0.08   & 0.25   & 0.24   & -      &  -      & 1.07   & -      & -      \\
NS               & -     & 2.26   &  1.66   & -      & 2.63   & -      & -      & -      & -      &  -      & -      & 2.66   & -      \\
OCS              & -     & 5.05   &  -      & -      & -      & -      & -      & -      & -      &  -      & 1.20   & -      & -      \\
SiO              & 2.94  & 27.77  &  5.83   & -      & 7.30   & -      & -      & -      & -      &  -      & 13.93  & -      & -      \\
$^{29}$SiO       & -     & 1.64   &  -      & -      & -      & -      & -      & -      & -      &  -      & -      & -      & -      \\
$^{30}$SiO       & -     & -      &  -      & -      & -      & -      & -      & -      & -      &  -      & -      & -      & -      \\
SO               & 11.81 & 129.66 &  32.39  & 2.91   & 42.62  & 0.97   & 0.98   & 0.77   & 1.61   &  4.28   & 65.82  & 8.00   & 9.64   \\
$^{34}$SO        & -     & 12.05  &  -      & -      & 0.85   & -      & -      & -      & -      &  -      & 1.16   & 0.09   & -      \\
SO$^{+}$         & -     & -      &  -      & -      & 0.59   & -      & -      & -      & -      &  -      & -      & 0.13   & 0.36   \\
SO$_{2}$         & 2.65  & 155.03 &  4.15   & -      & 6.57   & -      & -      & -      & -      &  -      & 35.27  & -      & -      \\
\hline    
sum : & $\sim319$ & $\sim1360$ & $\sim711$ & $\sim286$ & $\sim954$ & $\sim227$ & $\sim188$ & $\sim178$ & $\sim279$ & $\sim345$ & $\sim925$ 
& $\sim641$ & $\sim603$

\\
approx. detection \\
limit $\lbrack \mathrm{K~km ~ s}^{-1} \rbrack$ : & 0.17 & 0.29 & 0.22 & 0.16 & 0.20 & 0.15 
& 0.11 & 0.10 & 0.19 & 0.16 & 0.38 & 0.14 & 0.27
\\
\hline
\end{tabular}
\end{table*}

\subsection{Approximating the emission of KL and Orion South}\label{KL_emission}

While automated fitting procedures are problematic for high column density and high temperature 
environments with heavily overlapping lines, we can approximate
the emission around KL and Orion South from our existing regions. With the known pixel sizes of 
`high column density' and `high column density without KL and Orion South' we can gauge the emission of the 
region and compute its average. The result is not as robust as the others, as more lines are
expected for some species which are not accounted for in our routine (especially for SO$_{2}$, 
SO, and CH$_{3}$OH), and 
omitted complex organic molecules like CH$_{3}$OCH$_{3}$ presumably have non-negligible impact here.
However, the overall results, as presented in Table \ref{table_KL} and Fig. \ref{pie_KL}, are in 
general agreement with 
\citet{Schilke1997} for the 325 to 360 GHz frequency regime. The dominant species are in both cases
SO$_{2}$, followed by CO, SO, CH$_{3}$OH, and HCN.

\begin{table}
\tiny
\caption{Approximated averaged total intensities from the region around KL and Orion South}        
\label{table_KL}     
\centering 
\begin{tabular}{l@{\hskip 1cm} c}
\hline
\hline  
species & $\int T_{\mathrm{mb}}dv$ $\lbrack \mathrm{K~km ~ s}^{-1} \rbrack$   \\
\hline                       
\\
CO                  &   1464.9   \\
$^{13}$CO           &   199.6    \\
C$^{18}$O           &   31.9     \\
C$^{17}$O           &   7.8      \\
$^{13}$C$^{18}$O    &   0.3      \\
c-C$_{3}$H$_{2}$    &   9.9      \\
C$_{2}$H            &   88.2     \\
C$_{2}$D            &   1.5      \\
CF$^{+}$            &   0.7      \\
CH$_{3}$CCH         &   68.9     \\
CH$_{3}$CN          &   374.2    \\
CH$_{3}$OH          &   921.1    \\
CN                  &   149.8    \\
$^{13}$CN           &   1.0      \\
CS                  &   159.4    \\
$^{13}$CS           &   16.1     \\
C$^{34}$S           &   17.0     \\
C$^{33}$S           &   5.7      \\
HC$_{3}$N           &   195.5    \\
HCN                 &   480.3    \\
DCN                 &   14.5     \\
H$^{13}$CN          &   69.5     \\
HC$^{15}$N          &   24.6     \\
HCO                 &   -        \\
HCO$^{+}$           &   317.0    \\
DCO$^{+}$           &   4.1      \\
H$^{13}$CO$^{+}$    &   12.4     \\
HC$^{18}$O$^{+}$    &   1.4      \\
HC$^{17}$O$^{+}$    &   -        \\  
HCS$^{+}$           &   7.0      \\
HDO                 &   7.7      \\
H$_{2}$CCO          &   10.4     \\
H$_{2}$CO           &   343.3    \\
HDCO                &   22.9     \\
H$_{2}^{13}$CO      &   25.6     \\
H$_{2}$CS           &   113.1    \\
HDCS                &   -        \\
H$_{2}$C$^{34}$S    &   -        \\
H$_{2}$S            &   29.8     \\
HNC                 &   63.6     \\
DNC                 &   2.1      \\
HN$^{13}$C          &   2.0      \\
HNCO                &   31.6     \\
N$_{2}$H$^{+}$      &   16.7     \\
N$_{2}$D$^{+}$      &   0.1      \\
NO                  &   16.7     \\
NS                  &   9.3      \\
OCS                 &   65.3     \\
SiO                 &   289.8    \\
$^{29}$SiO          &   21.2     \\
$^{30}$SiO          &   -        \\
SO                  &   1291.6   \\
$^{34}$SO           &   156.0    \\
SO$^{+}$            &   -        \\
SO$_{2}$            &   1957.2   \\
\\
\hline    
total & 9120.2
\\
\hline    
\end{tabular}
\end{table}

\subsection{Correlations and line ratios on large scales}\label{correlation}

In the following analysis of correlations and line ratios, the integrated intensity for a given
species always refers to the sum of 
all of its considered transitions in the $1.3~\mathrm{mm}$ window listed in Table \ref{all_transitions}. 
For some species (CO, $^{13}$CO, C$^{18}$O, HCO$^{+}$, H$^{13}$CO$^{+}$, HCN, HNC, N$_{2}$H$^{+}$, and CS) 
this involves only one transition,
but includes several for others (C$_{2}$H, CN, CH$_{3}$OH, H$_{2}$CO, and SO). 
For the discussion of line ratios, we use simplified quantum numbers for species with one transition, and 
mark those species with multiple transitions with the letter $\Sigma$.
Not all energy levels, for example 
of  CH$_{3}$OH, will be populated in all regions, but its integrated intensity still carries 
meaning as a measure of the cooling in the examined frequency range. Furthermore, neighbouring transitions 
from complex species like CH$_{3}$OH will strongly overlap in the broad lines from extragalactic sources, 
thus also inherently limiting the transferability of results concerning single or a few transitions only.

Correlations between species do not only point to similarities in their physical and chemical behaviour, 
but also reduce the number of necessary transitions for gauging the conditions in a molecular cloud. If the 
emission
of two species correlates strongly it is possible, for example, to   limit the frequency coverage to one species and save
observation time.
We limit our analysis to 14 species that are typically used as tracers.

\begin{figure*}
\centering
    \includegraphics[width=0.76\textwidth]{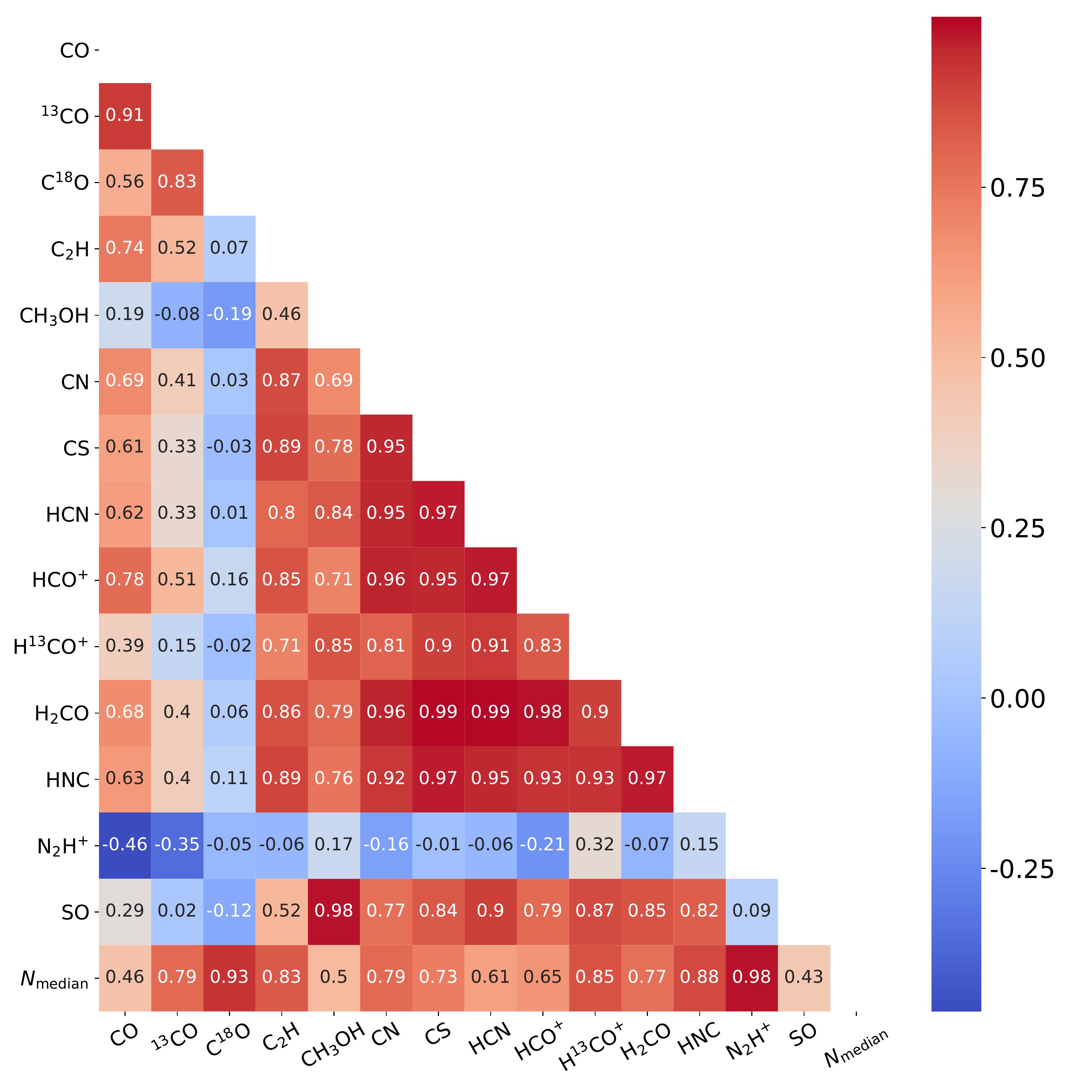}
    \caption{Pearson product-moment correlation coefficients between prevalent species (normalised
with the median column density) and between prevalent species and the median column density.}
    \label{table_coeff}
\end{figure*}

All emission correlates to the first order with column density (see also Fig.\,\ref{table_coeff}), which 
will thus influence the correlation between 
species. To reduce this effect in our analysis of the 13 regional spectra, we divided the total 
intensity for each species and region by the respective
median column density for the region (see Table \ref{table_regions}). 
This is not as good as a pixel-by-pixel normalisation, but it is  more feasible, especially for SO with its
numerous partially overlapping lines. Furthermore, the number of 
data points is small 
(13 regions), meaning we do not have enough statistical data for a truly reliable correlation coefficient. 
However, we mainly hope to distinguish 
between species with strong correlation and those without correlation. 
Linear correlations were measured with the Pearson correlation
coefficient\footnote{Calculated using scipy.stats.pearsonr from the SciPy library \citep{scipy} in Python.}
and are shown in Fig. \ref{table_coeff}, while selected plots are presented in Figs. \ref{fig:correlations} and
\ref{fig:correlations_density_temp}. 
Using our averaged (and therefore unresolved) regional data adds uncertainty for
the interpretation compared to a pixel-by-pixel analysis (e.g. the discussed column density value, which is 
only an approximation for the whole region).
On the other hand, 
our averaged data enables us to include species like SO and CH$_{3}$OH even in low column density and low-temperature environments.

\subsubsection{Correlations and tracers}

We expect chemical effects to influence correlation, and also different optical depths
of the species. 
We do not correct for the latter, as our aim is to stay as close to the data an observer might receive 
from an unresolved source, where 
the assumptions needed to correct for optical depth may add additional uncertainty. 

Strong correlations ($\gtrsim 0.90$) are found between the typical high-density tracers HCN, HCO$^{+}$, 
H$_{2}$CO, HNC, CS, but also CN.
There is generally little spread between the data points in the correlation plots, for example between   
H$_{2}$CO and HCO$^{+}$ (Fig. \ref{fig:corr_h2co_hco+}) or H$_{2}$CO and HCN
(Fig. \ref{fig:corr_hcn_h2co}), but the `\ion{H}{ii}' region with its high integrated intensities 
(after normalisation) constitutes a more isolated data point. 

To a lesser degree, correlations are also found between H$^{13}$CO$^{+}$ and the other high-density tracers 
(Fig. \ref{fig:corr_h13co+_hnc}). However, the integrated intensity of the former is often an order of
magnitude smaller than for the latter, making its observation more challenging.

Not only is CN correlated with $\mathrm{C}_{2}\mathrm{H}$, which is often associated with UV irradiation 
(\citealt{Nagy2015} and references therein),
but it is also correlated with high-density tracers (e.g. Fig. \ref{fig:corr_cn_cs},\ref{fig:corr_c2h_cn}).
Another strong correlation is found between the shock tracers (e.g. \citealt{Bachiller1997},
\citealt{Sakai2012}, \citealt{Nagy2015} and references therein, e.g. \citealt{Wakelam2004}) SO and
$\mathrm{CH}_{3}\mathrm{OH}$ (Fig. \ref{fig:corr_shock_tracer}).

For the evaluation of correlations between species and column density in this case we have to factor in that
the intensities are not normalised (with the median column density). Hence, it is possible
to see two species correlating strongly with column density, but not with each other after normalisation 
(see C$^{18}$O and N$_{2}$H$^{+}$). 
Of all the considered species, N$_{2}$H$^{+}$ shows the strongest correlation with the median column density 
(Fig. \ref{fig:corr_density_n2h+}), the outlier being the
non-detection ($<5\sigma$) in the `dense PDR' region, where N$_{2}$H$^{+}$ is most 
likely expected to be  destroyed. 
We also see a strong correlation between C$^{18}$O and the median column density (Fig. \ref{fig:corr_c18o_density}),
but the data points show a wider spread. 
When considering slope (and intercept) of the linear fit 
between species and median column density, we see that N$_{2}$H$^{+}$ reacts more sensitively to changes 
in column density.
From the more luminous 
high-density tracers, HNC may correlate the most with the median column density, in agreement with
\citet{Pety2016}.

CO and its isotopologues do not show clear correlations with other species (but they do with each other). This
may be explained to some degree by optical depth effects, meaning that CO mainly traces the surface 
of the cloud, while other species probe deeper layers. Especially for C$^{18}$O, which is unlikely to be 
optically thick, depletion may be an important factor. High-density tracers generally 
profit from higher (volume) densities, while C$^{18}$O may freeze out in these environments 
if the temperatures are low (as the chance to collide with and stick to dust grains is increased). 
N$_{2}$H$^{+}$ only correlates with the median column density.

\subsubsection{Line ratios}

A related important diagnostic, especially for extragalactic observations, but also for molecular 
clouds in the Milky Way, are line or integrated intensity ratios. They are more reliable tracers of
physical or chemical conditions than lines from a single 
species as calibration errors can cancel out for ratios, depending on the observation technique.
These ratios can also be examined with our data.

\begin{figure*}[ht]
%\centering
%\vspace{3cm}
    \includegraphics[width=\textwidth]{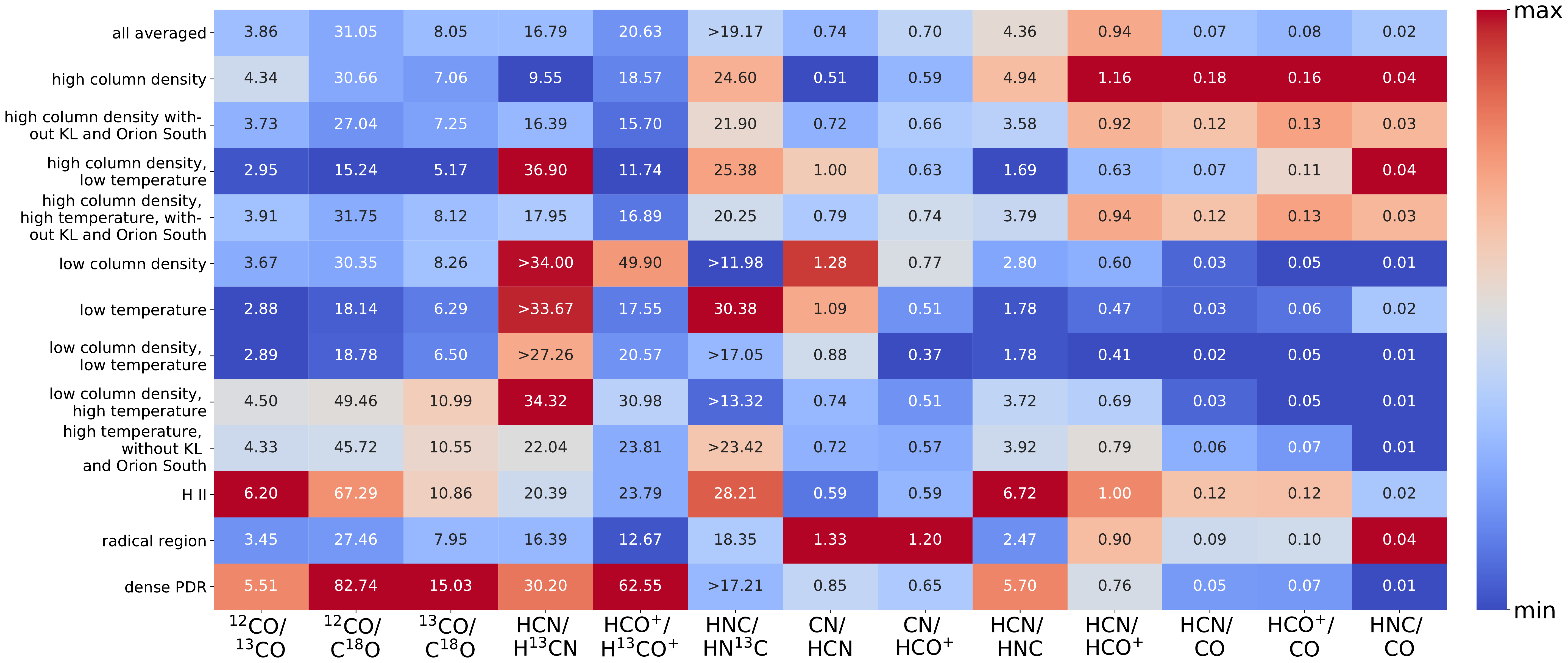}
    \caption{Integrated intensity ratios of selected species for the different regions.
    Lower limits are based on a hypothetical line with the median line width for the respective 
region and a peak intensity of five times the local noise level. Assuming an uncertainty of 30\% for the intensities,
all ratios have an uncertainty of $\sim42\%$. The maximum and minimum value of the colour bar are not defined globally,
but for every ratio (column) individually.}
\label{table_isotope_ratios}
\end{figure*}

We will examine some integrated intensity ratios
which have also been discussed in  \citet{Pety2016}, \citet{Gratier2017}, or
\citet{Bron2018}, among others, in the 3 mm window. The considered transitions are thus different, 
but general trends (e.g. HCN/HNC value increasing with temperature) may still 
be seen with our data. Selected integrated intensity ratios are compiled in Fig. \ref{table_isotope_ratios}.

Our sample supports the notion from \citet{Pety2016} that the CN/HCN ratio is not a reliable tracer
of UV illuminated gas. As the photodissociation of HCN produces CN, one might expect the highest ratio 
in the `dense PDR' region of our data
(CN$(\Sigma)$/HCN$(3-2) \approx  0.9$). Instead we find it in the `low column density' and 
`radical region' environments (both $\approx 1.3$). 
The lowest ratio is found for the `high column density' region ($\approx 0.5$).
The CN$(1-0)$/HCO$^{+}(1-0)$ ratio, discussed in \cite{Bron2018}, is suggested to 
help distinguish UV-illuminated gas from shielded gas,
with higher values associated with higher illumination. We find the lowest value
(CN$(2-1)$/HCO$^{+}(3-2) \approx 0.4$) 
for our `low column density, low temperature' region, the highest 
($\approx ~ 1.2$) in the `radical region', not with the expected `dense PDR' ($\approx ~ 0.7$). 
So at least
on our examined large scales and without correction for optical depth, using this ratio in the 
1.3\,mm window to trace UV illumination seems difficult too.

The ratio HCN/HNC should increase with temperature (\citealt{Pety2016} and references therein, e.g. 
\citealt{Graninger2014}) due to HNC reacting with H to form HCN at temperatures $\gtrsim 30~\mathrm{K}$. 
Indeed we find the three lowest values 
(HCN$(3-2)$/HNC$(3-2) = 1.7 - 1.8$) for our regions associated with the lowest 
median temperatures, while the highest two 
values of $5.7 - 6.7$ coincide with the highest temperatures, found in our case in the
`\ion{H}{ii}' and `dense PDR' regions. Larger deviations from this trend are found for  the 
`radical region', for example (see also Fig. \ref{fig:corr_temp_HCN-HNC}).

In their PCA of Orion B data in the 3\,mm window, \cite{Gratier2017} found that a higher 
N$_{2}$H$^{+}(\mathrm{J}=1 - 0, \mathrm{F}1=2 - 1,\mathrm{F}=3 - 2,)$/
CH$_{3}$OH$(\mathrm{J}=2 - 1, \mathrm{K}=0 - 0,\mathrm{A}^{+})$
ratio possibly highlights the chemistry of the 
densest cores. This cannot be meaningfully examined for the 1.3 mm window
with our spatially unresolved data, and we do not see a correlation with column density. 
While the emission of both species is higher in
regions of enhanced column density in our data, their ratio is not. The highest ratios are
not found in the regions encompassing the dense filament, but instead in the `low temperature' and
`low column density, low temperature' regions 
(N$_{2}$H$^{+}(3-2)$/CH$_{3}$OH$(\Sigma)\approx 3.8-4.1$), 
while the spectra from 
the high column density regions have notably lower values ($\approx 0.1$ for the `high column density', $\approx 2.4$
for the `high column density, low temperature', and $\approx 0.4$ for the `high column density, high temperature, 
without KL and Orion South' region). 
However, we see a potential correlation of N$_{2}$H$^{+}(3-2)$/CH$_{3}$OH$(\Sigma)$ with temperature 
(Fig. \ref{fig:corr_temp_N2H+-CH3OH}).

\subsubsection{Line luminosities}

\begin{figure*}[ht]
%\centering
%\vspace{3cm}
    \includegraphics[width=\textwidth]{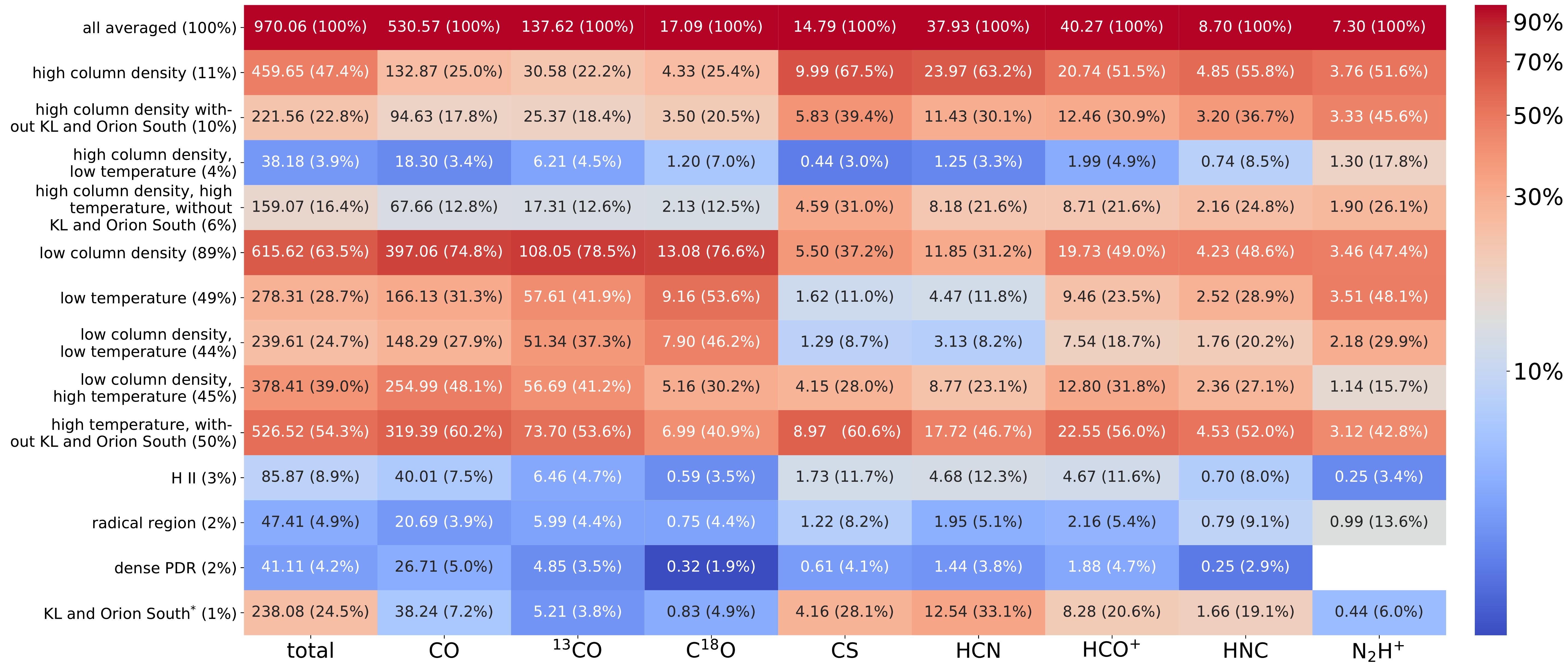}
    \caption{Total line luminosity (first column) and absolute and relative line luminosities 
    $\lbrack \mathrm{K~km ~ s}^{-1}\mathrm{pc}^{2} \rbrack$ 
    for selected species in the different regions. 
    A region's relative size is given in the brackets beside its name. Regions are generally 
    not disjointed, but overlap to varying degrees. Additionally, rounding uncertainties for both intensities and 
    region sizes accumulate and slightly different detection limits apply for each region. Hence the percentages 
    do not add up to 100\% and should be considered as estimates. $^{(*)}$ Approximated as described in Section
    \ref{KL_emission}.}
   \label{table_line_luminosities} 
\end{figure*}

For the interpretation of emission from unresolved and/or extragalactic sources, information on line luminosities 
$\lbrack \mathrm{K~km ~ s}^{-1}\mathrm{pc}^{2} \rbrack$ is important. 
Regions with low overall emission may still contribute considerably to the emission of some species if these regions
are spatially extended. Conversely, strong but compact emission may be diluted on larger scales.
Figure \ref{table_line_luminosities} lists the total luminosity (sum over all considered
species in the 1.3 mm window) and the values for selected species for all our examined regions, including the 
approximations for KL and Orion South (see also Section \ref{KL_emission}).

Despite its overall lower emission, the extended `low column density' region is the source of $\sim75\%$ of the CO, 
$^{13}$CO, and C$^{18}$O emission, while $\sim25\%$ can be allotted to the `high column density' region. For the high-density tracers HCO$^{+}$ and HNC, but also N$_{2}$H$^{+}$, the allocation between low and high column density is 
around $50\%/50\%$. For HCN and CS it is less of an even split between the two column density regimes and 
$\sim65\%$ originate from the high column density region. 

A lot of the emission from the high column density region can be
attributed to the environment of KL and Orion South: with a size of $<1\%$ of the total examined area, it emits
$\sim20\%$ of the HCO$^{+}$ and HNC, and $\sim30\%$ of HCN and CS. When considering all emission in the 1.3 mm window,
KL and Orion South are the source of roughly $25\%$.

%\begin{landscape}
%  \end{landscape}

\section{Discussion}

By averaging over the area, we highlight the individual emission profile
of the regional spectra, including lower intensity species, while ignoring their actual spatial extent
(which here mainly influences the respective noise level). This approach might help to characterise sources
for which the spatial resolution is not good enough to differentiate between distinct regions.
It is also a different approach compared to the analysis of the Orion B data set
\citep{Pety2016, Gratier2017,Bron2018}
in the $3 ~ \mathrm{mm}$ window, 
where the focus is more on selected, typically more luminous transitions, not necessarily on 
a complete inventory of species. We  discuss our findings further in this section and compare 
them with results from other authors.

We find that temperature has a significant impact on the total intensities of our selected regions;
considering column density alone is not sufficient.
This is illustrated most obviously in the cases of the `high column density, low temperature' and
`\ion{H}{ii}' regions.
The former has higher column densities but low overall emission, while the latter has lower 
median column density but a factor of $\approx 2.6$ higher total intensity (total intensity summed
over all considered species). 
We mostly consider temperatures and column densities, but it also seems instructive to keep the main
feedback processes in different regions in mind. While  the column densities in
OMC-2/3 are similar to values found in parts of OMC-1, the energy input in the former is mainly 
driven by outflows from
low-mass protostars, while in the latter it is shared between outflows and feedback from the \ion{H}{II}
region, as discussed in \citet{Berne2014} and inferred from CO and $^{13}\mathrm{CO} ~ 2-1$ emission. The 
kinetic energy in KL was found to be dominated by feedback of massive protostars (outflows,
jets, explosive motion). In contrast, the Bar exhibits very little outflow activity; its kinetic
energy is mainly caused by the expanding \ion{H}{II} region. 

We find very similar total CO intensities for  the `high column density' and `dense PDR' regions 
($\sim 393 ~ \mathrm{K~km ~ s}^{-1}$ and $~ \sim 392 ~ \mathrm{K~km ~ s}^{-1}$), but their total 
intensities summed  over all species vary by more than 50\% 
($\sim 1360 ~ \mathrm{K~km ~ s}^{-1}$ and $~ \sim 603 ~ \mathrm{K~km ~ s}^{-1}$). While the different
column densities and temperatures in the two regions still result in similar CO intensities, other
species react very differently and are much more enhanced in the `high column density' region. This
might again be partly due to opacity effects and CO only tracing the outer layers.

The median column densities for the `dense PDR' and `\ion{H}{ii}' regions are about a factor of 3 lower
than for `high column density' and `high column density, high temperature without KL and Orion South', but their 
median temperatures are the highest for all considered
regions. This might explain their high emission despite overall lower column densities.

\subsection{Correlations}
For the further examined most luminous species, strong correlations are typically found between  high-density tracers, 
excluding N$_{2}$H$^{+}$. This may be related to the optical depth effects mentioned
before, but also to the influence of temperature, as shown in the case of the warm `dense PDR' region,
where N$_{2}$H$^{+}$ does not emit over $5\sigma$ despite enhanced column densities. 
N$_{2}$H$^{+}$ shows no correlation with any other prevalent species in the 1.3 mm window,
but instead with column density.
If N$_{2}$H$^{+}$ data is not available,
HNC might overall be a better tracer of column density than  HCN or HCO$^{+}$, for example.
HNC shows similar intensities to N$_{2}$H$^{+}$, however, so it should not be a question of observation time.
That HCN does not exclusively trace dense gas was also shown in \citet{Kauffmann2017} for
the $1-0$ transition in Orion A, where HCN was found to trace lower densities $\sim 10^{3} ~ 
\mathrm{cm}^{-3}$ in cold sections of the cloud. Additionally, they found that the cold dense gas
emits too little HCN to explain the luminosities observed in extragalactic sources. This is 
consistent with the below average emission we see for the OMC-2/3 region in our data (`high column density,
low temperature'). 

Different papers discuss additional mechanisms which may excite `dense gas' tracers. As elaborated 
in \cite{Goldsmith2017}, electron excitation may be important for high-dipole moment molecules in regions where the
fraction of ionised carbon is significant. The low-J transitions of HCN (but also HCO$^{+}$, CN, and CS) could thus be
observed in lower density environments and may not qualify as indicators of high density. Another mechanism could be
radiative trapping (\citealt {Shirley2015, Pety2016}). It is argued that 
fundamental lines of HCN, HNC, and HCO$^{+}$ could be excited in regions well below their critical density as the 
latter is computed assuming optically thin emission only.
Both mechanisms could explain our observations. Particularly in OMC-1, where C$^{+}$ is abundant (\citealt{Pabst2019}), 
the emission of  HCN, for example,  may originate in part from electron excitation. Additionally, some transitions may have a 
high optical depth, increasing contributions from photon-trapping. 
An alternative approach to the detection of low-lying rotational lines is discussed in \cite{Liszt2016}. 
They discuss the observability of transitions for HCO$^{+}$, HNC, and CS in regions where the density 
is far below the critical density. The authors find, in the limit of weak collisional excitation, that there is 
a column density (not a volume density) that will produce a given output.

We find that the typical UV tracers CN and C$_{2}$H  correlate not only with each other, but also 
strongly with high-density tracers. This means that tracing UV illumination on large scales could be challenging,
as enhanced CN or C$_{2}$H emission might not necessarily be indicative of higher UV illumination.

\FloatBarrier
\subsection{Emission on larger scales}
The stark difference in the emission profile of KL and Orion South (see Section \ref{KL_emission})
compared to the other regions was also found for some species in \citet{Ungerechts1997}, where the 
integrated intensity maps of  SO or HC$_{3}$N showed distributions strongly peaked towards KL. 
While we find that line emission from KL and Orion South has a noticeable impact on spectra averaged 
over larger scales (compare  Fig. \ref{fig:pie_hd} and \ref{fig:pie_hd_wo_kl}, but also appreciate 
that  SiO and SO$_{2}$ are detectable in the averaged spectrum despite not being detected for the 
majority of regions), this influence of high column density regions on the averaged spectrum was not seen by
\citet{Watanabe2017} for the spectrum of W51 in the 3 mm window. Their spatial coverage of 
$39 ~ \mathrm{pc} \times 39 ~ \mathrm{pc}$ is significantly larger than for our data set, so emission
from high-density regions is expected to be smeared out more, such that they find a dominance of the 
quiescent material over the averaged spectrum.

In their analysis of a line survey of the central parts of the starburst galaxy M82 in the 1.3 mm and 2 mm window, 
\citet{Aladro2011} found that the physical processes are dominated by PDRs. Arguing that feedback from young 
OB stars leaves an imprint on molecular composition in the form of an overabundance of  CO$^{+}$,
HCO, c-C$_{3}$H$_{2}$, and CH$_{3}$CCH, for example,  they find M82 to match these criteria. 
While we can confirm the high c-C$_{3}$H$_{2}$ emission in our dense PDR region, CH$_{3}$CCH is below 
the $5\sigma$ limit there. This is also interesting because their spatial resolution varies between 
$158~\mathrm{pc}$ and $333~\mathrm{pc}$, which is 2 to 3 orders of magnitude 
higher than our averaged spectra, but we observe CH$_{3}$CCH emission as quite compact and thus strongly diluted 
on larger scales. The noise levels are comparable in both studies, $\sim3-8~\mathrm{mK}$ for M82, around 
$8-20~\mathrm{mK}$ median noise for the different regions in our Orion data.

\subsection{Line ratios on larger scales}
Line ratios seem to be more ambiguous on large scales, where both CN$(\Sigma)$/HCN$(3-2)$ and 
CN$(\Sigma)$/HCO$^{+}(3-2)$ do not clearly
highlight regions with enhanced UV irradiation. We can confirm a correlation between 
HCN$(3-2)$/HNC$(3-2)$ and 
temperature. On even larger scales, however, as examined by  \citet{Meier2005}, among others,  for the nuclear region
of IC 342 in the 3 mm window ($\sim 50 ~ \mathrm{pc}$ resolution), the ratio seems to be fairly constant
and not related to (kinetic) temperature. Their value of HCN$(1-0)$/HNC$(1-0) \approx  1-2$ 
is found in our data set 
for those regions associated with low temperature (see Fig. \ref{fig:corr_temp_HCN-HNC}). 
The considered transitions are different to ours, but both species still have comparable upper
energies, such that a comparison seems meaningful.

Extragalactic studies like that of \citet{JimenezDonaire2017}, where the $1-0$ transitions of HCN, HCO$^{+}$,
HNC, and some of their isotopologues are mapped for six nearby galaxies (a few hundred pc to 
$\sim1~\mathrm{kpc}$ resolution), use spectral stacking to
maximise the signal-to-noise ratio for the examined lines. The resulting line ratios of  
HCN$(1-0)$/HCO$^{+}(1-0)$ 
are higher than seen for our regions 
(HCN$(1-0)$/HCO$^{+}(1-0) = 1.0-1.7$ vs. 
HCN$(3-2)$/HCO$^{+}(3-2) = 0.4-1.2$, see also Fig. \ref{table_isotope_ratios}). 
Instead, our ratio values are comparable to those found by \citet{Harada2018} in the two nuclei of Merger NGC 3256 (likewise examining the $1-0$
transitions, 
$\sim300~\mathrm{pc}$ resolution), although the ratio is critically discussed in 
this context mainly as a diagnostic for AGN and/or starburst galaxies.
Dense gas tracers ($1-0$ transitions) for nine nearby massive spiral galaxies were further examined for
the EMPIRE survey (\citealt{Jimenez2019}), a continuation of these authors' 2017 work. Their Table 4 
shows the 
dense gas line ratios averaged over their galaxy sample, separated into the center (inner $30\arcsec$, 
$1-2~\mathrm{kpc}$ resolution) and the disc. Their values of $\sim0.018-0.034$ for the 
HCN$(1-0)$/CO$(1-0)$ ratio are 
found in our $1.3~\mathrm{mm}$ data for regions associated with low column density and low temperature, while 
higher column density 
regions have a higher value (e.g. HCN$(3-2)$/CO$(2-1) = 0.18$ for the 
`high column density' environment). Our 
HCO$^{+}$/CO values are 
always higher than theirs (HCO$^{+}(3-2)$/CO$(2-1) \sim 0.050-0.156$ vs. 
HCO$^{+}(1-0)$/CO$(1-0) \sim0.014-0.025$), the most similar again associated with 
lower column density regions. Their values for HNC$(1-0)$/CO$(1-0)$ ($\sim0.010-0.014$)
can also be found in our data, where we find deviations
towards higher values to be associated with higher column density regions
(e.g. HNC$(3-2)$/CO$(2-1) \sim 0.04$ for the `high column density' environment).

The correlation of N$_{2}$H$^{+}(3-2)$/CH$_{3}$OH$(\Sigma)$ with temperature we found on larger scales
might be explained both with 
the temperature sensitivity of N$_{2}$H$^{+}$ and the association of CH$_{3}$OH with shocks. If 
temperatures rise, more CO can enter the gas phase and subsequently destroy N$_{2}$H$^{+}$, while more
CH$_{3}$OH evaporates.

\section{Summary}
We have conducted an imaging line survey of OMC-1 to OMC-3 from 200.2 to 281.8 GHz and examined the 
emission of distinct regions. These were selected to represent regimes of low or high column density 
and differing temperature, but also to cover the influence of UV irradiation. By contrasting their
emission with each other, we aim to provide templates for the interpretation of other
more distant, spatially unresolved sources. Comparing spectra from these unresolved sources with our 
different templates might reveal similarities that can help to further characterise the distant object.
The transitions of the 29 species (55 isotopologues) listed in Table \ref{all_transitions} were considered
for the analysis. The integrated intensity of a given species is  considered here to be the sum over all of their 
respective transitions.
From our analysis we would like to highlight the following results:
   \begin{enumerate}
      \item Line emission from an Orion KL-like source can contribute significantly to spectra averaged over
      larger regions, both in terms of averaged total intensity and chemical diversity. In terms of line 
      luminosities, KL and Orion South contribute around $25\%$ of all emission in the 1.3 mm window in the 
      area of OMC-1 to OMC-3.
      \item Regions like OMC-2/3, with pre-stellar Class 0 and Class I objects and enhanced column density
      but low temperature, have a much lower total intensity. Their signatures (e.g. high N$_{2}$H$^{+}$
      emission coincident with a low HCN$(3-2)$/HNC$(3-2)$ ratio) would be difficult 
      to pick up in a non-resolved source.
      \item While the  contribution of CO to the share of the total intensity can vary more strongly in the 
      examined 1.3 mm window,
      HCO$^{+}(3-2)$ contributes $\sim 3\%-6\%$ in all cases (average $\approx 4\%$). This seems 
      to hold true even in the case of the emission around KL and Orion South.
      \item The emissions of the high-density tracers HCN, HCO$^{+}$, H$_{2}$CO, HNC, CS, but also CN are 
      strongly correlated with each other, but not with N$_{2}$H$^{+}$. Of all the examined 
      species, N$_{2}$H$^{+}$ shows the strongest correlation with column density. If 
      N$_{2}$H$^{+}$ observations are not available, HNC seems to trace the column density most reliably.
      \item Around $50\%$ of the line luminosity of HCO$^{+}$ and HNC in the 1.3 mm window comes from lower
      column density material, for CS and HCN $\sim35\%$.
      \item The ratios HCN$(3-2)$/HNC$(3-2)$ and 
      N$_{2}$H$^{+}(3-2)$/CH$_{3}$OH$(\Sigma)$ may be related to temperature.
      \item Identifying UV illuminated material on large scales seems to be challenging, as both
       CN$(\Sigma)$/HCN$(3-2)$ and 
       CN$(\Sigma)$/HCO$^{+}(3-2)$ show ambiguous results.
   \end{enumerate}
   
\begin{acknowledgements}
We thank the anonymous referee for their careful reading of the manuscript and helpful comments that improved the article. 
DC acknowledges support by the Deutsche Forschungsgemeinschaft, 
DFG through project number SFB956C.
\end{acknowledgements}

%\FloatBarrier
%-------------------------------------------------------------------
%\begin{thebibliography}{}
\bibliographystyle{aa}
\bibliography{lit}

%\end{thebibliography}

%%%%%%%%%%%%%%%%%%%%%%%%%%%%%%%%%%%%%%%%%

\begin{appendix}
\FloatBarrier
\section{Tables}
\onecolumn

\small
\begin{longtab}
\begin{longtable}{llcccccl}
\hline\hline  

\textbf{Species} & \textbf{Isotopologue} & \textbf{Frequency} & \textbf{Transition} & \textbf{Energy} & \textbf{G$_{\mathrm{up}}$}
& \textbf{A$_{\mathrm{ij}}$} & \textbf{Comment} \\
& & \textbf{[MHz]} & & \textbf{[K]} & & \textbf{[s$^{-1}$]} & \\ \\
\endfirsthead

\hline (Continuation) & & & & & & &\\
\textbf{Species} & \textbf{Isotopologue} & \textbf{Frequency} & \textbf{Transition} & \textbf{Energy} & \textbf{G$_{\mathrm{up}}$}
& \textbf{A$_{\mathrm{ij}}$} & \textbf{Comment} \\ 
\hline  

\endhead

\hline  & & & & & & & \textbf{Continued on next page}\\  
\endfoot

%\hline \hline
\endlastfoot

\hline
c-C$_{3}$H$_{2}$ & c-C$_{3}$H$_{2}$ & 204788.926 & $4_{2,2,0}-3_{3,1,0}$ & 28.8 & 9 & 1.37e-04 & ignored, S/N not good  \\
 &  & 216278.756 & $3_{3,0,0}-2_{2,1,0}$ & 19.5 & 21 & 2.81e-04 &  \\
 &  & 217822.148 & $6_{1,6,0}-5_{0,5,0}$ & 38.6 & 39 & 5.93e-04 &  \\
 &  & 217822.148 & $6_{0,6,0}-5_{1,5,0}$ & 38.6 & 13 & 5.93e-04 &  \\
 &  & 217940.046 & $5_{1,4,0}-4_{2,3,0}$ & 35.4 & 33 & 4.43e-04 &  \\
 &  & 218160.442 & $5_{2,4,0}-4_{1,3,0}$ & 35.4 & 11 & 4.44e-04 &  \\
 &  & 227169.127 & $4_{3,2,0}-3_{2,1,0}$ & 29.1 & 27 & 3.42e-04 &  \\
 &  & 249054.368 & $5_{2,3,0}-4_{3,2,0}$ & 41.0 & 33 & 4.57e-04 &  \\
 &  & 251314.337 & $7_{0,7,0}-6_{1,6,0}$ & 50.7 & 45 & 9.35e-04 &  \\
 &  & 251314.343 & $7_{1,7,0}-6_{0,6,0}$ & 50.7 & 15 & 9.35e-04 &  \\
 &  & 251508.691 & $6_{1,5,0}-5_{2,4,0}$ & 47.5 & 13 & 7.42e-04 &  \\
 &  & 251527.302 & $6_{2,5,0}-5_{1,4,0}$ & 47.5 & 39 & 7.42e-04 &  \\
 &  & 254987.640 & $5_{3,3,0}-4_{2,2,0}$ & 41.1 & 11 & 5.17e-04 & ignored, S/N not good \\
 &  & 260479.746 & $5_{3,2,0}-4_{4,1,0}$ & 44.7 & 33 & 1.77e-04 & ignored, blended with SiO \\
 &  & 265759.438 & $4_{4,1,0}-3_{3,0,0}$ & 32.2 & 27 & 7.99e-04 &  \\
 \hline
CCH & CCH & 261978.120 & $3_{4,3}-2_{3,3}$ & 25.1 & 7 & 1.96e-06 &  \\
 &  & 262004.260 & $3_{4,4}-2_{3,3}$ & 25.1 & 9 & 5.32e-05 &  \\
 &  & 262006.482 & $3_{4,3}-2_{3,2}$ & 25.1 & 7 & 5.12e-05 &  \\
 &  & 262064.986 & $3_{3,3}-2_{2,2}$ & 25.2 & 7 & 4.89e-05 &  \\
 &  & 262067.469 & $3_{3,2}-2_{2,1}$ & 25.2 & 5 & 4.47e-05 &  \\
 &  & 262078.935 & $3_{3,2}-2_{2,2}$ & 25.2 & 5 & 6.02e-06 &  \\
 &  & 262208.614 & $3_{3,3}-2_{3,3}$ & 25.2 & 7 & 3.96e-06 &  \\
 &  & 262250.929 & $3_{3,2}-2_{3,2}$ & 25.2 & 5 & 2.27e-06 &  \\
 & CCD & 216372.830 & $3_{4,5}-2_{3,4}$ & 20.8 & 10 & 3e-05 &  \\
 &  & 216373.320 & $3_{4,3}-2_{3,2}$ & 20.8 & 6 & 2.67e-05 &  \\
 &  & 216373.320 & $3_{4,4}-2_{3,3}$ & 20.8 & 8 & 2.76e-05 &  \\
 &  & 216428.320 & $3_{3,4}-2_{2,3}$ & 20.8 & 8 & 2.78e-05 &  \\
 &  & 216428.320 & $3_{3,3}-2_{2,2}$ & 20.8 & 6 & 2.34e-05 &  \\
 &  & 216428.760 & $3_{3,2}-2_{2,1}$ & 20.8 & 4 & 2.1e-05 &  \\
 \hline
CF$^{+}$ & CF$^{+}$ & 205170.520 & $2_{0}-1_{0}$ & 14.8 & 5 & 4.62e-05 & may be blended with CH$_{3}$CHO
\\
\hline
CH$_{3}$CCH & CH$_{3}$CCH & 205018.080 & $12_{4}-11_{4}$ & 179.2 & 25 & 2.41e-05 & blended with CH$_{3}$OCH$_{3}$ \\
 &  & 205045.401 & $12_{3}-11_{3}$ & 128.8 & 50 & 2.54e-05 &  \\
 &  & 205065.015 & $12_{2}-11_{2}$ & 92.8 & 25 & 2.63e-05 &  \\
 &  & 205076.775 & $12_{1}-11_{1}$ & 71.2 & 25 & 2.69e-05 &  \\
 &  & 205080.660 & $12_{0}-11_{0}$ & 64.0 & 25 & 2.71e-05 &  \\
 &  & 222099.151 & $13_{4}-12_{4}$ & 189.8 & 27 & 3.13e-05 &  \\
 &  & 222128.808 & $13_{3}-12_{3}$ & 139.4 & 54 & 3.27e-05 &  \\
 &  & 222150.008 & $13_{2}-12_{2}$ & 103.4 & 27 & 3.37e-05 &  \\
 &  & 222162.729 & $13_{1}-12_{1}$ & 81.8 & 27 & 3.44e-05 &  \\
 &  & 222166.970 & $13_{0}-12_{0}$ & 74.6 & 27 & 3.46e-05 &  \\
 &  & 239211.216 & $14_{3}-13_{3}$ & 150.9 & 58 & 4.13e-05 &  \\
 &  & 239234.011 & $14_{2}-13_{2}$ & 114.9 & 29 & 4.24e-05 &  \\
 &  & 239247.727 & $14_{1}-13_{1}$ & 93.3 & 29 & 4.31e-05 &  \\
 &  & 239252.297 & $14_{0}-13_{0}$ & 86.1 & 29 & 4.33e-05 &  \\
 &  & 256292.638 & $15_{3}-14_{3}$ & 163.2 & 62 & 5.12e-05 &  \\
 &  & 256317.078 & $15_{2}-14_{2}$ & 127.2 & 31 & 5.24e-05 &  \\
 &  & 256331.746 & $11_{3}-10_{3}$ & 105.6 & 31 & 5.31e-05 &  \\
 &  & 256336.636 & $15_{0}-14_{0}$ & 98.4 & 31 & 5.34e-05 &  \\
 &  & 273373.006 & $16_{3}-15_{3}$ & 176.3 & 66 & 6.26e-05 &  \\
 &  & 273399.067 & $16_{2}-15_{2}$ & 140.3 & 33 & 6.39e-05 &  \\
 &  & 273414.707 & $16_{1}-15_{1}$ & 118.7 & 33 & 6.46e-05 &  \\
 &  & 273419.921 & $16_{0}-15_{0}$ & 111.5 & 33 & 6.49e-05 &  \\
 \hline
CH$_{3}$CN & CH$_{3}$CN & 202215.371 & $11_{6}-10_{6}$ & 315.3 & 92 & 4.97e-04 &  \\
 &  & 202258.154 & $11_{5}-10_{5}$ & 236.8 & 46 & 5.62e-04 &  \\
 &  & 202293.183 & $11_{4}-10_{4}$ & 172.6 & 46 & 6.15e-04 &  \\
 &  & 202320.443 & $11_{3}-10_{3}$ & 122.6 & 92 & 6.56e-04 &  \\
 &  & 202339.921 & $11_{2}-10_{2}$ & 86.8 & 46 & 6.86e-04 &  \\
 &  & 202351.612 & $11_{1}-10_{1}$ & 65.4 & 46 & 7.03e-04 &  \\
 &  & 202355.509 & $11_{0}-10_{0}$ & 58.3 & 46 & 7.09e-04 &  \\
 &  & 220594.423 & $12_{6}-11_{6}$ & 325.9 & 100 & 6.92e-04 &  \\
 &  & 220641.084 & $12_{5}-11_{5}$ & 247.4 & 50 & 7.63e-04 &  \\
 &  & 220679.287 & $12_{4}-11_{4}$ & 183.1 & 50 & 8.21e-04 &  \\
 &  & 220709.016 & $12_{3}-11_{3}$ & 133.2 & 100 & 8.66e-04 &  \\
 &  & 220730.261 & $12_{2}-11_{2}$ & 97.4 & 50 & 8.98e-04 &  \\
 &  & 220743.011 & $12_{1}-11_{1}$ & 76.0 & 50 & 9.18e-04 &  \\
 &  & 220747.261 & $12_{0}-11_{0}$ & 68.9 & 50 & 9.24e-04 &  \\
 &  & 238972.389 & $13_{6}-12_{6}$ & 337.4 & 108 & 9.26e-04 &  \\
 &  & 239022.924 & $13_{5}-12_{5}$ & 258.9 & 54 & 1e-03 &  \\
 &  & 239064.299 & $13_{4}-12_{4}$ & 194.6 & 54 & 1.07e-03 &  \\
 &  & 239096.497 & $13_{3}-12_{3}$ & 144.6 & 108 & 1.12e-03 &  \\
 &  & 239119.504 & $13_{2}-12_{2}$ & 108.9 & 54 & 1.15e-03 &  \\
 &  & 239133.313 & $13_{1}-12_{1}$ & 87.5 & 54 & 1.17e-03 &  \\
 &  & 239137.916 & $13_{0}-12_{0}$ & 80.3 & 54 & 1.18e-03 &  \\
 &  & 257349.179 & $14_{6}-13_{6}$ & 349.7 & 116 & 1.2e-03 &  \\
 &  & 257403.584 & $14_{5}-13_{5}$ & 271.2 & 58 & 1.29e-03 &  blended with CH$_{3}$OH\\
 &  & 257448.128 & $14_{4}-13_{4}$ & 207.0 & 58 & 1.35e-03 &  \\
 &  & 257482.791 & $14_{3}-13_{3}$ & 157.0 & 116 & 1.41e-03 &  \\
 &  & 257507.561 & $14_{2}-13_{2}$ & 121.3 & 58 & 1.45e-03 &  \\
 &  & 257522.427 & $14_{1}-13_{1}$ & 99.8 & 58 & 1.47e-03 &  \\
 &  & 257527.383 & $14_{0}-13_{0}$ & 92.7 & 58 & 1.48e-03 &  \\
 &  & 275724.702 & $15_{6}-14_{6}$ & 363.0 & 124 & 1.53e-03 &  \\
 &  & 275782.974 & $15_{5}-14_{5}$ & 284.5 & 62 & 1.62e-03 &  \\
 &  & 275830.683 & $15_{4}-14_{4}$ & 220.2 & 62 & 1.69e-03 &  \\
 &  & 275867.810 & $15_{3}-14_{3}$ & 170.2 & 124 & 1.75e-03 &  \\
 &  & 275894.340 & $15_{2}-14_{2}$ & 134.5 & 62 & 1.79e-03 &  \\
 &  & 275910.263 & $15_{1}-14_{1}$ & 113.1 & 62 & 1.81e-03 &  \\
 &  & 275915.571 & $15_{0}-14_{0}$ & 105.9 & 62 & 1.82e-03 &  \\
 \hline
CH$_{3}$OH & CH$_{3}$OH & 201071.847 & $8_{4,5,0}-9_{3,6,0}$ & 163.9 & 68 & 9.12e-06 &  \\
 &  & 201088.939 & $8_{4,4,0}-9_{3,7,0}$ & 163.9 & 68 & 9.12e-06 &  \\
 &  & 200820.674 & $15_{3,12,2}-14_{4,11,2}$ & 341.2 & 124 & 1.27e-05 &  \\
 &  & 201445.493 & $5_{2,3,0}-6_{1,6,0}$ & 72.5 & 44 & 1.3e-05 &  \\
 &  & 201996.513 & $18_{1,17,1}-18_{0,18,1}$ & 417.9 & 148 & 2.92e-05 &  \\
 &  & 205791.270 & $1_{1,1,0}-2_{0,2,0}$ & 16.8 & 12 & 3.36e-05 &  \\
 &  & 206001.302 & $12_{5,7,1}-13_{4,9,1}$ & 317.1 & 100 & 1.04e-05 &  \\
 &  & 209518.804 & $19_{1,18,1}-19_{0,19,1}$ & 462.0 & 156 & 3.13e-05 & ignored, blended with CH$_{3}$OCH$_{3}$ \\
 &  & 213377.528 & $13_{6,8,1}-14_{5,9,1}$ & 389.9 & 108 & 1.07e-05 &  \\
 &  & 213427.061 & $1_{1,0,1}-0_{0,0,1}$ & 23.4 & 12 & 3.37e-05 &  \\
 &  & 215302.206 & $6_{1,6,3}-7_{2,6,3}$ & 373.8 & 52 & 4.18e-05 &  \\
 &  & 216945.521 & $5_{1,4,1}-4_{2,3,1}$ & 55.9 & 44 & 1.21e-05 &  \\
 &  & 217299.205 & $6_{1,5,3}-7_{2,5,3}$ & 373.9 & 52 & 4.28e-05 &  \\
 &  & 218440.063 & $4_{2,3,1}-3_{1,2,1}$ & 45.5 & 36 & 4.69e-05 &  \\
 &  & 220078.561 & $8_{0,8,1}-7_{1,6,1}$ & 96.6 & 68 & 2.52e-05 &  \\
 &  & 220401.317 & $10_{5,6,2}-11_{4,8,2}$ & 251.6 & 84 & 1.12e-05 & ignored, blended with $^{13}$CO \\
 &  & 227814.528 & $16_{1,16,0}-15_{2,13,0}$ & 327.2 & 132 & 2.18e-05 &  \\
 &  & 229589.056 & $15_{4,11,1}-16_{3,14,1}$ & 374.4 & 124 & 2.08e-05 &  \\
 &  & 229758.756 & $8_{1,8,2}-7_{0,7,1}$ & 89.1 & 68 & 4.19e-05 &  \\
 &  & 230027.047 & $3_{2,1,2}-4_{1,4,2}$ & 39.8 & 28 & 1.49e-05 &  \\
 &  & 231281.110 & $10_{2,9,0}-9_{3,6,0}$ & 165.3 & 84 & 1.83e-05 &  \\
 &  & 232418.521 & $10_{2,8,0}-9_{3,7,0}$ & 165.4 & 84 & 1.87e-05 &  \\
 &  & 232783.446 & $18_{3,16,0}-17_{4,13,0}$ & 446.5 & 148 & 2.17e-05 &  \\
 &  & 232945.797 & $10_{3,7,2}-11_{2,9,2}$ & 190.4 & 84 & 2.13e-05 &  \\
 &  & 233795.666 & $18_{3,15,0}-17_{4,14,0}$ & 446.6 & 148 & 2.2e-05 &  \\
 &  & 234683.370 & $4_{2,3,0}-5_{1,4,0}$ & 60.9 & 36 & 1.87e-05 &  \\
 &  & 234698.519 & $5_{4,2,2}-6_{3,3,2}$ & 122.7 & 44 & 6.34e-06 &  \\
 &  & 236936.089 & $14_{1,13,0}-13_{2,12,0}$ & 260.2 & 116 & 3.11e-05 &  \\
 &  & 239746.219 & $5_{1,5,0}-4_{1,4,0}$ & 49.1 & 44 & 5.66e-05 &  \\
 &  & 240241.490 & $5_{3,3,1}-6_{2,5,1}$ & 82.5 & 44 & 1.44e-05 &  \\
 &  & 240960.557 & $5_{1,5,3}-4_{1,4,3}$ & 360.0 & 44 & 5.76e-05 &  \\
 &  & 241159.199 & $5_{4,2,4}-4_{4,1,4}$ & 398.1 & 44 & 2.15e-05 &  \\
 &  & 241166.580 & $5_{3,2,4}-4_{3,1,4}$ & 452.1 & 44 & 3.86e-05 &  \\
 &  & 241179.886 & $5_{3,3,5}-4_{3,2,5}$ & 357.4 & 44 & 3.83e-05 &  \\
 &  & 241184.189 & $5_{4,1,5}-4_{4,0,5}$ & 440.1 & 44 & 2.16e-05 &  \\
 &  & 241187.428 & $5_{2,4,5}-4_{2,3,5}$ & 399.3 & 44 & 5.07e-05 &  \\
 &  & 241192.856 & $5_{2,4,3}-4_{2,3,3}$ & 333.4 & 44 & 5.03e-05 &  \\
 &  & 241196.430 & $5_{2,3,3}-4_{2,2,3}$ & 333.4 & 44 & 5.03e-05 &  \\
 &  & 241198.285 & $5_{3,3,3}-4_{3,2,3}$ & 430.8 & 44 & 3.83e-05 &  \\
 &  & 241198.291 & $5_{3,2,3}-4_{3,1,3}$ & 430.8 & 44 & 3.83e-05 &  \\
 &  & 241203.706 & $5_{1,5,4}-4_{1,4,4}$ & 326.2 & 44 & 5.75e-05 &  \\
 &  & 241206.035 & $5_{0,5,4}-4_{0,4,4}$ & 335.3 & 44 & 6e-05 &  \\
 &  & 241210.764 & $5_{2,3,4}-4_{2,2,4}$ & 434.6 & 44 & 5.04e-05 &  \\
 &  & 241238.144 & $5_{1,4,5}-4_{1,3,5}$ & 448.1 & 44 & 5.75e-05 &  \\
 &  & 241267.862 & $5_{0,5,3}-4_{0,4,3}$ & 458.4 & 44 & 6e-05 &  \\
 &  & 241441.270 & $5_{1,4,3}-4_{1,3,3}$ & 360.0 & 44 & 5.79e-05 &  \\
 &  & 241700.159 & $5_{0,5,1}-4_{0,4,1}$ & 47.9 & 44 & 6.04e-05 &  \\
 &  & 241767.234 & $5_{1,5,2}-4_{1,4,2}$ & 40.4 & 44 & 5.81e-05 &  \\
 &  & 241791.352 & $5_{0,5,0}-4_{0,4,0}$ & 34.8 & 44 & 6.05e-05 &  \\
 &  & 241806.524 & $5_{4,2,0}-4_{4,1,0}$ & 115.2 & 44 & 2.18e-05 &  \\
 &  & 241806.525 & $5_{4,1,0}-4_{4,0,0}$ & 115.2 & 44 & 2.18e-05 &  \\
 &  & 241813.255 & $5_{4,2,2}-4_{4,1,2}$ & 122.7 & 44 & 2.18e-05 &  \\
 &  & 241829.629 & $5_{4,1,1}-4_{4,0,1}$ & 130.8 & 44 & 2.19e-05 &  \\
 &  & 241832.718 & $5_{3,3,0}-4_{3,2,0}$ & 84.6 & 44 & 3.87e-05 &  \\
 &  & 241833.106 & $5_{3,2,0}-4_{3,1,0}$ & 84.6 & 44 & 3.87e-05 &  \\
 &  & 241842.284 & $5_{2,4,0}-4_{2,3,0}$ & 72.5 & 44 & 5.11e-05 &  \\
 &  & 241843.604 & $5_{3,3,1}-4_{3,2,1}$ & 82.5 & 44 & 3.88e-05 &  \\
 &  & 241852.299 & $5_{3,2,2}-4_{3,1,2}$ & 97.5 & 44 & 3.89e-05 &  \\
 &  & 241879.025 & $5_{1,4,1}-4_{1,3,1}$ & 55.9 & 44 & 5.96e-05 &  \\
 &  & 241887.674 & $5_{2,3,0}-4_{2,2,0}$ & 72.5 & 44 & 5.12e-05 &  \\
 &  & 241904.147 & $5_{2,3,2}-4_{2,2,2}$ & 60.7 & 44 & 5.09e-05 &  \\
 &  & 241904.643 & $5_{2,4,1}-4_{2,3,1}$ & 57.1 & 44 & 5.03e-05 &  \\
 &  & 242446.084 & $14_{1,14,2}-13_{2,11,2}$ & 248.9 & 116 & 2.29e-05 &  \\
 &  & 243915.788 & $5_{1,4,0}-4_{1,3,0}$ & 49.7 & 44 & 5.97e-05 &  \\
 &  & 244337.983 & $9_{1,9,4}-8_{0,8,4}$ & 395.6 & 76 & 4.06e-05 & ignored, S/N not good \\
 &  & 246873.301 & $19_{3,16,0}-19_{2,17,0}$ & 490.7 & 156 & 8.27e-05 &  \\
 &  & 247161.950 & $16_{2,15,1}-15_{3,13,1}$ & 338.1 & 132 & 2.57e-05 &  \\
 &  & 247228.587 & $4_{2,2,0}-5_{1,5,0}$ & 60.9 & 36 & 2.12e-05 &  \\
 &  & 247610.918 & $18_{3,15,0}-18_{2,16,0}$ & 446.6 & 148 & 8.29e-05 &  \\
 &  & 248282.424 & $17_{3,14,0}-17_{2,15,0}$ & 404.8 & 140 & 8.3e-05 &  \\
 &  & 248885.468 & $16_{3,13,0}-16_{2,14,0}$ & 365.4 & 132 & 8.32e-05 &  \\
 &  & 249192.836 & $16_{3,13,2}-15_{4,12,2}$ & 378.3 & 132 & 2.54e-05 &  \\
 &  & 249419.924 & $15_{3,12,0}-15_{2,13,0}$ & 328.3 & 124 & 8.32e-05 &  \\
 &  & 249443.301 & $7_{4,4,0}-8_{3,5,0}$ & 145.3 & 60 & 1.48e-05 &  \\
 &  & 249451.842 & $7_{4,3,0}-8_{3,6,0}$ & 145.3 & 60 & 1.48e-05 &  \\
 &  & 249887.467 & $14_{3,11,0}-14_{2,12,0}$ & 293.5 & 116 & 8.32e-05 &  \\
 &  & 250291.181 & $13_{3,10,0}-13_{2,11,0}$ & 261.0 & 108 & 8.3e-05 &  \\
 &  & 250506.853 & $11_{0,11,0}-10_{1,10,0}$ & 153.1 & 92 & 8.46e-05 &  \\
 &  & 250635.200 & $12_{3,9,0}-12_{2,10,0}$ & 230.8 & 100 & 8.28e-05 &  \\
 &  & 250924.398 & $11_{3,8,0}-11_{2,9,0}$ & 203.0 & 92 & 8.24e-05 &  \\
 &  & 251164.108 & $10_{3,7,0}-10_{2,8,0}$ & 177.5 & 84 & 8.18e-05 &  \\
 &  & 251359.888 & $9_{3,6,0}-9_{2,7,0}$ & 154.2 & 76 & 8.09e-05 &  \\
 &  & 251517.309 & $8_{3,5,0}-8_{2,6,0}$ & 133.4 & 68 & 7.96e-05 &  \\
 &  & 251641.787 & $7_{3,4,0}-7_{2,5,0}$ & 114.8 & 60 & 7.76e-05 &  \\
 &  & 251738.437 & $6_{3,3,0}-6_{2,4,0}$ & 98.5 & 52 & 7.46e-05 &  \\
 &  & 251811.956 & $5_{3,2,0}-5_{2,3,0}$ & 84.6 & 44 & 6.97e-05 &  \\
 &  & 251866.524 & $4_{3,1,0}-4_{2,2,0}$ & 73.0 & 36 & 6.1e-05 &  \\
 &  & 251890.886 & $5_{3,3,0}-5_{2,4,0}$ & 84.6 & 44 & 6.97e-05 &  \\
 &  & 251895.728 & $6_{3,4,0}-6_{2,5,0}$ & 98.5 & 52 & 7.47e-05 &  \\
 &  & 251900.452 & $4_{3,2,0}-4_{2,3,0}$ & 73.0 & 36 & 6.1e-05 &  \\
 &  & 251905.729 & $3_{3,0,0}-3_{2,1,0}$ & 63.7 & 28 & 4.36e-05 &  \\
 &  & 251917.065 & $3_{3,1,0}-3_{2,2,0}$ & 63.7 & 28 & 4.36e-05 &  \\
 &  & 251923.701 & $7_{3,5,0}-7_{2,6,0}$ & 114.8 & 60 & 7.78e-05 &  \\
 &  & 251984.837 & $8_{3,6,0}-8_{2,7,0}$ & 133.4 & 68 & 7.98e-05 &  \\
 &  & 252090.409 & $9_{3,7,0}-9_{2,8,0}$ & 154.2 & 76 & 8.13e-05 &  \\
 &  & 252252.849 & $10_{3,8,0}-10_{2,9,0}$ & 177.5 & 84 & 8.25e-05 &  \\
 &  & 252485.675 & $11_{3,9,0}-11_{2,10,0}$ & 203.0 & 92 & 8.34e-05 &  \\
 &  & 252803.388 & $12_{3,10,0}-12_{2,11,0}$ & 230.8 & 100 & 8.42e-05 &  \\
 &  & 253221.376 & $13_{3,11,0}-13_{2,12,0}$ & 261.0 & 108 & 8.49e-05 &  \\
 &  & 253755.809 & $14_{3,12,0}-14_{2,13,0}$ & 293.5 & 116 & 8.56e-05 &  \\
 &  & 254015.377 & $2_{0,2,1}-1_{1,1,2}$ & 20.1 & 20 & 1.9e-05 &  \\
 &  & 254419.419 & $11_{5,6,1}-12_{4,8,1}$ & 289.2 & 92 & 1.79e-05 &  \\
 &  & 254423.520 & $15_{3,13,0}-15_{2,14,0}$ & 328.3 & 124 & 8.63e-05 &  \\
 &  & 255241.888 & $16_{3,14,0}-16_{2,15,0}$ & 365.4 & 132 & 8.71e-05 &  \\
 &  & 256228.714 & $17_{3,15,0}-17_{2,16,0}$ & 404.8 & 140 & 8.8e-05 &  \\
 &  & 257402.086 & $18_{3,16,0}-18_{2,17,0}$ & 446.5 & 148 & 8.9e-05 & ignored, blended with CH$_{3}$CN \\
 &  & 258780.248 & $19_{3,17,0}-19_{2,18,0}$ & 490.6 & 156 & 9.01e-05 &  \\
 &  & 261704.409 & $12_{6,7,1}-13_{5,8,1}$ & 359.8 & 100 & 1.78e-05 &  \\
 &  & 261805.675 & $2_{1,1,1}-1_{0,1,1}$ & 28.0 & 20 & 5.57e-05 &  \\
 &  & 263793.875 & $5_{1,5,3}-6_{2,5,3}$ & 360.0 & 44 & 8.22e-05 & ignored, blended with HCCCN \\
 &  & 265224.426 & $5_{1,4,3}-6_{2,4,3}$ & 360.0 & 44 & 8.33e-05 &  \\
 &  & 265289.562 & $6_{1,5,1}-5_{2,4,1}$ & 69.8 & 52 & 2.58e-05 &  \\
 &  & 266838.148 & $5_{2,4,1}-4_{1,3,1}$ & 57.1 & 44 & 7.74e-05 &  \\
 &  & 267403.471 & $9_{0,9,1}-8_{1,7,1}$ & 117.5 & 76 & 4.67e-05 &  \\
 &  & 267406.071 & $17_{1,17,0}-16_{2,14,0}$ & 366.3 & 140 & 3.51e-05 &  \\
 &  & 268743.954 & $9_{5,5,2}-10_{4,7,2}$ & 228.4 & 76 & 1.76e-05 &  \\
 &  & 278304.512 & $9_{1,9,2}-8_{0,8,1}$ & 110.0 & 76 & 7.69e-05 &  \\
 &  & 278342.261 & $2_{2,0,2}-3_{1,3,2}$ & 32.9 & 20 & 1.65e-05 &  \\
 &  & 278599.037 & $14_{4,10,1}-15_{3,13,1}$ & 339.6 & 116 & 3.6e-05 & ignored, S/N not good \\
 &  & 279351.887 & $11_{2,10,0}-10_{3,7,0}$ & 190.9 & 92 & 3.45e-05 &  \\
 &  & 280679.621 & $19_{3,17,0}-18_{4,14,0}$ & 490.6 & 156 & 3.87e-05 & ignored, S/N not good \\
 &  & 281000.109 & $11_{2,9,0}-10_{3,8,0}$ & 190.9 & 92 & 3.53e-05 &  \\
 \hline
CN & CN & 226287.418 & $2_{0,2,1}-1_{0,2,1}$ & 16.3 & 2 & 1.03e-05 &  \\
 &  & 226298.943 & $2_{0,2,1}-1_{0,2,2}$ & 16.3 & 2 & 8.23e-06 &  \\
 &  & 226303.037 & $2_{0,2,2}-1_{0,2,1}$ & 16.3 & 4 & 4.17e-06 &  \\
 &  & 226314.540 & $2_{0,2,2}-1_{0,2,2}$ & 16.3 & 4 & 9.91e-06 &  \\
 &  & 226332.499 & $2_{0,2,2}-1_{0,2,3}$ & 16.3 & 4 & 4.56e-06 &  \\
 &  & 226341.930 & $2_{0,2,3}-1_{0,2,2}$ & 16.3 & 6 & 3.16e-06 &  \\
 &  & 226359.871 & $2_{0,2,3}-1_{0,2,3}$ & 16.3 & 6 & 1.61e-05 &  \\
 &  & 226616.571 & $2_{0,2,1}-1_{0,1,2}$ & 16.3 & 2 & 1.07e-05 &  \\
 &  & 226632.190 & $2_{0,2,2}-1_{0,1,2}$ & 16.3 & 4 & 4.26e-05 &  \\
 &  & 226659.558 & $2_{0,2,3}-1_{0,1,2}$ & 16.3 & 6 & 9.47e-05 & ignored, frequency gap \\
 &  & 226663.693 & $2_{0,2,1}-1_{0,1,1}$ & 16.3 & 2 & 8.47e-05 &  \\
 &  & 226679.311 & $2_{0,2,2}-1_{0,1,1}$ & 16.3 & 4 & 5.27e-05 &  \\
 &  & 226874.191 & $2_{0,3,3}-1_{0,2,2}$ & 16.3 & 6 & 9.62e-05 & partially blanked (*) \\
 &  & 226874.781 & $2_{0,3,4}-1_{0,2,3}$ & 16.3 & 8 & 1.14e-04 & partially blanked (*)  \\
 &  & 226875.896 & $2_{0,3,2}-1_{0,2,1}$ & 16.3 & 4 & 8.59e-05 & partially blanked (*)  \\
 &  & 226887.420 & $2_{0,3,2}-1_{0,2,2}$ & 16.3 & 4 & 2.73e-05 &  \\
 &  & 226892.128 & $2_{0,3,3}-1_{0,2,3}$ & 16.3 & 6 & 1.81e-05 &  \\
 &  & 226905.357 & $2_{0,3,2}-1_{0,2,3}$ & 16.3 & 4 & 1.13e-06 & ignored, S/N not good \\
 & $^{13}$CN & 217467.150 & $2_{3,3,4}-1_{2,2,3}$ & 15.7 & 9 & 1.01e-04 &  \\
  \hline
CO & CO & 230538.000 & 2-1 & 16.6 & 5 & 6.91e-07 &  \\
 & $^{13}$CO & 220398.684 & 2-1 & 15.9 & 5 & 6.08e-07 &  \\
 & C$^{18}$O & 219560.357 & 2-1 & 15.8 & 5 & 6.01e-07 &  \\
 & C$^{17}$O & 224714.385 & 2-1 & 16.2 & 5 & 6.43e-07 &  \\
 & $^{13}$C$^{18}$O & 209419.138 & $2_{2}-1_{1}$ & 15.1 & 4 & 4.36e-07 &  \\
 &  & 209419.172 & $2_{3}-1_{2}$ & 15.1 & 6 & 5.23e-07 &  \\
 \hline
CS & CS & 244935.644 & 5-4 & 35.3 & 11 & 3e-04 &  \\
 & $^{13}$CS & 231220.996 & 5-4 & 33.3 & 11 & 2.52e-04 &  \\
 &  & 277455.481 & 6-5 & 46.6 & 13 & 4.42e-04 &  \\
 & C$^{34}$S & 241016.194 & 5-4 & 34.7 & 11 & 2.86e-04 &  \\
 & C$^{33}$S & 242913.610 & $5_{0}-4_{0}$ & 35.0 & 44 & 2.91e-04 &  \\
 \hline
HCCCN & HCCCN & 209230.234 & 23-22 & 120.5 & 141 & 7.24e-04 &  \\
 &  & 218324.788 & 24-23 & 131.0 & 147 & 8.23e-04 &  \\
 &  & 227418.906 & 25-24 & 141.9 & 153 & 9.31e-04 &  \\
 &  & 236512.777 & 26-25 & 153.2 & 159 & 1.05e-03 &  \\
 &  & 245606.308 & 27-26 & 165.0 & 165 & 1.17e-03 &  \\
 &  & 254699.500 & 28-27 & 177.3 & 171 & 1.31e-03 &  \\
 &  & 263792.308 & 29-28 & 189.9 & 177 & 1.46e-03 &  blended with CH$_{3}$OH\\
 &  & 272884.734 & 30-29 & 203.0 & 183 & 1.61e-03 &  \\
 \hline
HCN & HCN & 265886.180 & 3-2 & 25.5 & 21 & 8.42e-04 &  \\
 & DCN & 217238.400 & $3_{2}-2_{1}$ & 20.9 & 5 & 3.83e-04 &  \\
 &  & 217238.631 & $3_{2}-2_{2}$ & 20.9 & 5 & 7.08e-05 &  \\
 &  & 217238.631 & $3_{3}-2_{3}$ & 20.9 & 7 & 5.06e-05 &  \\
 &  & 217238.631 & $3_{3}-2_{2}$ & 20.9 & 7 & 4.05e-04 &  \\
 &  & 217238.631 & $3_{4}-2_{3}$ & 20.9 & 9 & 4.55e-04 &  \\
 & H$^{13}$CN & 259011.821 & $3_{2}-2_{2}$ & 24.9 & 5 & 1.2e-04 &  \\
 &  & 259011.821 & $3_{3}-2_{3}$ & 24.9 & 7 & 8.58e-05 &  \\
 &  & 259011.821 & $3_{2}-2_{1}$ & 24.9 & 5 & 6.48e-04 &  \\
 &  & 259011.821 & $3_{3}-2_{2}$ & 24.9 & 7 & 6.86e-04 &  \\
 &  & 259011.821 & $3_{4}-2_{3}$ & 24.9 & 9 & 7.72e-04 &  \\
 & HC$^{15}$N & 258157.100 & 3-2 & 24.8 & 7 & 7.65e-04 &  \\
 \hline
HCO & HCO & 260060.329 & $3_{0,3,4,4}-2_{0,2,3,3}$ & 25.0 & 9 & 1.63e-04 &  \\
 &  & 260082.192 & $3_{0,3,4,3}-2_{0,2,3,2}$ & 25.0 & 7 & 1.61e-04 &  \\
 &  & 260133.586 & $3_{0,3,3,3}-2_{0,2,2,2}$ & 25.0 & 7 & 1.45e-04 &  \\
 &  & 260155.769 & $3_{0,3,3,2}-2_{0,2,2,1}$ & 25.0 & 5 & 1.37e-04 &  \\
 \hline
HCO$^{+}$ & HCO$^{+}$ & 267557.626 & 3-2 & 25.7 & 7 & 1.45e-03 & blended with SO$_{2}$ and OCS \\
 & DCO$^{+}$ & 216112.582 & 3-2 & 20.7 & 7 & 7.66e-04 &  \\
 & H$^{13}$CO$^{+}$ & 260255.339 & 3-2 & 25.0 & 7 & 1.34e-03 &  \\
 & HC$^{18}$O$^{+}$ & 255479.389 & 3-2 & 24.5 & 7 & 1.27e-03 &  \\
 & HC$^{17}$O$^{+}$ & 261164.920 & 3-2 & 25.1 & 42 & 1.35e-03 &  \\
 \hline
HCS$^{+}$ & HCS$^{+}$ & 213360.650 & 5-4 & 30.7 & 11 & 1.97e-04 &  \\
 &  & 256027.100 & 6-5 & 43.0 & 13 & 3.46e-04 &  \\
\hline
HDO & HDO & 225896.720 & $3_{1,2}-2_{2,1}$ & 167.6 & 7 & 1.32e-05 &  \\
 &  & 241561.550 & $2_{1,1}-2_{1,2}$ & 95.2 & 5 & 1.19e-05 &  \\
 &  & 266161.070 & $2_{2,0}-3_{1,3}$ & 157.2 & 5 & 1.75e-05 &  \\
 \hline
H$_{2}$CCO & H$_{2}$CCO & 202014.311 & $10_{0,10}-9_{0,9}$ & 53.3 & 21 & 9.24e-05 & ignored, S/N not good \\
 &  & 203940.225 & $10_{1,9}-9_{1,8}$ & 66.9 & 63 & 9.41e-05 &  \\
 &  & 220177.569 & $11_{1,11}-10_{1,10}$ & 76.5 & 69 & 1.19e-04 &  \\
 &  & 222197.635 & $11_{0,11}-10_{0,10}$ & 64.0 & 23 & 1.24e-04 &  \\
 &  & 224327.250 & $11_{1,10}-10_{1,9}$ & 77.7 & 69 & 1.26e-04 &  \\
 &  & 240185.794 & $12_{1,12}-11_{1,11}$ & 88.0 & 75 & 1.55e-04 &  \\
 &  & 242375.735 & $12_{0,12}-11_{0,11}$ & 75.6 & 25 & 1.61e-04 & ignored, S/N not good \\
 &  & 244712.269 & $12_{1,11}-11_{11,0}$ & 89.4 & 75 & 1.64e-04 &  \\
 &  & 260191.982 & $13_{1,13}-12_{1,12}$ & 100.5 & 81 & 1.98e-04 &  \\
 &  & 262548.207 & $13_{0,13}-12_{0,12}$ & 88.2 & 27 & 2.05e-04 & ignored, S/N not good \\
 &  & 265095.049 & $13_{1,12}-12_{1,11}$ & 102.1 & 81 & 2.1e-04 &  \\
 \hline
H$_{2}$CO & H$_{2}$CO & 211211.468 & $3_{1,3}-2_{1,2}$ & 32.1 & 21 & 2.27e-04 &  \\
 &  & 216568.651 & $9_{1,8}-9_{1,9}$ & 174.0 & 57 & 7.22e-06 &  \\
 &  & 218222.192 & $3_{0,3}-2_{0,2}$ & 21.0 & 7 & 2.82e-04 &  \\
 &  & 218475.632 & $3_{2,2}-2_{2,1}$ & 68.1 & 7 & 1.57e-04 &  \\
 &  & 218760.066 & $3_{2,1}-2_{2,0}$ & 68.1 & 7 & 1.58e-04 &  \\
 &  & 225697.775 & $3_{1,2}-2_{1,1}$ & 33.4 & 21 & 2.77e-04 &  \\
 &  & 264270.140 & $10_{1,9}-10_{1,10}$ & 209.9 & 63 & 1.08e-05 &  \\
 &  & 281526.929 & $4_{1,4}-3_{1,3}$ & 45.6 & 27 & 5.88e-04 &  \\
 & HDCO & 201341.350 & $3_{1,2}-2_{1,1}$ & 27.3 & 7 & 1.96e-04 &  \\
 &  & 246924.600 & $4_{1,4}-3_{1,3}$ & 37.6 & 9 & 3.96e-04 &  \\
 &  & 256585.430 & $4_{0,4}-3_{0,3}$ & 30.8 & 9 & 4.74e-04 &  \\
 &  & 257748.760 & $4_{2,3}-3_{2,2}$ & 62.8 & 9 & 3.6e-04 &  \\
 &  & 259034.910 & $4_{2,2}-3_{2,1}$ & 62.9 & 9 & 3.66e-04 &  \\
 &  & 268292.020 & $4_{1,3}-3_{1,2}$ & 40.2 & 9 & 5.08e-04 &  \\
 & H$_{2}^{13}$CO & 206131.626 & $3_{1,3}-2_{1,2}$ & 31.6 & 21 & 2.11e-04 &  \\
 &  & 212811.184 & $3_{0,3}-2_{0,2}$ & 20.4 & 7 & 2.61e-04 &  \\
 &  & 219908.525 & $3_{1,2}-2_{1,1}$ & 32.9 & 21 & 2.56e-04 &  \\
 &  & 274762.112 & $4_{1,4}-3_{1,3}$ & 44.8 & 27 & 5.47e-04 &  \\
 \hline
H$_{2}$CS & H$_{2}$CS & 202923.515 & $6_{1,6}-5_{1,5}$ & 47.3 & 39 & 1.18e-04 &  \\
 &  & 205987.391 & $6_{0,6}-5_{0,5}$ & 34.6 & 13 & 1.27e-04 &  \\
 &  & 206053.584 & $6_{2,5}-5_{2,4}$ & 87.3 & 13 & 1.13e-04 &  \\
 &  & 206158.016 & $6_{2,4}-5_{2,3}$ & 87.3 & 13 & 1.13e-04 & ignored, blended with SO \\
 &  & 209200.101 & $6_{1,5}-5_{1,4}$ & 48.3 & 39 & 1.3e-04 &  \\
 &  & 236726.770 & $7_{1,7}-6_{1,6}$ & 58.6 & 45 & 1.91e-04 &  \\
 &  & 240266.320 & $7_{0,7}-6_{0,6}$ & 46.1 & 15 & 2.04e-04 &  \\
 &  & 240381.750 & $7_{2,6}-6_{2,5}$ & 98.9 & 15 & 1.88e-04 &  \\
 &  & 240548.229 & $7_{2,5}-6_{2,4}$ & 98.9 & 15 & 1.88e-04 &  \\
 &  & 244047.840 & $7_{1,6}-6_{1,5}$ & 60.0 & 45 & 2.1e-04 &  \\
 &  & 270520.740 & $8_{1,8}-7_{1,7}$ & 71.6 & 51 & 2.9e-04 &  \\
 &  & 274520.870 & $8_{0,8}-7_{0,7}$ & 59.3 & 17 & 3.07e-04 &  \\
 &  & 274702.055 & $8_{2,7}-7_{2,6}$ & 112.0 & 17 & 2.89e-04 &  \\
 &  & 274952.473 & $8_{2,6}-7_{2,5}$ & 112.1 & 17 & 2.9e-04 &  \\
 &  & 278886.400 & $8_{1,7}-7_{1,6}$ & 73.4 & 51 & 3.17e-04 &  \\
 & HDCS & 212648.339 & $7_{1,7}-6_{1,6}$ & 49.8 & 15 & 1.39e-04 & \\
 &  & 216662.429 & $7_{0,7}-6_{0,6}$ & 41.6 & 15 & 1.5e-04 &  \\
 & H$_{2}$C$^{34}$S & 202492.418 & $6_{0,6}-5_{0,5}$ & 34.0 & 13 & 1.21e-04 &  \\
 \hline
H$_{2}$S & H$_{2}$S & 216710.435 & $2_{2,0}-2_{1,1}$ & 84.0 & 5 & 4.83e-05 &  \\
\hline
HNC & HNC & 271981.142 & 3-2 & 26.1 & 7 & 9.34e-04 &  \\
 & DNC & 228910.489 & 3-2 & 22.0 & 7 & 5.57e-04 &  \\
 & HN$^{13}$C & 261263.310 & 3-2 & 25.1 & 7 & 6.48e-04 &  \\
 \hline
HNCO & HNCO & 218981.009 & $10_{1,10}-9_{1,9}$ & 101.1 & 21 & 1.42e-04 &  \\
 &  & 219798.274 & $10_{0,10}-9_{0,9}$ & 58.0 & 21 & 1.47e-04 &  \\
 &  & 220584.751 & $10_{1,9}-9_{1,8}$ & 101.5 & 21 & 1.45e-04 &  \\
 &  & 240875.727 & $11_{1,11}-10_{1,10}$ & 112.6 & 23 & 1.9e-04 &  \\
 &  & 241774.032 & $11_{0,11}-10_{0,10}$ & 69.6 & 23 & 1.96e-04 &  \\
 &  & 242639.704 & $11_{1,10}-10_{1,9}$ & 113.1 & 23 & 1.95e-04 &  \\
 &  & 262769.477 & $12_{1,12}-11_{1,11}$ & 125.3 & 25 & 2.48e-04 &  \\
 &  & 263748.625 & $12_{0,12}-11_{0,11}$ & 82.3 & 25 & 2.56e-04 &  \\
 &  & 264693.655 & $12_{1,11}-11_{1,10}$ & 125.9 & 25 & 2.54e-04 &  \\
 \hline
N$_{2}$H$^{+}$ & N$_{2}$H$^{+}$ & 279511.701 & 3-2 & 26.8 & 63 & 1.26e-03 &  \\
 & N$_{2}$D$^{+}$ & 231321.665 & 3-2 & 22.2 & 21 & 7.14e-04 &  \\
 \hline
NO & NO & 250436.848 & $3_{1,3,4}-2_{-1,2,3}$ & 19.2 & 8 & 1.84e-06 &  \\
 &  & 250440.659 & $3_{1,3,3}-2_{-1,2,2}$ & 19.2 & 6 & 1.55e-06 &  \\
 &  & 250448.530 & $3_{1,3,2}-2_{-1,2,1}$ & 19.2 & 4 & 1.38e-06 &  \\
 &  & 250796.436 & $3_{-1,3,4}-2_{1,2,3}$ & 19.3 & 8 & 1.85e-06 &  \\
 &  & 250815.594 & $3_{-1,3,3}-2_{1,2,2}$ & 19.3 & 6 & 1.55e-06 &  \\
 &  & 250816.954 & $3_{-1,3,2}-2_{1,2,1}$ & 19.3 & 4 & 1.39e-06 &  \\
 \hline
NS & NS & 207436.246 & $5_{1,-1,6}-4_{1,1,5}$ & 27.6 & 12 & 1.51e-04 &  \\
 &  & 207436.246 & $5_{1,-1,5}-4_{1,1,4}$ & 27.6 & 10 & 1.44e-04 &  \\
 &  & 207438.692 & $5_{1,-1,4}-4_{1,1,3}$ & 27.6 & 8 & 1.42e-04 &  \\
 &  & 207834.866 & $5_{1,1,6}-4_{1,-1,5}$ & 27.7 & 12 & 1.52e-04 &  \\
 &  & 207838.365 & $5_{1,1,5}-4_{1,-1,4}$ & 27.7 & 10 & 1.45e-04 &  \\
 &  & 207838.365 & $5_{1,1,4}-4_{1,-1,3}$ & 27.7 & 8 & 1.43e-04 &  \\
 &  & 253570.476 & $6_{1,1,7}-5_{1,-1,6}$ & 39.8 & 14 & 2.83e-04 &  \\
 &  & 253570.476 & $6_{1,1,6}-5_{1,-1,5}$ & 39.8 & 12 & 2.73e-04 &  \\
 &  & 253572.148 & $6_{1,1,5}-5_{1,-1,4}$ & 39.8 & 10 & 2.71e-04 &  \\
 &  & 253968.393 & $6_{1,-1,7}-5_{1,1,6}$ & 39.9 & 14 & 2.84e-04 &  \\
 &  & 253970.581 & $6_{1,-1,6}-5_{1,1,5}$ & 39.9 & 12 & 2.75e-04 &  \\
 &  & 253970.581 & $6_{1,-1,5}-5_{1,1,4}$ & 39.9 & 10 & 2.73e-04 &  \\
 \hline
OCS & OCS & 206745.161 & 17-16 & 89.3 & 35 & 2.55e-05 &  \\
 &  & 218903.356 & 18-17 & 99.8 & 37 & 3.04e-05 &  \\
 &  & 231060.983 & 19-18 & 110.9 & 39 & 3.58e-05 &  \\
 &  & 243218.040 & 20-19 & 122.6 & 41 & 4.18e-05 &  \\
 &  & 255374.461 & 21-20 & 134.8 & 43 & 4.84e-05 &  \\
 &  & 267530.239 & 22-21 & 147.7 & 45 & 5.57e-05 & ignored, blended with HCO$^{+}$ \\
 &  & 279685.318 & 23-22 & 161.1 & 47 & 6.37e-05 &  \\
 \hline
SiO & SiO & 217104.980 & 5-4 & 31.3 & 11 & 5.21e-04 &  \\
 &  & 260518.020 & 6-5 & 43.8 & 13 & 9.15e-04 & blended with c-C$_{3}$H$_{2}$\\
 & $^{29}$SiO & 214385.036 & 5-4 & 30.9 & 11 & 5e-04 &  \\
 &  & 257254.227 & 6-5 & 43.2 & 13 & 8.78e-04 &  \\
 & $^{30}$SiO & 211852.797 & 5-4 & 30.5 & 11 & 4.83e-04 &  \\
 &  & 254215.845 & 6-5 & 42.7 & 13 & 8.47e-04 &  \\
 \hline
SO & SO & 206176.005 & $5_{4}-4_{3}$ & 38.6 & 9 & 1.03e-04 & blended with H$_{2}$CS \\
 &  & 214357.039 & $8_{7}-7_{7}$ & 81.2 & 15 & 3.42e-06 &  \\
 &  & 215220.653 & $5_{5}-4_{4}$ & 44.1 & 11 & 1.22e-04 &  \\
 &  & 219949.442 & $5_{6}-4_{5}$ & 35.0 & 13 & 1.36e-04 &  \\
 &  & 236452.325 & $2_{1}-1_{2}$ & 15.8 & 3 & 1.45e-06 & ignored, S/N not good\\
 &  & 246404.687 & $3_{2}-2_{3}$ & 21.1 & 5 & 1.03e-06 & ignored, S/N not good \\
 &  & 251825.770 & $6_{5}-5_{4}$ & 50.7 & 11 & 1.96e-04 &  \\
 &  & 254573.500 & $9_{8}-8_{8}$ & 99.7 & 17 & 4.32e-06 &  \\
 &  & 258255.813 & $6_{6}-5_{5}$ & 56.5 & 13 & 2.16e-04 &  \\
 &  & 261843.684 & $6_{7}-5_{6}$ & 47.6 & 15 & 2.33e-04 &  \\
 & $^{34}$SO & 201846.573 & $5_{4}-4_{3}$ & 38.1 & 9 & 9.66e-05 &  \\
 &  & 211013.673 & $5_{5}-4_{4}$ & 43.5 & 11 & 1.15e-04 & ignored, frequency gap \\
 &  & 215839.436 & $5_{6}-4_{5}$ & 34.4 & 13 & 1.29e-04 &  \\
 &  & 246663.638 & $6_{5}-5_{4}$ & 49.9 & 11 & 1.84e-04 &  \\
 &  & 253208.020 & $6_{6}-5_{5}$ & 55.7 & 13 & 2.04e-04 &  \\
 &  & 256877.456 & $6_{7}-5_{6}$ & 46.7 & 15 & 2.2e-04 &  \\
 \hline
SO$^{+}$ & SO$^{+}$ & 208590.016 & $5_{1,5}-4_{-1,4}$ & 26.7 & 10 & 4.7e-05 & blended with CH$_{3}$OCH$_{3}$ \\
 &  & 208965.420 & $5_{-1,5}-4_{1,4}$ & 26.8 & 10 & 4.72e-05 & blended with C$_{2}$H$_{3}$CN \\
 &  & 254977.935 & $6_{-1,6}-5_{1,5}$ & 38.9 & 12 & 8.77e-05 & blended with C$_{2}$H$_{5}$CN \\
 &  & 255353.237 & $6_{1,6}-5_{-1,5}$ & 39.0 & 12 & 8.81e-05 &  \\
 \hline
SO$_{2}$ & SO$_{2}$ & 200809.180 & $16_{1,15}-16_{0,16}$ & 130.7 & 33 & 4.7e-05 &  \\
 &  & 203391.550 & $12_{0,12}-11_{1,11}$ & 70.1 & 25 & 8.8e-05 &  \\
 &  & 204246.760 & $18_{3,15}-18_{2,16}$ & 180.6 & 37 & 9.27e-05 &  \\
 &  & 204384.300 & $7_{4,4}-8_{3,5}$ & 65.5 & 15 & 1.11e-05 &  \\
 &  & 205300.570 & $11_{2,10}-11_{1,11}$ & 70.2 & 23 & 5.32e-05 &  \\
 &  & 208700.320 & $3_{2,2}-2_{1,1}$ & 15.3 & 7 & 6.72e-05 &  \\
 &  & 209936.050 & $12_{5,7}-13_{4,10}$ & 133.0 & 25 & 1.59e-05 &  \\
 &  & 213068.400 & $26_{3,23}-26_{2,24}$ & 350.8 & 53 & 1.16e-04 &  \\
 &  & 214689.380 & $16_{3,13}-16_{2,14}$ & 147.8 & 33 & 9.9e-05 &  \\
 &  & 214728.330 & $17_{6,12}-18_{5,13}$ & 229.0 & 35 & 1.89e-05 &  \\
 &  & 216643.300 & $22_{2,20}-22_{1,21}$ & 248.4 & 45 & 9.27e-05 &  \\
 &  & 221965.210 & $11_{1,11}-10_{0,10}$ & 60.4 & 23 & 1.14e-04 &  \\
 &  & 223883.569 & $6_{4,2}-7_{3,5}$ & 58.6 & 13 & 1.16e-05 &  \\
 &  & 224264.811 & $20_{2,18}-19_{3,17}$ & 207.8 & 41 & 3.94e-05 &  \\
 &  & 225153.702 & $13_{2,12}-13_{1,13}$ & 93.0 & 27 & 6.52e-05 &  \\
 &  & 226300.027 & $14_{3,11}-14_{2,12}$ & 119.0 & 29 & 1.07e-04 &  \\
 &  & 229347.628 & $11_{5,7}-12_{4,8}$ & 122.0 & 23 & 1.91e-05 &  \\
 &  & 234421.586 & $16_{6,10}-17_{5,13}$ & 213.3 & 33 & 2.35e-05 &  \\
 &  & 235151.720 & $4_{2,2}-3_{1,3}$ & 19.0 & 9 & 7.69e-05 &  \\
 &  & 236216.685 & $16_{1,15}-15_{2,14}$ & 130.7 & 33 & 7.5e-05 &  \\
 &  & 237068.870 & $12_{3,9}-12_{2,10}$ & 94.0 & 25 & 1.14e-04 &  \\
 &  & 240942.791 & $18_{1,17}-18_{0,18}$ & 163.1 & 37 & 7.02e-05 &  \\
 &  & 241615.798 & $5_{2,4}-4_{1,3}$ & 23.6 & 11 & 8.45e-05 &  \\
 &  & 243087.647 & $5_{4,2}-6_{3,3}$ & 53.1 & 11 & 1.03e-05 &  \\
 &  & 244254.218 & $14_{0,14}-13_{1,13}$ & 93.9 & 29 & 1.64e-04 &  \\
 &  & 245563.423 & $10_{3,7}-10_{2,8}$ & 72.7 & 21 & 1.19e-04 &  \\
 &  & 248057.401 & $15_{2,14}-15_{1,15}$ & 119.3 & 31 & 8.06e-05 &  \\
 &  & 248830.824 & $10_{5,5}-11_{4,8}$ & 111.9 & 21 & 2.19e-05 &  \\
 &  & 251199.675 & $13_{1,13}-12_{0,12}$ & 82.2 & 27 & 1.76e-04 &  \\
 &  & 251210.586 & $8_{3,5}-8_{2,6}$ & 55.2 & 17 & 1.2e-04 &  \\
 &  & 253956.567 & $15_{6,10}-16_{5,11}$ & 198.6 & 31 & 2.82e-05 & ignored, possibly artifact \\
 &  & 254280.536 & $6_{3,3}-6_{2,4}$ & 41.4 & 13 & 1.14e-04 &  \\
 &  & 254283.319 & $24_{2,22}-24_{1,23}$ & 292.7 & 49 & 1.33e-04 &  \\
 &  & 255553.303 & $4_{3,1}-4_{2,2}$ & 31.3 & 9 & 9.28e-05 &  \\
 &  & 255958.044 & $3_{3,1}-3_{2,2}$ & 27.6 & 7 & 6.63e-05 &  \\
 &  & 256246.946 & $5_{3,3}-5_{2,4}$ & 35.9 & 11 & 1.07e-04 &  \\
 &  & 257099.966 & $7_{3,5}-7_{2,6}$ & 47.8 & 15 & 1.22e-04 &  \\
 &  & 258942.199 & $9_{3,7}-9_{2,8}$ & 63.5 & 19 & 1.32e-04 &  \\
 &  & 262256.905 & $11_{3,9}-11_{2,10}$ & 82.8 & 23 & 1.41e-04 &  \\
 &  & 267537.450 & $13_{3,11}-13_{2,12}$ & 105.8 & 27 & 1.51e-04 & ignored, blended with HCO$^{+}$ \\
 &  & 268168.334 & $9_{5,5}-10_{4,6}$ & 102.7 & 19 & 2.39e-05 &  \\
 &  & 271529.015 & $7_{2,6}-6_{1,5}$ & 35.5 & 15 & 1.11e-04 &  \\
 &  & 273462.668 & $14_{6,8}-15_{5,11}$ & 184.8 & 29 & 3.3e-05 &  \\
 &  & 273752.961 & $17_{2,16}-17_{1,17}$ & 149.2 & 35 & 9.97e-05 &  \\
 &  & 275240.182 & $15_{3,13}-15_{2,14}$ & 132.5 & 31 & 1.64e-04 &  \\
 &  & 280807.280 & $26_{4,22}-26_{3,23}$ & 364.3 & 53 & 2.33e-04 &  \\
 \hline
\caption{All considered transitions for our fitting procedure. The comments are based on the spectrum
of the high column density region. Notes concerning overlap between lines, especially with complex organic
molecules, are thus likely not relevant on larger scales or for low column density/low temperature regions. 
Lines ignored due to e.g. locally too poor
S/N are marked as such. In case of a strong overlap and one line being significantly stronger than the 
other (e.g. overlap with HCO$^{+}$), the area is contributed to the stronger line only. (*) Three CN 
transitions are strongly blended and part of this line falls into a gap in 
the frequency coverage. A Gauss fit is still possible, and we made sure that the fitted line has 
no peak intensity larger than the visible part of the line. }
\label{all_transitions}
\end{longtable}
\end{longtab}
{\addtocounter{table}{-1}}

\begin{table*}
\tiny
\caption{Uncertainties for the averaged total intensities $\Delta\int T_{\mathrm{mb}}dv$ $\lbrack \mathrm{K~km ~ s}^{-1} \rbrack$ from Table \ref{table_all_total_intensities}.}   
\label{table_all_total_intensities_errors}     
\centering 
\begin{tabular}{l | c c c c c c c c c c c c c}
\hline\hline   % table heading
\\
species 
& \rotatebox{90}{all averaged} 
&\rotatebox{90}{high column density} 
& \rotatebox{90}{high column density without}
\rotatebox{90}{KL and Orion South} 
& \rotatebox{90}{high column density,} \rotatebox{90}{low temperature} 
& \rotatebox{90}{high column density,} \rotatebox{90}{high temperature,}
\rotatebox{90}{without KL and Orion South}
& \rotatebox{90}{low column density} 
& \rotatebox{90}{low temperature} 
&\rotatebox{90}{low column density,} \rotatebox{90}{low temperature}
& \rotatebox{90}{low column density,} \rotatebox{90}{high temperature}
& \rotatebox{90}{high temperature}\rotatebox{90}{without KL and Orion South}
& \rotatebox{90}{\ion{H}{ii}} & \rotatebox{90}{radical region} & \rotatebox{90}{dense PDR}
\\
\hline                       
\\
CO                 &   0.01   &  0.03  &  0.12    &   0.06   &  0.03  &  0.60    & 0.01    &  0.01   &   7.13    & 0.64  &   0.01  &   0.01 & 0.03  \\         
$^{13}$CO          &   0.01   &  0.01  &  <0.01   &   0.01   &  0.01  &  <0.01   & 0.02    &  0.11   &   <0.01   & 0.02  &   0.01  &   0.01 & <0.01 \\         
C$^{18}$O          &   0.08   &  <0.01 &  0.01    &   0.03   &  0.04  &  0.02    & 0.28    &  0.02   &   0.07    & 0.01  &   0.05  &   0.03 & 0.25  \\         
C$^{17}$O          &   0.14   &  0.11  &  0.15    &   0.01   &  0.35  &  0.15    & 0.03    &  0.02   &   0.02    & 0.03  &   0.76  &   0.35 & 0.29  \\         
$^{13}$C$^{18}$O   &   -      &  0.02  &  0.01    &   0.02   &  0.01  &       -  & 0.02    &      -  &         - &    -  &    -    &   0.02 &   -   \\        
c-C$_{3}$H$_{2}$   &   0.03   &  0.36  &  0.21    &   0.04   &  0.27  &       -  &      -  &      -  &   0.03    & 0.07  &   0.07  &   0.19 & 0.26  \\         
C$_{2}$H           &   0.11   &  0.24  &  0.18    &   0.11   &  0.39  &  0.09    & 0.05    &  0.05   &   0.14    & 0.08  &   0.05  &   0.15 & 0.11  \\         
C$_{2}$D           &   -      &  0.05  &  0.05    &   0.04   &  0.05  &       -  &      -  &      -  &         - &    -  &    -    &   0.06 &   -   \\        
CF$^{+}$           &   0.01   &  0.02  &  0.02    &       -  &  0.02  &  0.02    & 0.01    &      -  &         - &    -  &    -    &     -  &   -   \\        
CH$_{3}$CCH        &   0.02   &  0.39  &  0.25    &   0.07   &  0.37  &       -  &      -  &      -  &         - & 0.03  &   0.26  &   0.29 &   -   \\        
CH$_{3}$CN         &   -      &  0.93  &  0.08    &       -  &  0.08  &       -  &      -  &      -  &         - & 1.27  &   0.10  &     -  &   -   \\        
CH$_{3}$OH         &   0.28   &  3.38  &  0.49    &   0.25   &  14.13 &  0.05    & 0.03    &  0.03   &   0.05    & 0.08  &   1.61  &   6.32 & 0.05  \\         
CN                 &   0.41   &  0.62  &  0.48    &   0.20   &  0.33  &  0.48    & 0.13    &  0.09   &   0.11    & 0.44  &   0.31  &   0.91 & 0.40  \\         
$^{13}$CN          &   -      &  0.02  &  0.02    &       -  &  0.09  &       -  &      -  &      -  &         - &    -  &    -    &   0.03 &   -   \\        
CS                 &   0.31   &  0.03  &  0.43    &   0.04   &  0.05  &  0.07    & 0.03    &  0.02   &   0.02    & 0.25  &   0.37  &   0.02 & 0.27  \\         
$^{13}$CS          &   0.02   &  0.25  &  0.10    &       -  &  0.20  &       -  &      -  &      -  &         - &    -  &   0.07  &   0.04 & 0.06  \\         
C$^{34}$S          &   0.02   &  0.10  &  0.10    &       -  &  <0.01 &       -  &      -  &      -  &   0.02    & 0.02  &   0.03  &   0.02 & 0.26  \\         
C$^{33}$S          &   -      &  0.10  &  0.02    &       -  &  0.09  &       -  &      -  &      -  &         - &    -  &   0.03  &   0.02 & 0.03  \\         
HC$_{3}$N          &   -      &  0.75  &  0.19    &       -  &  0.32  &       -  &      -  &      -  &         - &    -  &   0.32  &   0.07 & 0.03  \\         
HCN                &   0.08   &  0.01  &  0.01    &   0.18   &  <0.01 &  0.02    & 0.07    &  0.07   &   0.03    & 0.07  &   0.13  &   0.01 & <0.01 \\         
DCN                &   0.11   &  0.06  &  0.08    &   0.02   &  0.08  &       -  &      -  &      -  &         - & 0.06  &   0.04  &   0.02 & 0.04  \\         
H$^{13}$CN         &   0.02   &  0.14  &  0.06    &   0.02   &  0.07  &       -  &      -  &      -  &   0.01    & 0.05  &   0.05  &   0.16 & 0.03  \\         
HC$^{15}$N         &   -      &  0.36  &  0.05    &       -  &  0.05  &       -  &      -  &      -  &         - &    -  &   0.03  &   0.02 &   -   \\        
HCO                &   -      &  0.01  &  0.01    &       -  &  0.04  &       -  &      -  &      -  &         - &    -  &    -    &   0.02 & 0.04  \\         
HCO$^{+}$          &   0.18   &  0.03  &  0.01    &   0.09   &  0.03  &  0.34    & 0.04    &  0.30   &   0.03    & 0.09  &   0.32  &   0.07 & 0.58  \\         
DCO$^{+}$          &   0.01   &  0.04  &  0.04    &   0.02   &  0.07  &  0.01    & 0.01    &  0.01   &         - &    -  &    -    &   0.02 &   -   \\        
H$^{13}$CO$^{+}$   &   0.03   &  0.03  &  0.02    &   0.02   &  0.09  &  0.01    & 0.01    &  0.01   &   0.01    & 0.08  &   0.16  &  <0.01 & 0.03  \\         
HC$^{18}$O$^{+}$   &   -      &  0.02  &  0.02    &       -  &  0.02  &       -  &      -  &      -  &         - &    -  &   0.02  &   0.02 &   -   \\        
HC$^{17}$O$^{+}$   &   -      &    -   &       -  &       -  &     -  &       -  &      -  &      -  &         - &    -  &    -    &     -  &   -   \\        
HCS$^{+}$          &   0.01   &  0.11  &  0.11    &       -  &  0.14  &       -  &      -  &      -  &         - & 0.02  &   0.05  &   0.03 &   -   \\        
HDO                &   -      &  0.03  &       -  &       -  &     -  &       -  &      -  &      -  &         - &    -  &    -    &     -  &   -   \\        
H$_{2}$CCO         &   -      &  0.06  &       -  &       -  &     -  &       -  &      -  &      -  &         - &    -  &    -    &   0.03 &   -   \\        
H$_{2}$CO          &   0.64   &  0.97  &  1.41    &   0.47   &  1.06  &  0.35    & 0.33    &  0.24   &   0.14    & 0.42  &   0.59  &   0.16 & 1.13  \\         
HDCO               &   -      &  0.11  &  0.02    &   0.02   &  0.10  &       -  &      -  &      -  &         - &    -  &   0.02  &   0.06 &   -   \\        
H$_{2}^{13}$CO     &   -      &  0.10  &  0.02    &   0.02   &  0.08  &       -  &      -  &      -  &         - &    -  &   0.05  &   0.02 &   -   \\        
H$_{2}$CS          &   0.08   &  0.85  &  0.76    &       -  &  1.09  &       -  &      -  &      -  &   0.03    & 0.12  &   0.42  &   0.27 & 0.02  \\         
HDCS               &   -      &    -   &       -  &       -  &  0.03  &       -  &      -  &      -  &         - &    -  &    -    &   0.04 &   -   \\        
H$_{2}$C$^{34}$S   &   -      &    -   &       -  &       -  &     -  &       -  &      -  &      -  &         - &    -  &    -    &     -  &   -   \\        
H$_{2}$S           &   -      &  0.15  &       -  &       -  &  0.04  &       -  &      -  &      -  &         - &    -  &    -    &     -  &   -   \\        
HNC                &   0.02   &  0.06  &  <0.01   &   0.03   &  0.02  &  0.08    & 0.08    &  0.02   &   0.02    & 0.09  &   0.03  &   0.03 & 0.03  \\         
DNC                &   0.01   &  0.10  &  0.02    &   0.02   &  0.07  &       -  & 0.01    &      -  &         - &    -  &    -    &   0.02 &   -   \\        
HN$^{13}$C         &   -      &  0.09  &  0.13    &   0.02   &  0.08  &       -  & 0.01    &      -  &         - &    -  &   0.02  &   0.02 &   -   \\        
HNCO               &   -      &  0.11  &       -  &       -  &  0.03  &       -  &      -  &      -  &         - &    -  &    -    &     -  &   -   \\        
N$_{2}$H$^{+}$     &   0.11   &  0.29  &  0.32    &   0.11   &  0.14  &  0.06    & 0.08    &  0.08   &   0.03    & 0.01  &   0.20  &   0.16 &   -   \\        
N$_{2}$D$^{+}$     &   -      &  0.01  &  0.01    &   0.03   &     -  &       -  &      -  &      -  &         - &    -  &    -    &     -  &   -   \\        
NO                 &   0.01   &  0.05  &  0.05    &   0.05   &  0.08  &  0.01    & 0.03    &  0.03   &         - &    -  &   0.05  &     -  &   -   \\        
NS                 &   -      &  0.19  &  0.18    &       -  &  0.18  &       -  &      -  &      -  &         - &    -  &    -    &   0.13 &   -   \\        
OCS                &   -      &  0.35  &       -  &       -  &     -  &       -  &      -  &      -  &         - &    -  &   0.10  &     -  &   -   \\        
SiO                &   0.08   &  0.35  &  0.30    &       -  &  0.26  &       -  &      -  &      -  &         - &    -  &   0.54  &     -  &   -   \\        
$^{29}$SiO         &   -      &  0.06  &       -  &       -  &     -  &       -  &      -  &      -  &         - &    -  &    -    &     -  &   -   \\        
$^{30}$SiO         &   -      &    -   &       -  &       -  &     -  &       -  &      -  &      -  &         - &    -  &    -    &     -  &   -   \\        
SO                 &   0.30   &  1.04  &  0.84    &   0.17   &  1.00  &  0.14    & 0.07    &  0.05   &   0.09    & 0.18  &   1.20  &   0.14 & 0.18  \\         
$^{34}$SO          &   -      &  0.81  &       -  &       -  &  0.06  &       -  &      -  &      -  &         - &    -  &   0.04  &   0.01 &   -   \\        
SO$^{+}$           &   -      &    -   &       -  &       -  &  0.05  &       -  &      -  &      -  &         - &    -  &    -    &   0.02 & 0.04  \\         
SO$_{2}$           &   0.07   &  4.28  &  0.16    &       -  &  0.38  &       -  &      -  &      -  &         - &    -  &   1.48  &     -  &   -   \\  \\
\hline
\end{tabular}
\tablefoot{Errors refer to   the Gaussian line fits
and do not account for the calibration uncertainties discussed in Section \ref{obs_calis}. 
Errors add up for species with many
transitions and are usually larger in cases of strong overlap between lines. 
However, they are generally $2\%-6\%$.}
\end{table*}

\begin{table*}
\tiny
\caption{Overlap between regions. To be read from left to right, e.g. $\sim39\%$ of the `high column density' region 
overlaps with the `low temperature' region. Conversely, $\sim9\%$ of the `low temperature' region overlaps with the 
`high column density' region.
}   
\label{table_overlaps}     
\centering 
\begin{tabular}{l | c c c c c c c c c c c c}
\hline\hline   % table heading
\\
&\rotatebox{90}{high column density} 
& \rotatebox{90}{high column density without}
\rotatebox{90}{KL and Orion South} 
& \rotatebox{90}{high column density,} \rotatebox{90}{low temperature} 
& \rotatebox{90}{high column density,} \rotatebox{90}{high temperature,}
\rotatebox{90}{without KL and Orion South}
& \rotatebox{90}{low column density} 
& \rotatebox{90}{low temperature} 
&\rotatebox{90}{low column density,} \rotatebox{90}{low temperature}
& \rotatebox{90}{low column density,} \rotatebox{90}{high temperature}
& \rotatebox{90}{high temperature}\rotatebox{90}{without KL and Orion South}
& \rotatebox{90}{\ion{H}{ii}} & \rotatebox{90}{radical region} & \rotatebox{90}{dense PDR}
 
\\
\hline                       
\\
high column density & -       &   0.924   &  0.393  & 0.494   & -      & 0.393  & -     & -      & 0.494  &   0.085  &  0.159  &  -      \\ \\

high column density without & 1.000   &   -       &  0.426  & 0.535   & -      & 0.426  & -     & -      & 0.535  &   0.063  &  0.172  &  -      \\
KL and Orion South & & & & & & \\ \\

high column density, & 1.000   &   1.000   &  -      & -       & -      & 1.000  & -     & -      & -      &   -      &  -      &  -      \\
low temperature & & & & & & \\ \\

high column density,& 1.000   &   1.000   &  -      & -       & -      & -      & -     & -      & 1.000  &   0.086  &  0.321  &  -      \\
high temperature, & & & & & & \\
without KL and Orion South & & & & & & \\ \\

low column density & -       &   -       &  -      & -       & -      & 0.498  & 0.498 & 0.502  & 0.502  &   0.024  &  0.008  &  0.025  \\ \\

low temperature & 0.090   &   0.090   &  0.090  & -       & 0.910  & -      & 0.910 & -      & -      &   -      &  -      &  -      \\ \\

low column density, & -       &   -       &  -      & -       & 1.000  & 1.000  & -     & -      & -      &   -      &  -      &  -      \\
low temperature & & & & & & \\ \\

low column density, & -       &   -       &  -      & -       & 1.000  & -      & -     & -      & 1.000  &   0.047  &  0.015  &  0.050  \\
high temperature & & & & & & \\ \\

high temperature & 0.109   &   0.109   &  -      & 0.109   & 0.891  & -      & -     & 0.891  & -      &   0.051  &  0.049  &  0.045  \\
without KL and Orion South & & & & & & \\ \\

\ion{H}{ii} & 0.311   &   0.211   &  -      & 0.156   & 0.689  & -      & -     & 0.689  & 0.844  &   -      &  -      &  0.256  \\ \\

radical region & 0.722   &   0.722   &  -      & 0.722   & 0.278  & -      & -     & 0.278  & 1.000  &   -      &  -      &  -      \\ \\

dense PDR & -       &   -       &  -      & -       & 1.000  & -      & -     & 1.000  & 1.000  &   0.348  &  -      &  -      \\
\hline  

\end{tabular}
\tablefoot{The majority of these overlaps are discussed in this paper as they correspond to regions themselves (e.g.
the overlap between the `high column density' and `low temperature' regions is identical to the region `high column density,
low temperature'). The exceptions to this are the three regions that are not based on column density or temperature thresholds, 
namely `\ion{H}{ii}', `radical region', and `dense PDR'.}
\end{table*}

\FloatBarrier

\section{Additional figures}

\begin{figure*}[ht]
\centering
        \includegraphics[width=\textwidth]{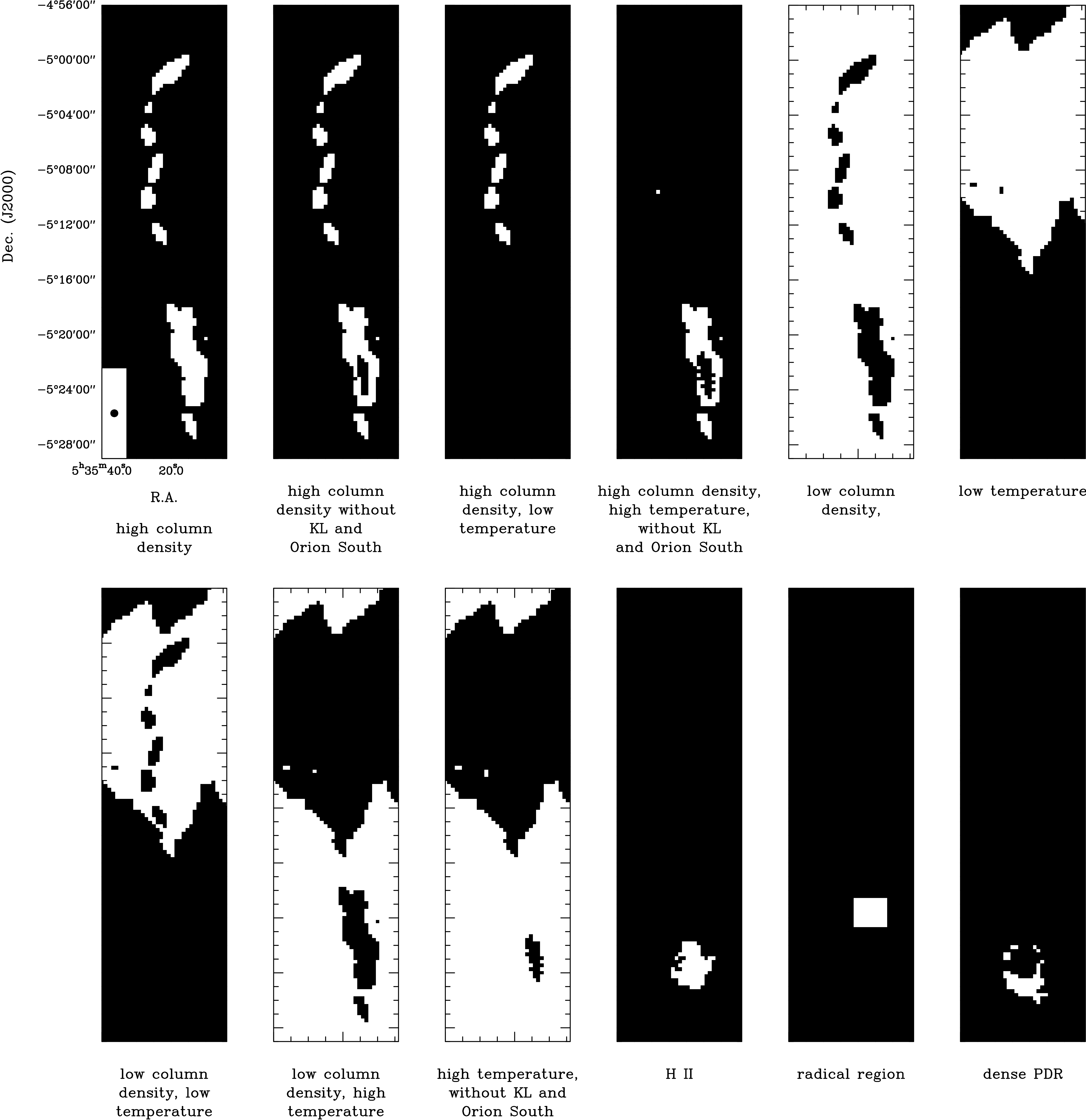}
        \caption{ Masks obtained with the parameters listed in Table \ref{table_regions}. 
        The images are intended as a visual aid and give an impression of the spatial extension
        of the considered regions. White pixels are part of the region, black pixels are not.}
        \label{fig:masks}
\end{figure*}

\begin{figure*}
    \centering

    \begin{subfigure}[t]{\textwidth}
    \centering
        \includegraphics[width=11cm, angle=90]{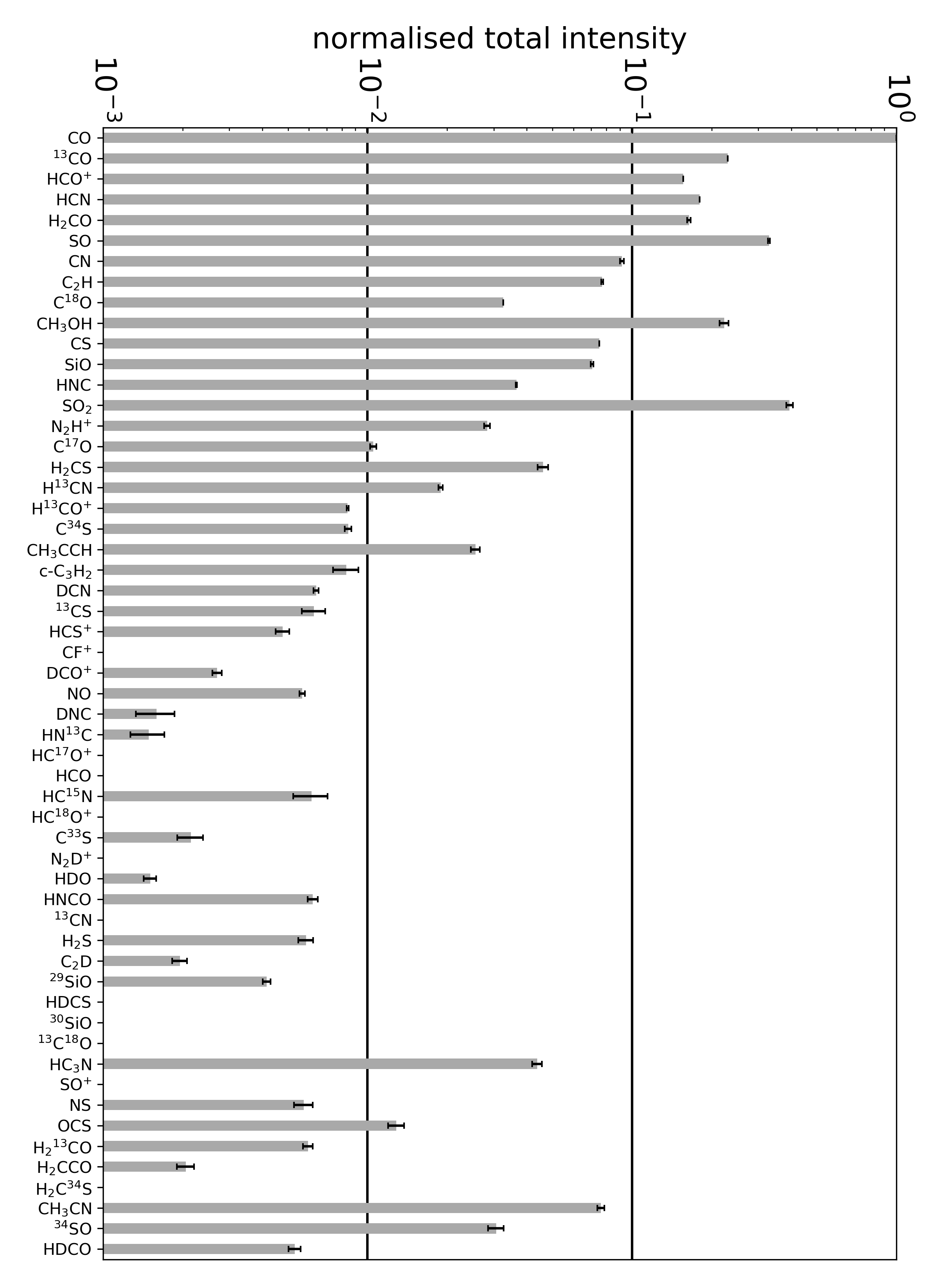}
        \caption{ high column density}
        \label{fig:ti_hd}
    \end{subfigure}
        ~  
    \begin{subfigure}[t]{\textwidth}
    \centering
        \includegraphics[width=11cm, angle=90]{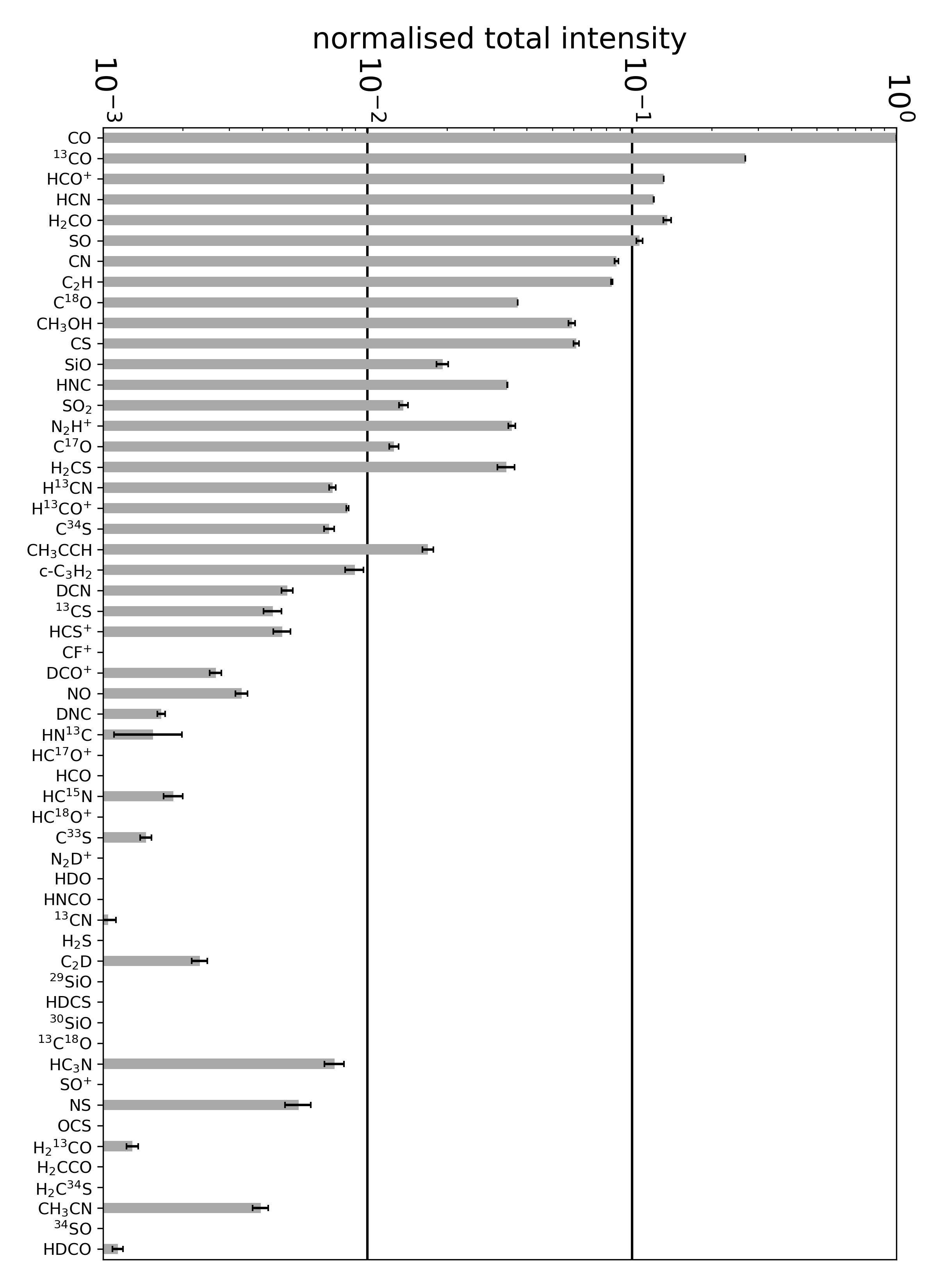}
        \caption{ high column density, without KL and Orion South}
        \label{fig:ti_hd_wo_KL}
    \end{subfigure} 
    
  \end{figure*}
  \begin{figure*}\ContinuedFloat     
    
    \begin{subfigure}[t]{\textwidth}
    \centering
        \includegraphics[width=11cm, angle=90]{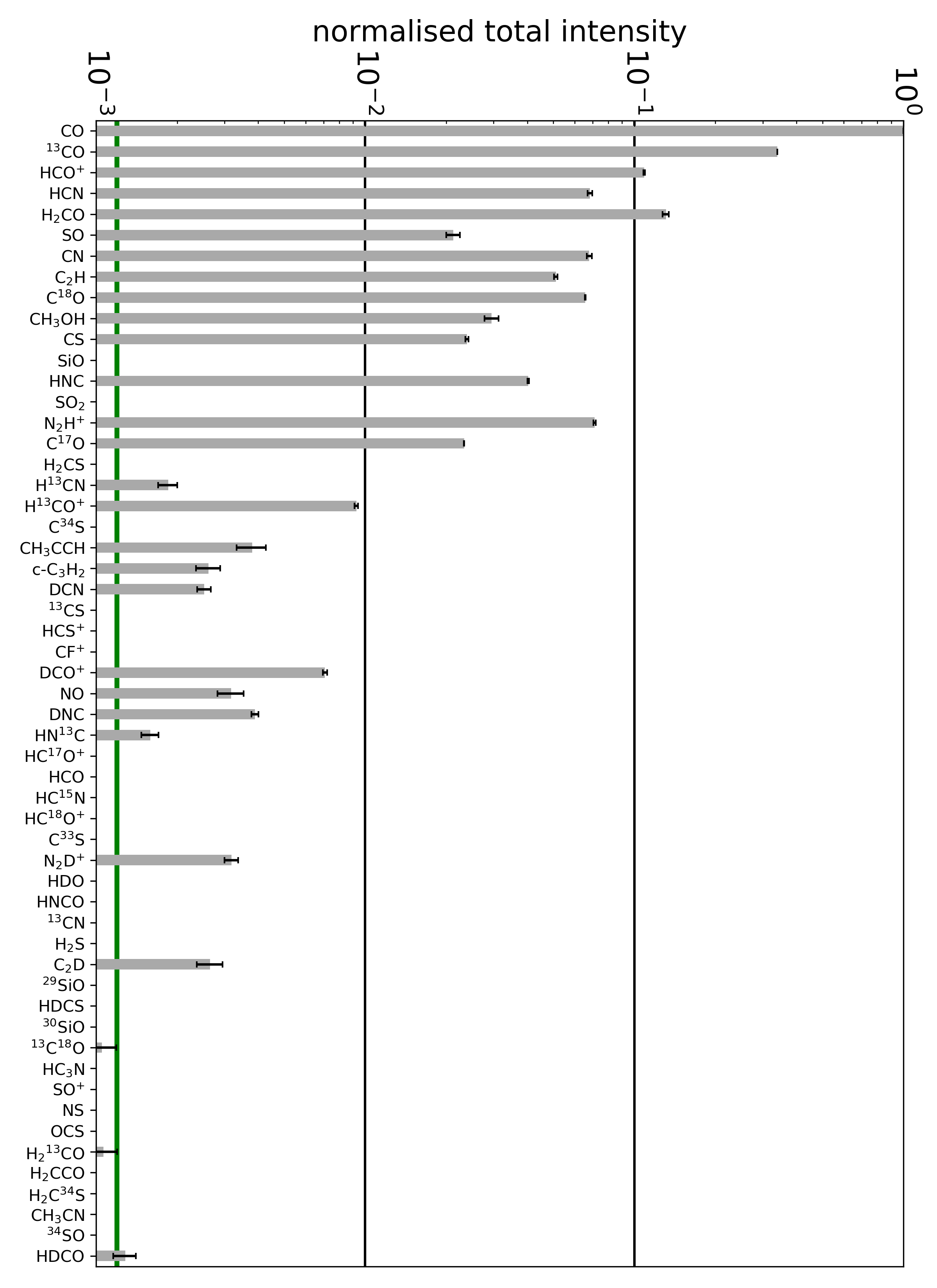}
        \caption{ high column density, low temperature}
        \label{fig:ti_lt_hd}
    \end{subfigure}
    ~
    \begin{subfigure}[t]{\textwidth}
    \centering
         \includegraphics[width=11cm, angle=90]{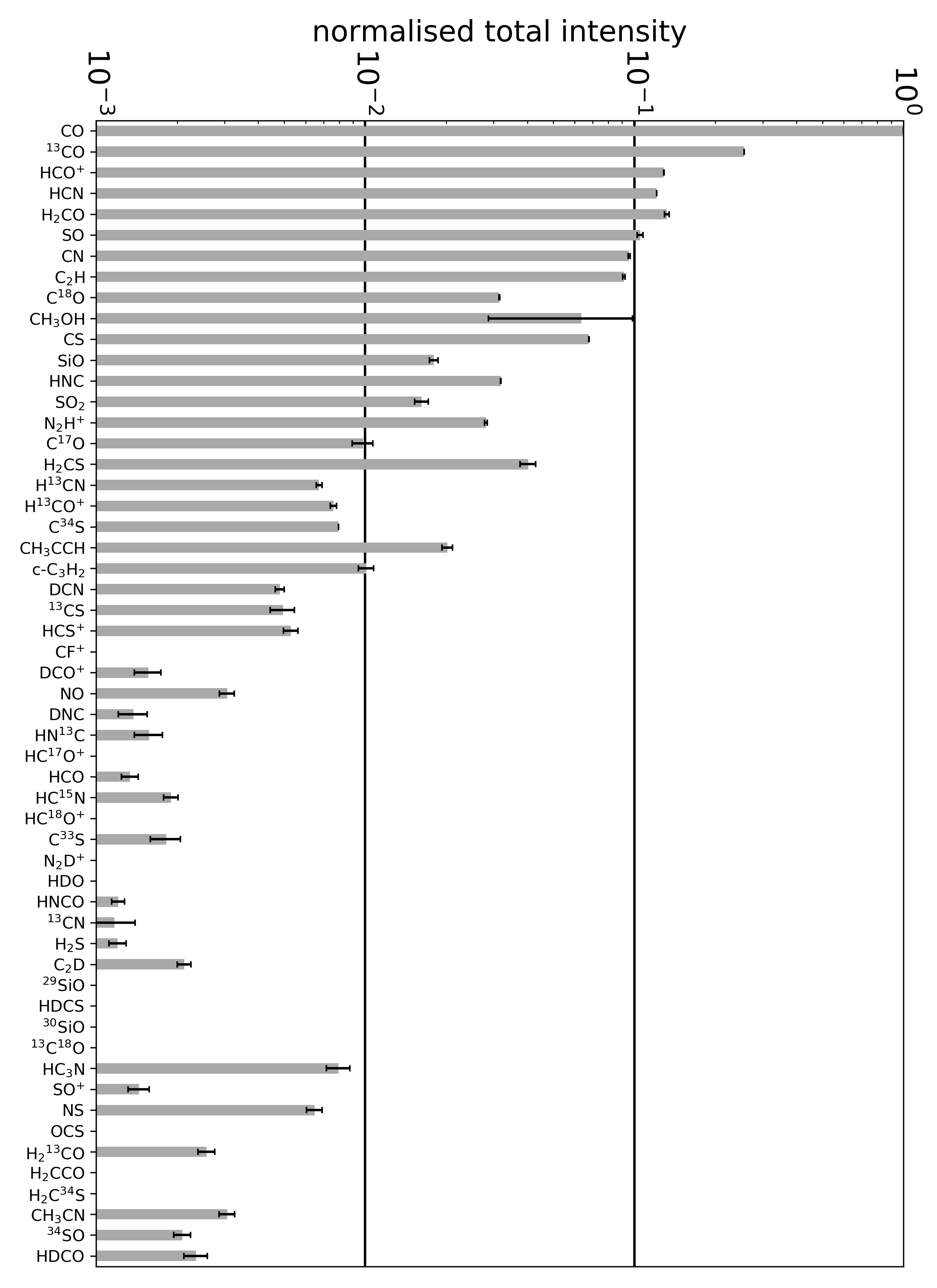}
         \caption{ high column density, high temperature, without KL and Orion South}
         \label{fig:ti_ht_hd_wo_KL}
     \end{subfigure}    

  \end{figure*}
  \begin{figure*}\ContinuedFloat  

        \begin{subfigure}[t]{\textwidth}
        \centering
        \includegraphics[width=11cm, angle=90]{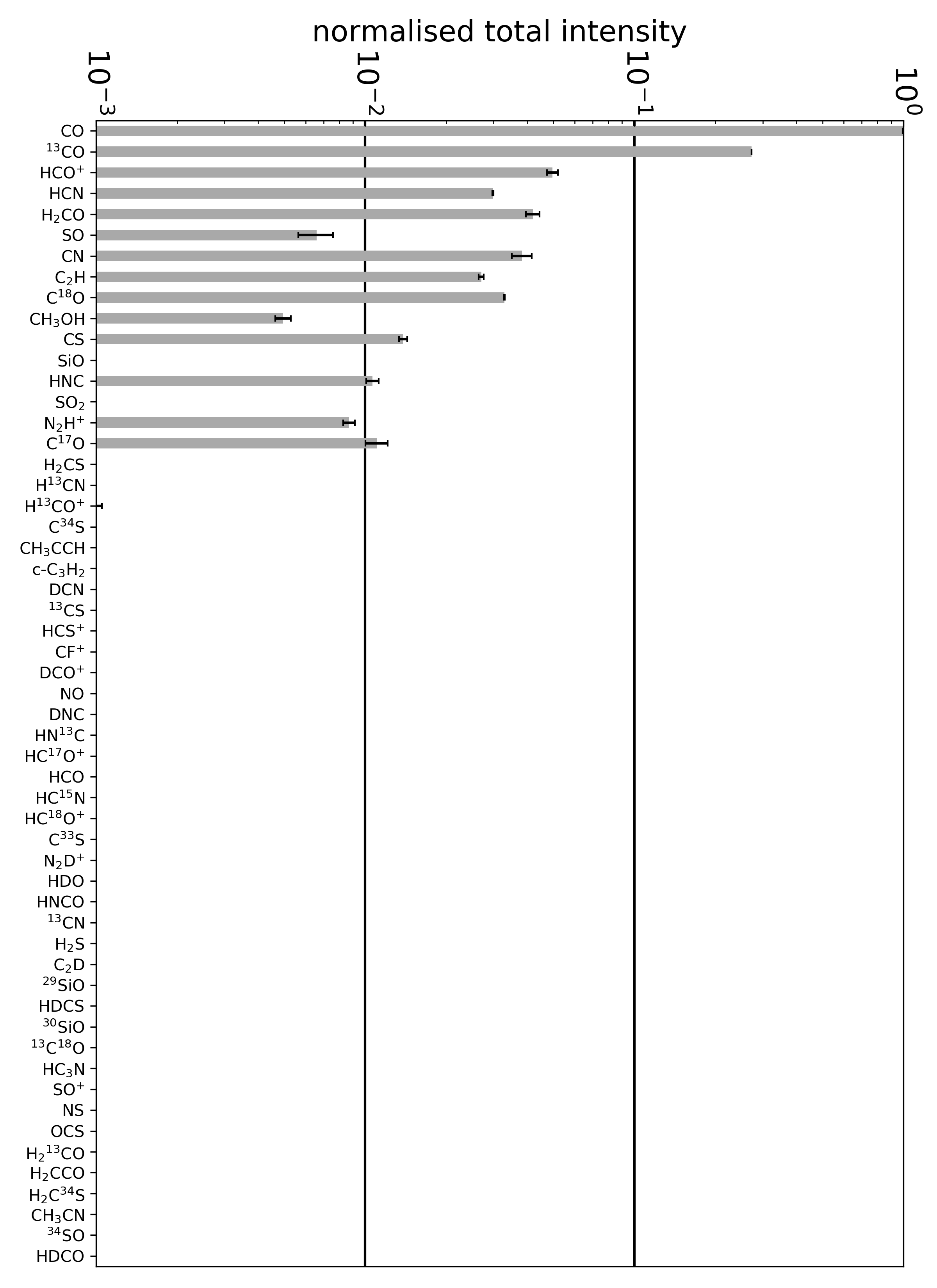}
        \caption{ low column density}
        \label{fig:ti_ld}
    \end{subfigure}
    ~
     \begin{subfigure}[t]{\textwidth}
     \centering
         \includegraphics[width=11cm, angle=90]{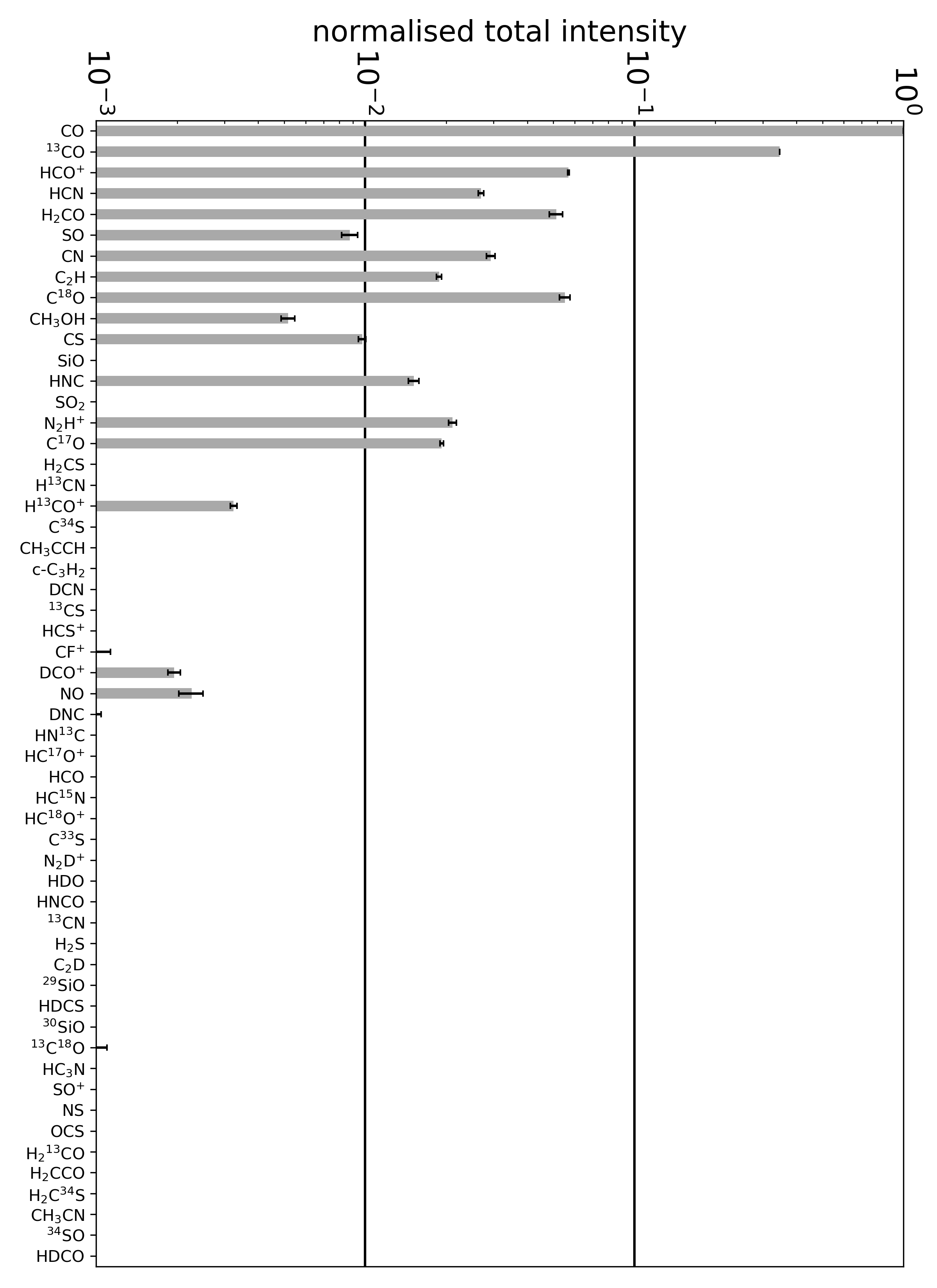}
         \caption{ low temperature}
         \label{fig:ti_lt}
     \end{subfigure}
     
\end{figure*}
  \begin{figure*}\ContinuedFloat      

    \begin{subfigure}[t]{\textwidth}
    \centering
        \includegraphics[width=11cm, angle=90]{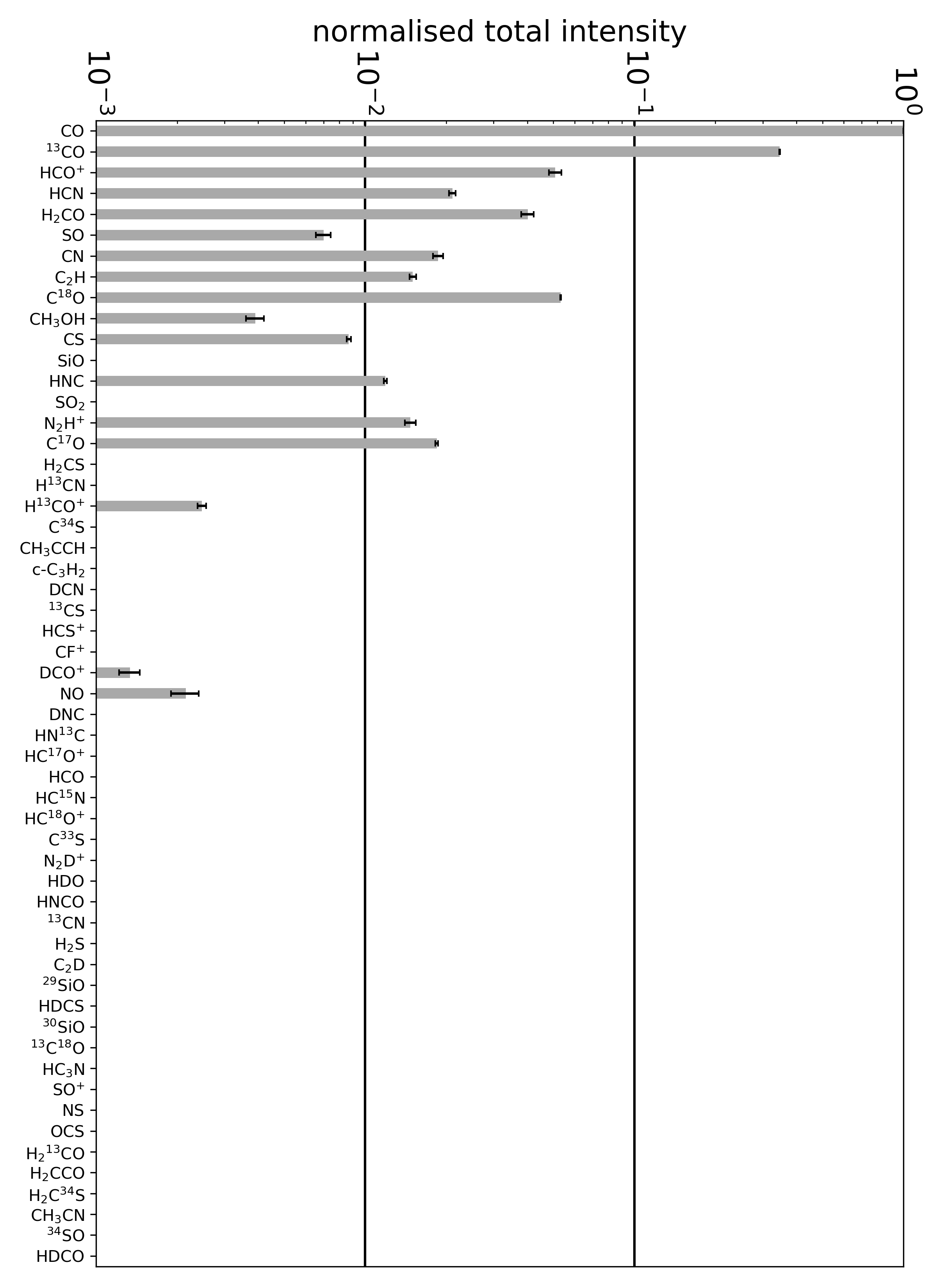}
        \caption{ low column density, low temperature}
        \label{fig:ti_lt_ld}
    \end{subfigure} 
    ~
    \begin{subfigure}[t]{\textwidth}
    \centering
        \includegraphics[width=11cm, angle=90]{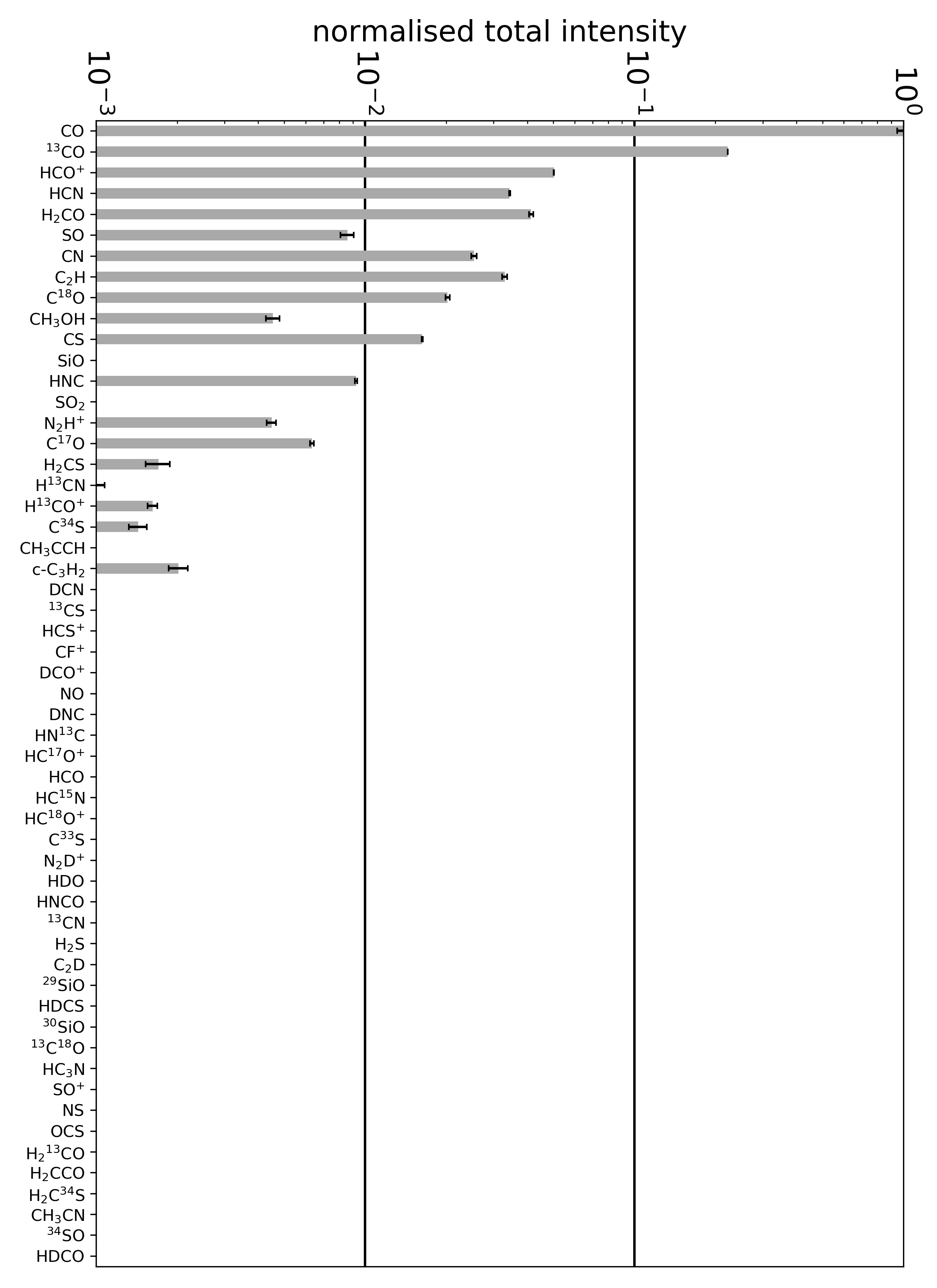}
        \caption{ low column density, high temperature}
        \label{fig:ti_ht_ld}
    \end{subfigure}
  
  \end{figure*}
  \begin{figure*}\ContinuedFloat    

     \begin{subfigure}[t]{\textwidth}
     \centering
         \includegraphics[width=11cm, angle=90]{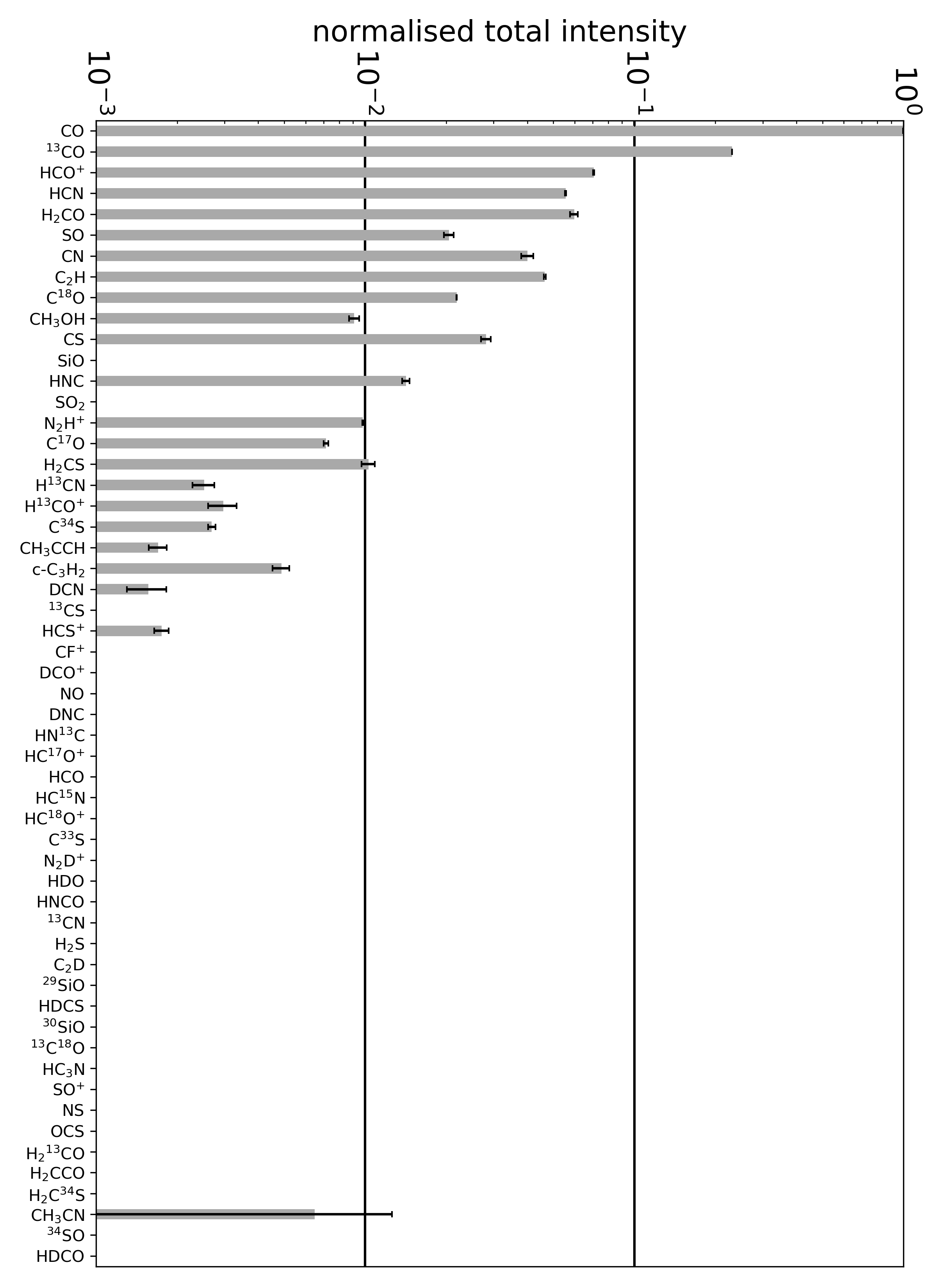}
         \caption{ high temperature, without KL and Orion South}
         \label{fig:ti_ht_wo_KL}
     \end{subfigure}
     ~
     \begin{subfigure}[t]{\textwidth}
     \centering
        \includegraphics[width=11cm, angle=90]{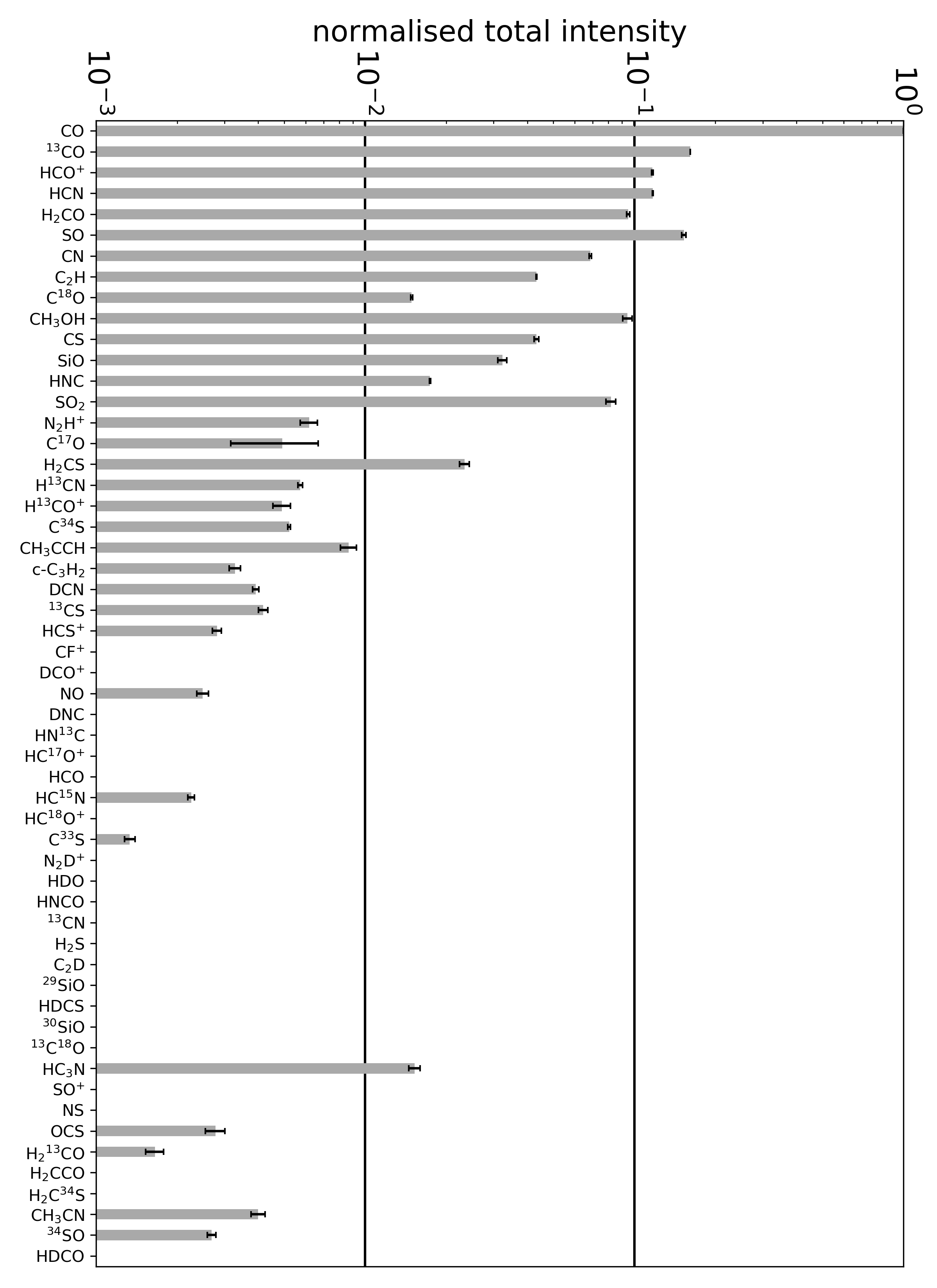}
        \caption{\ion{H}{ii}}
        \label{fig:ti_halpha}
    \end{subfigure}
    
 \end{figure*}
  \begin{figure*}\ContinuedFloat    
   \begin{subfigure}[t]{\textwidth}
   \centering
        \includegraphics[width=11cm, angle=90]{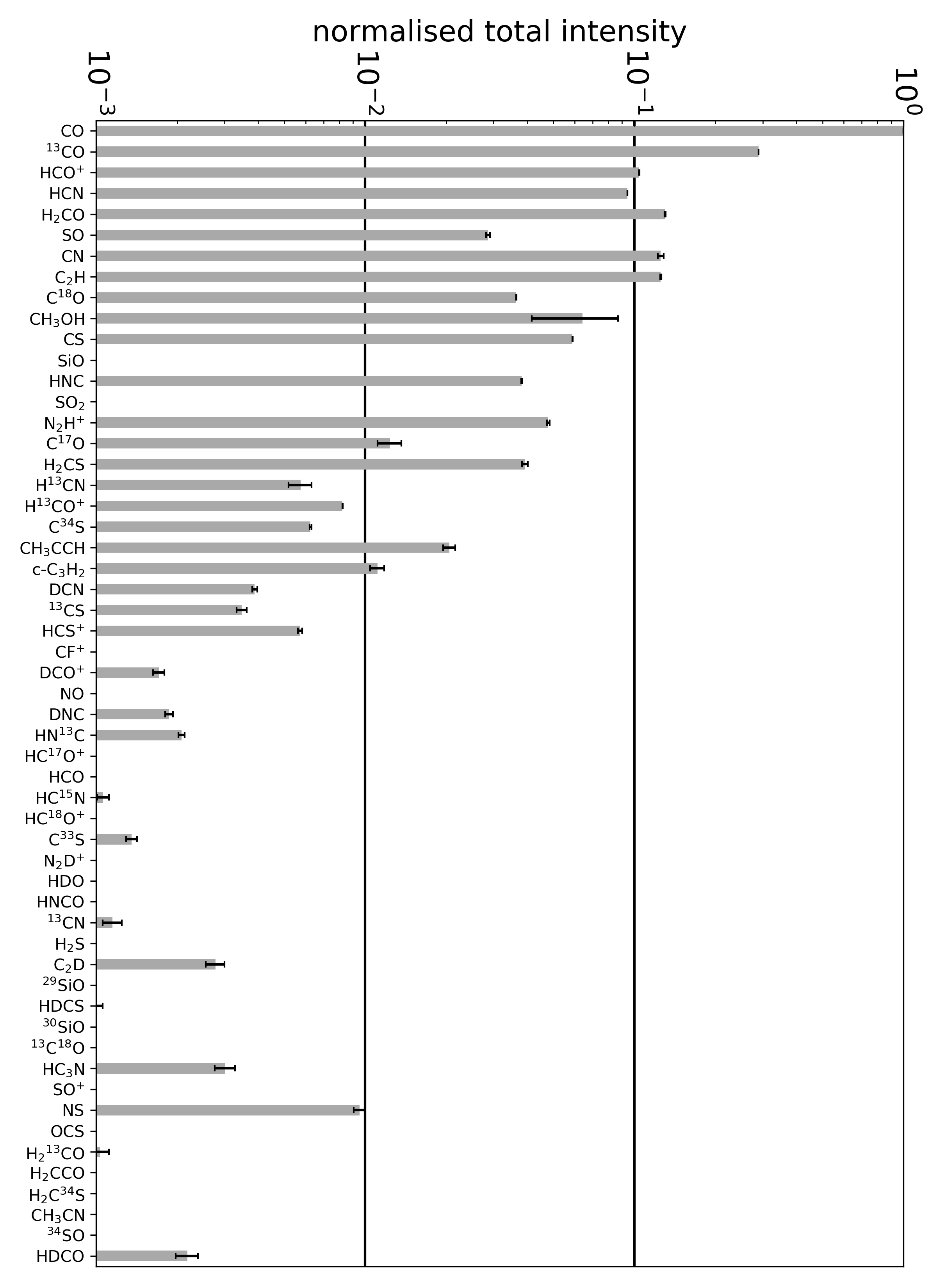}
        \caption{ radical region}
        \label{fig:ti_radical}
    \end{subfigure}
    ~
    \begin{subfigure}[t]{\textwidth}
    \centering
        \includegraphics[width=11cm, angle=90]{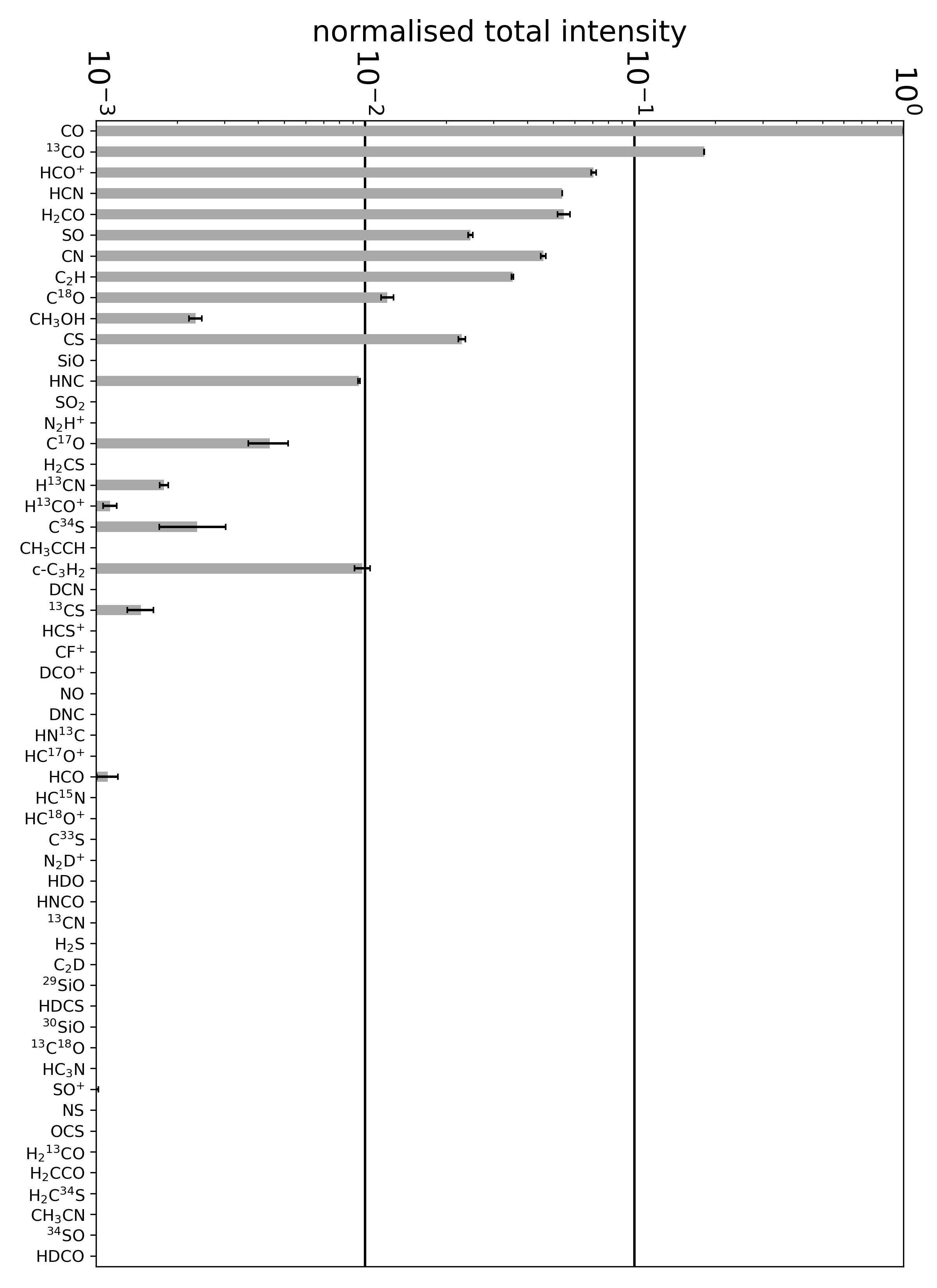}
        \caption{ dense PDR}
        \label{fig:ti_bar}
    \end{subfigure}    
    
    \caption{Total intensities of all considered species for the regions described in Table \ref{table_regions},
    normalised by the respective total CO intensity. Due to the noise
level of the region and its total CO intensity, a normalised total intensity of $0.1 \%$ is not detectable 
for region (c). Therefore, a green line approximating the detection limit for a $5\sigma$ feature
(assuming the median line width of the region and $\sigma_{\mathrm{median}}$) is added in this case. Detections
below that limit indicate that the local noise is below $\sigma_{\mathrm{median}}$. Error bars refer to the fit 
uncertainties.}
\label{fig:total_intensities}
\end{figure*}

\begin{figure*}
\centering
    \includegraphics[width=0.76\textwidth]{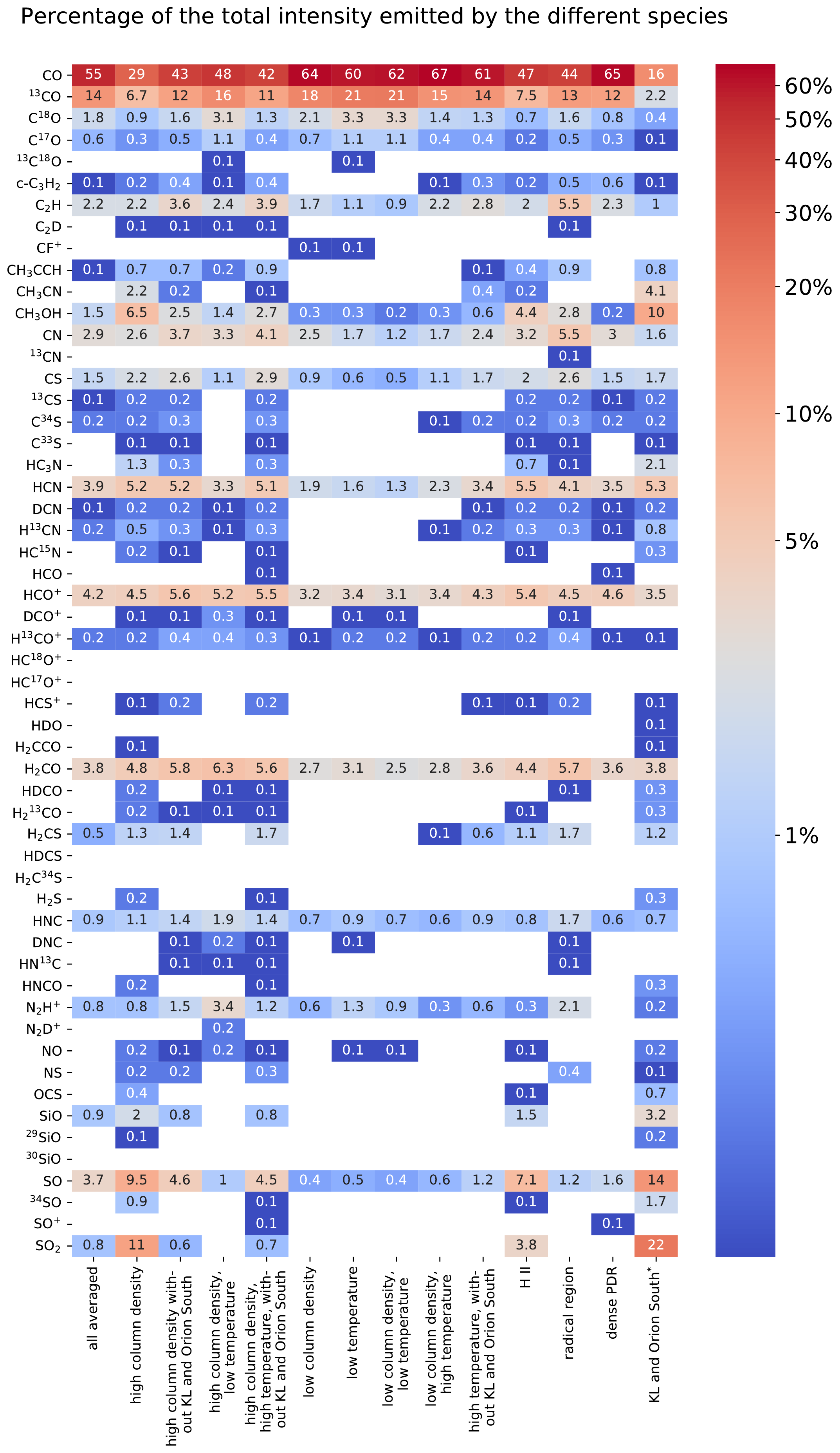}
    \caption{Share of each species to the total intensity of each region. For each species (rows), the colour bar
    offers a visual aid to quickly identify regions with lower or higher values. For each region (columns), the 
    colour bar helps to assess the influence of different species. Shares under 0.1\% are blank. 
    $^{(*)}$ The values for `KL and Orion South' are approximated as discussed in Section
    \ref{KL_emission}.}
    \label{fig_share_heatmap}
\end{figure*}

\begin{figure*}
    \centering

    \begin{subfigure}[t]{0.32\textwidth}
        \includegraphics[width=\textwidth]{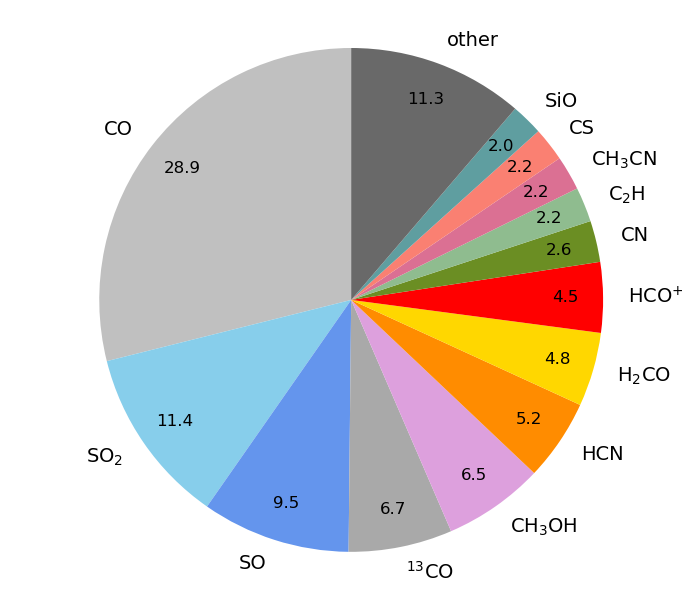}
        \caption{ high column density}
        \label{fig:pie_hd}
    \end{subfigure}
    ~
    \begin{subfigure}[t]{0.32\textwidth}
        \includegraphics[width=\textwidth]{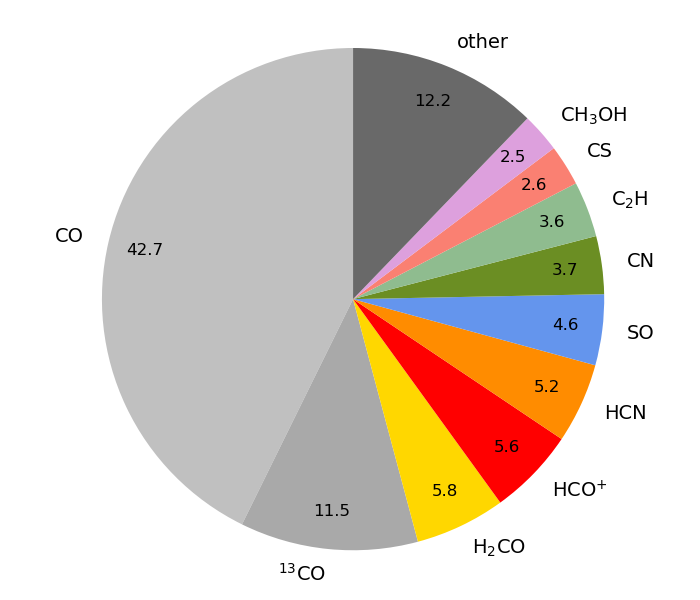}
        \caption{ high column density, without KL and Orion South}
        \label{fig:pie_hd_wo_kl}
    \end{subfigure}
    ~
    \begin{subfigure}[t]{0.32\textwidth}
        \includegraphics[width=\textwidth]{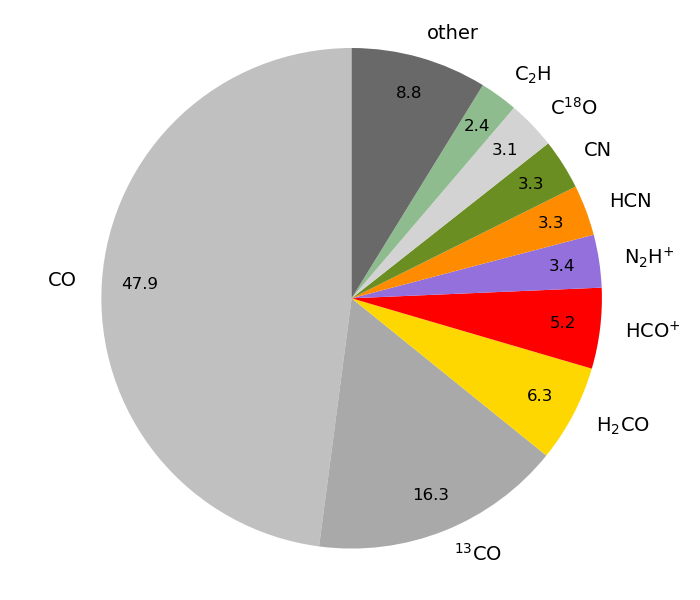}
        \caption{ high column density, low temperature}
        \label{fig:pie_lt_hd}
    \end{subfigure}

    \begin{subfigure}[t]{0.32\textwidth}
         \includegraphics[width=\textwidth]{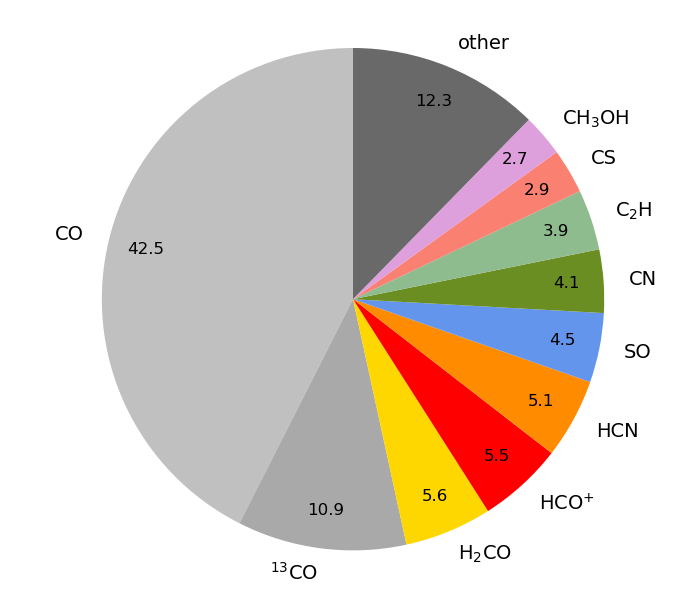}
         \caption{ high column density, high temperature, without KL and Orion South}
         \label{fig:pie_ht_hd_wo_KL}
     \end{subfigure}    
     ~
     \begin{subfigure}[t]{0.32\textwidth}
         \includegraphics[width=\textwidth]{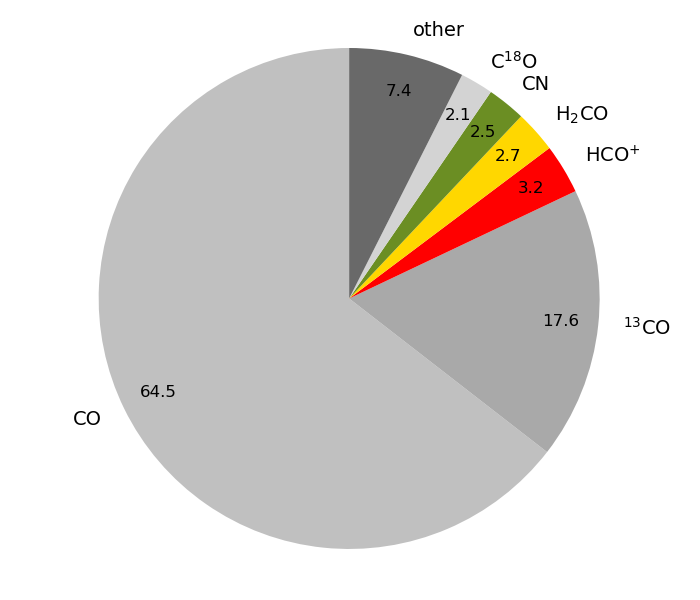}
         \caption{ low column density}
         \label{fig:pie_low_density}
     \end{subfigure}
     ~
     \begin{subfigure}[t]{0.32\textwidth}
         \includegraphics[width=\textwidth]{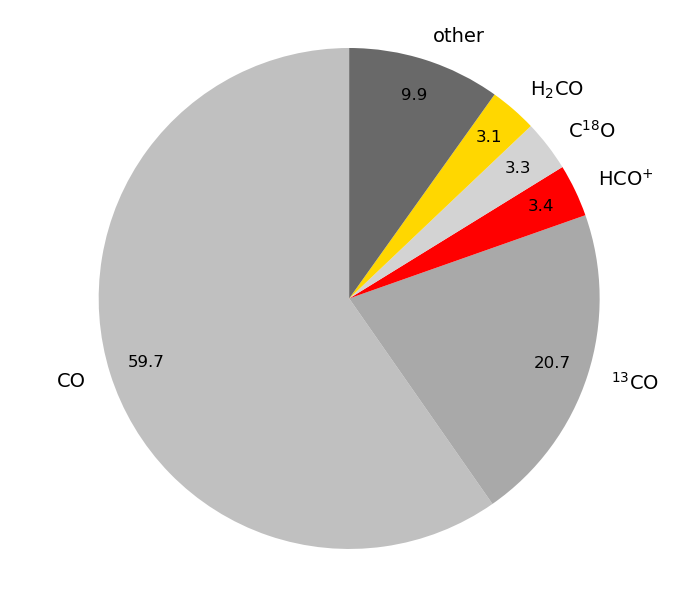}
         \caption{ low temperature}
         \label{fig:pie_lt}
     \end{subfigure}
    
    \begin{subfigure}[t]{0.32\textwidth}
        \includegraphics[width=\textwidth]{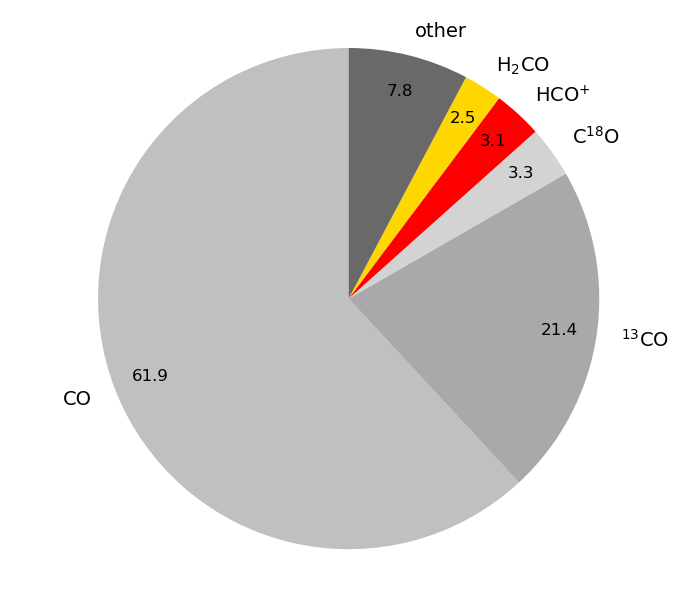}
        \caption{ low column density, low temperature}
        \label{fig:pie_lt_ld}
    \end{subfigure}     
     ~
    \begin{subfigure}[t]{0.32\textwidth}
        \includegraphics[width=\textwidth]{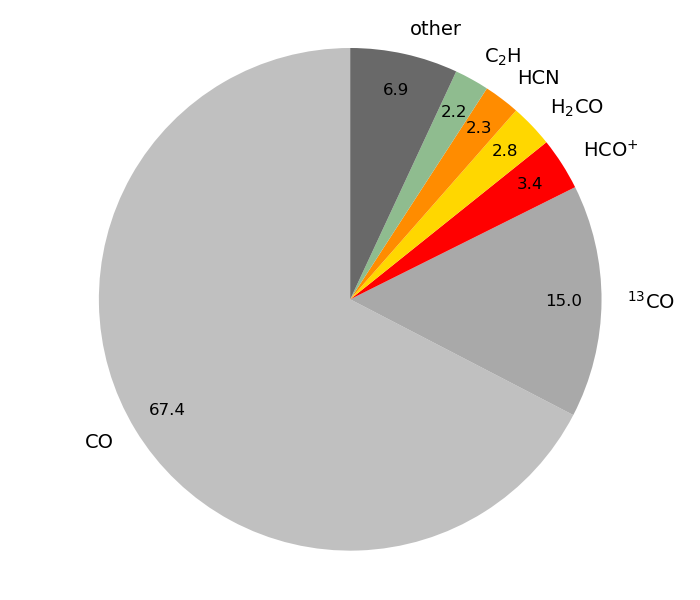}
        \caption{ low column density, high temperature}
        \label{fig:pie_ht_ld}
    \end{subfigure}
    ~
     \begin{subfigure}[t]{0.32\textwidth}
         \includegraphics[width=\textwidth]{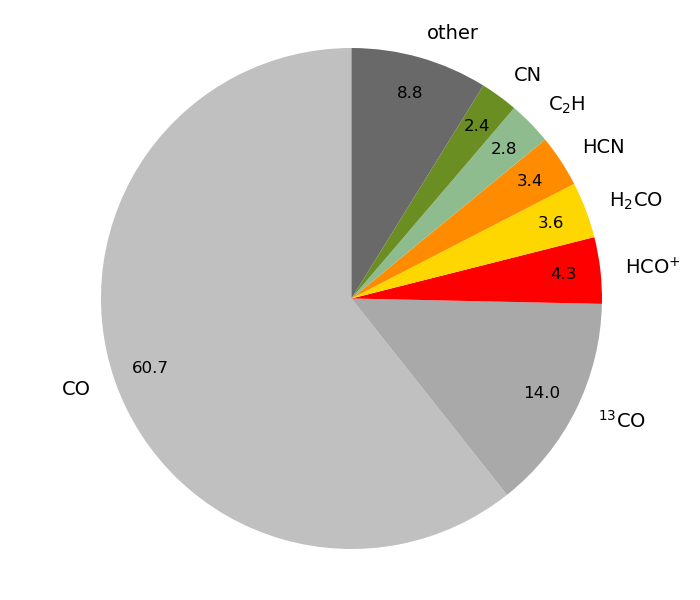}
         \caption{ high temperature, without KL and Orion South}
         \label{fig:pie_ht_wo_KL}
     \end{subfigure}
    
     \begin{subfigure}[t]{0.32\textwidth}
        \includegraphics[width=\textwidth]{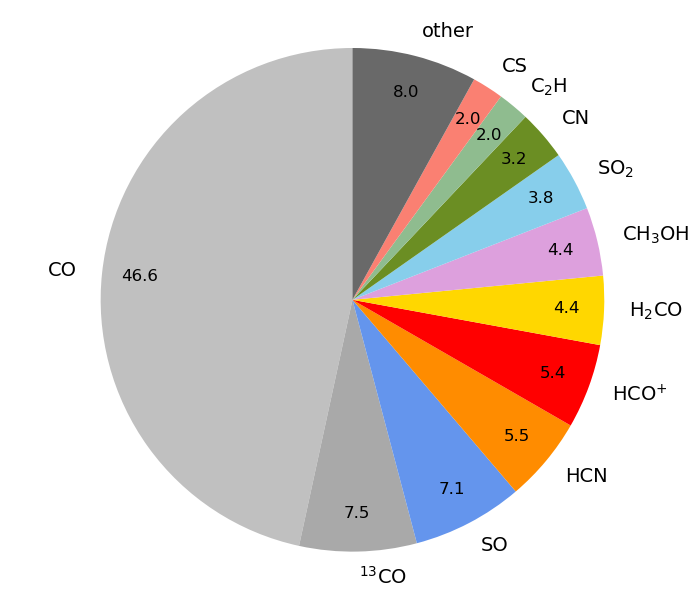}
        \caption{\ion{H}{ii}}
        \label{fig:pie_halpha}
    \end{subfigure}
    ~
    \begin{subfigure}[t]{0.32\textwidth}
        \includegraphics[width=\textwidth]{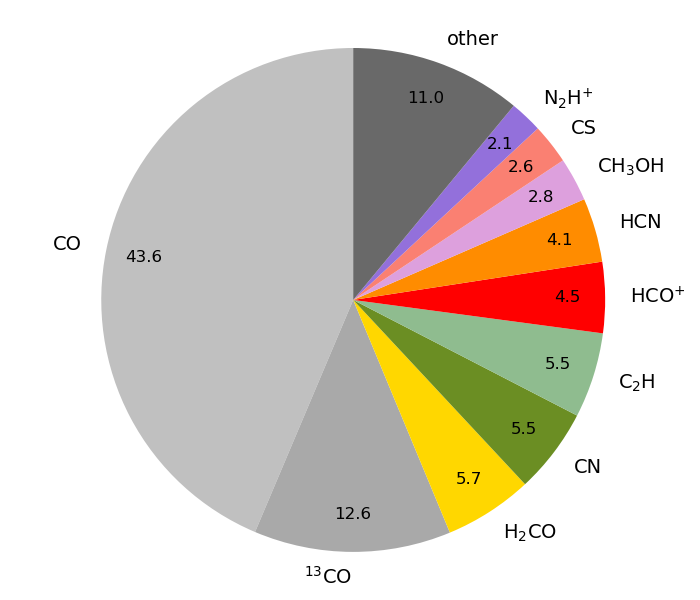}
        \caption{ radical region}
        \label{fig:pie_radical}
    \end{subfigure} 
    ~
    \begin{subfigure}[t]{0.32\textwidth}
        \includegraphics[width=\textwidth]{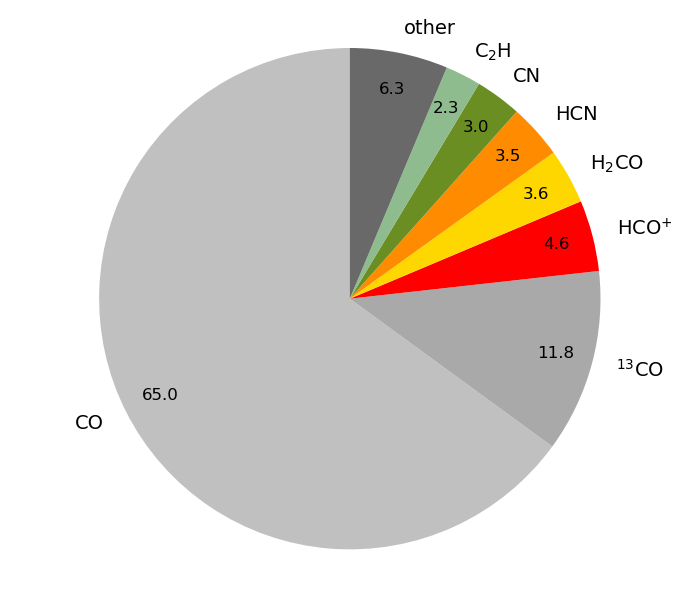}
        \caption{ dense PDR}
        \label{fig:pie_bar}
    \end{subfigure}

    \caption{Pie charts depicting the share of the total intensity 
    emitted by different species for all regions from Table \ref{table_regions} 
    (see Fig. \ref{all_averaged_charts} for the plot of the averaged data). Shares under $2\%$ are summed  
    under `other'.}\label{pie_charts}
\end{figure*}

\begin{figure}
\centering
    \includegraphics[width=0.49\textwidth]{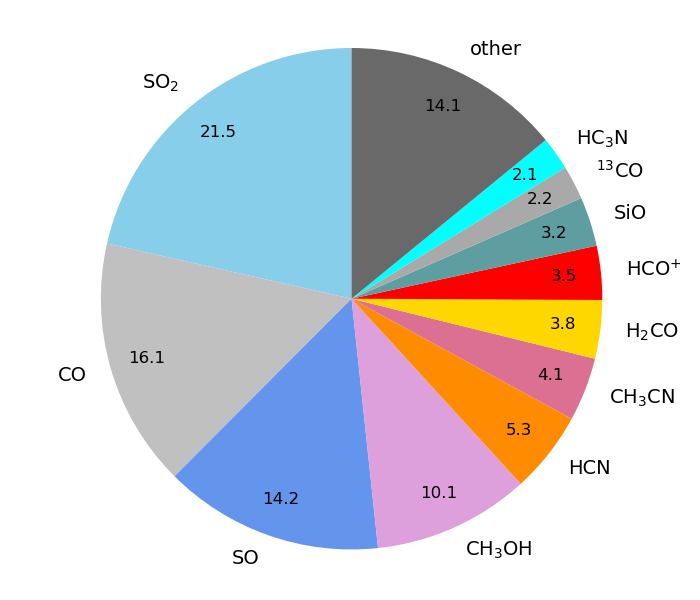}
    \caption{Approximated emission profile from the region around KL and Orion South.}\label{pie_KL}
\end{figure}
  
\begin{figure*}
    \begin{subfigure}[t]{0.49\textwidth}
         \centering
         \includegraphics[width=\textwidth]{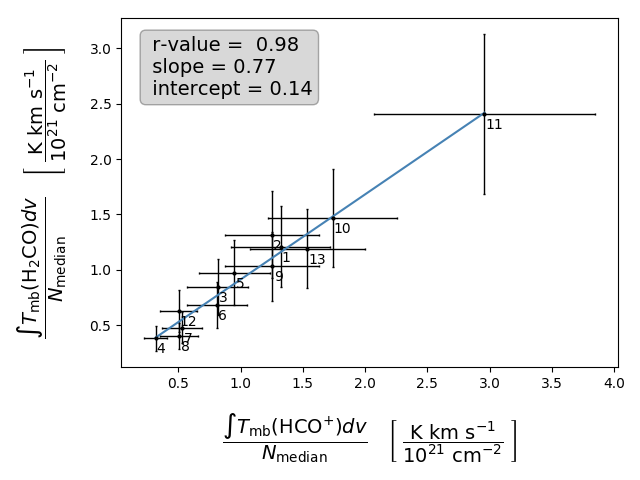}
         \caption{}
         \label{fig:corr_h2co_hco+}
     \end{subfigure}    
     ~
     ~
     \begin{subfigure}[t]{0.49\textwidth}
         \centering
         \includegraphics[width=\textwidth]{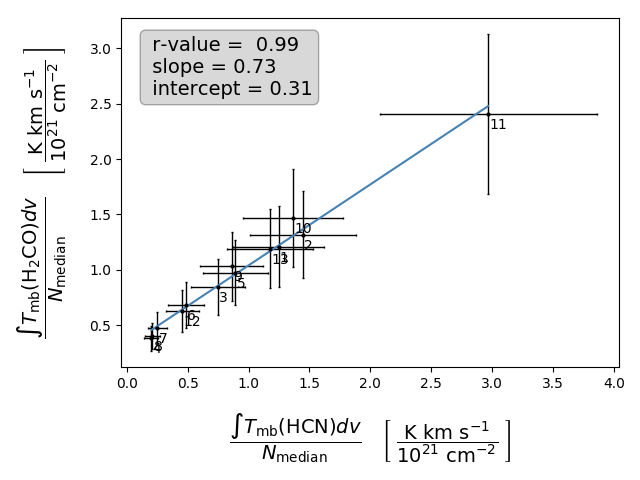}
         \caption{}
         \label{fig:corr_hcn_h2co}
     \end{subfigure}
     
     \begin{subfigure}[t]{0.49\textwidth}
         \centering
         \includegraphics[width=\textwidth]{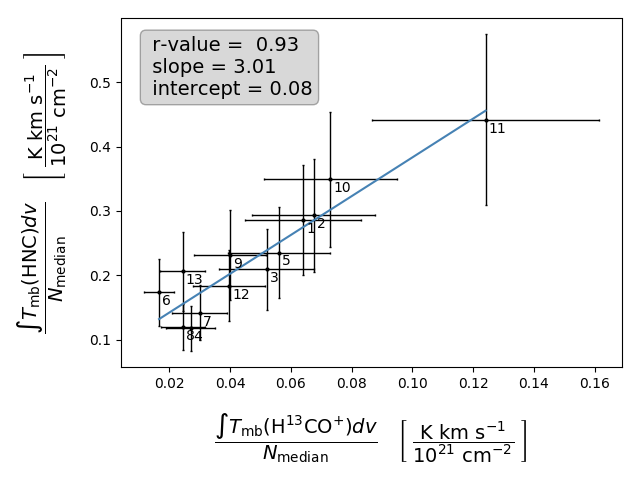}
         \caption{}
         \label{fig:corr_h13co+_hnc}
     \end{subfigure}
     ~
     ~
     \begin{subfigure}[t]{0.49\textwidth}
         \centering
         \includegraphics[width=\textwidth]{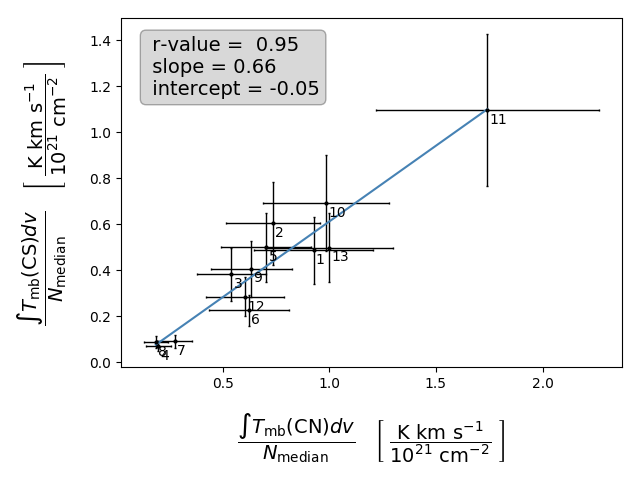}
         \caption{}
         \label{fig:corr_cn_cs}
     \end{subfigure}

      \begin{subfigure}[t]{0.49\textwidth}
         \centering
         \includegraphics[width=\textwidth]{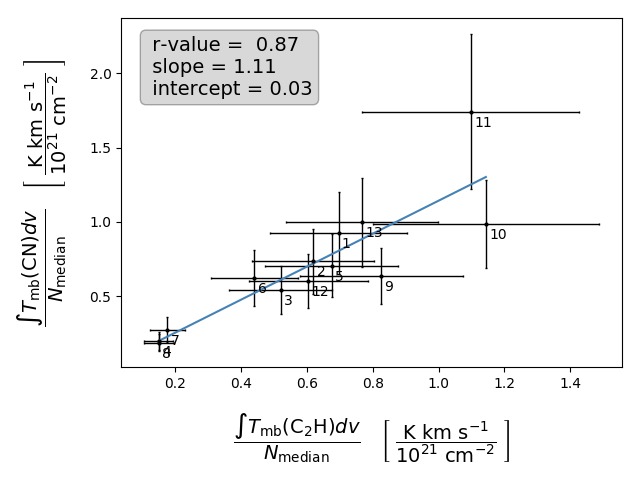}
         \caption{}
         \label{fig:corr_c2h_cn}
     \end{subfigure}
     ~
     ~
     \begin{subfigure}[t]{0.49\textwidth}
         \centering
         \includegraphics[width=\textwidth]{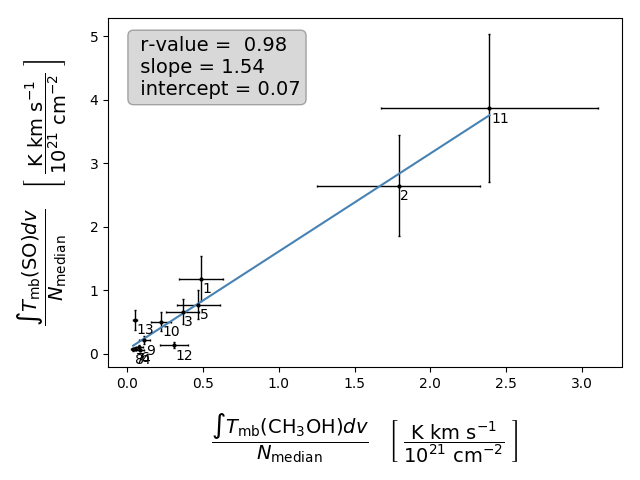}
         \caption{}
         \label{fig:corr_shock_tracer}
     \end{subfigure}         
    \caption{Examples of correlation plots between species. The numbers refer to the
    region identifiers listed in Table \ref{table_regions}. Due to the small number of
    samples, the r-value should be viewed with caution; it only gives   an
    indication of possible correlation. The error bars indicate the assumed 
    30\% uncertainty.}\label{fig:correlations}
\end{figure*}

\begin{figure*}
    \begin{subfigure}[t]{0.49\textwidth}
         \centering
         \includegraphics[width=\textwidth]{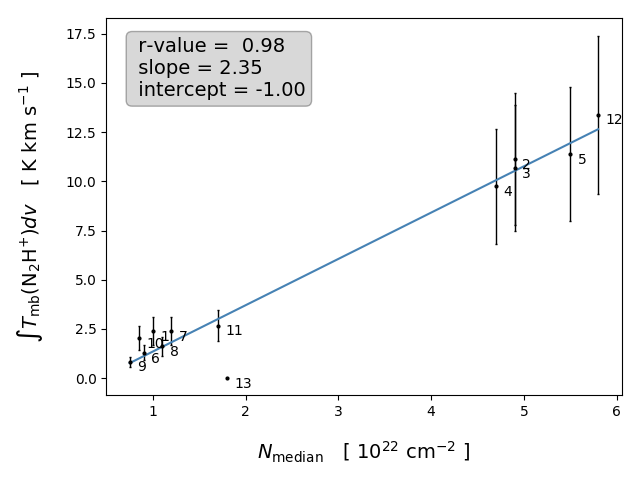}
         \caption{}
         \label{fig:corr_density_n2h+}
     \end{subfigure}    
     ~
     ~
     \begin{subfigure}[t]{0.49\textwidth}
         \centering
         \includegraphics[width=\textwidth]{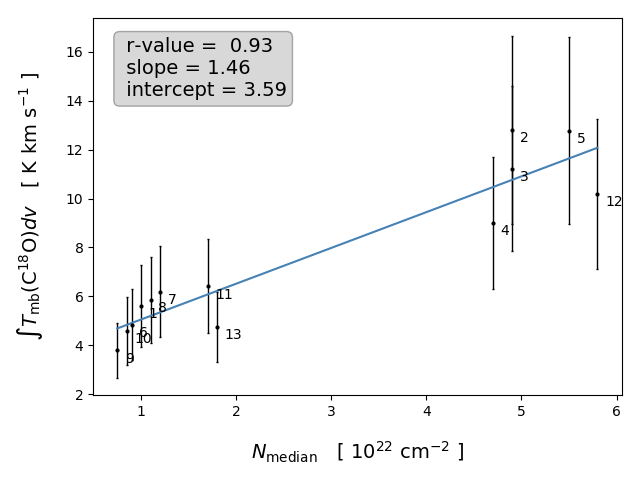}
         \caption{}
         \label{fig:corr_c18o_density}
     \end{subfigure}
     
     \begin{subfigure}[t]{0.49\textwidth}
         \centering
         \includegraphics[width=\textwidth]{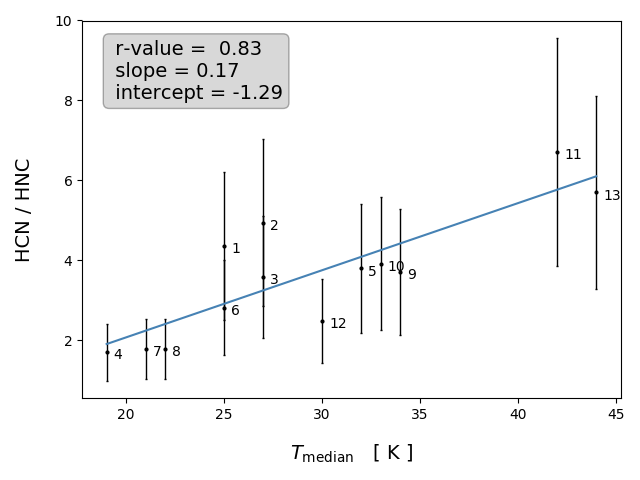}
         \caption{}
         \label{fig:corr_temp_HCN-HNC}
     \end{subfigure}
     ~
     ~
     \begin{subfigure}[t]{0.49\textwidth}
         \centering
         \includegraphics[width=\textwidth]{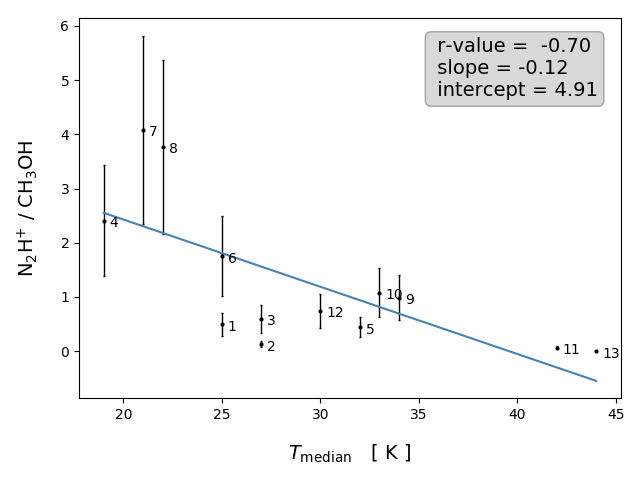}
         \caption{}
         \label{fig:corr_temp_N2H+-CH3OH}
     \end{subfigure}     
    \caption{Examples of correlation plots between species and physical parameters. 
    The numbers refer to the region identifiers listed in Table \ref{table_regions}. Due to the small number of
    samples, the r-value should be viewed with caution; it only  gives   an
    indication of possible correlation. The error bars indicate the assumed 
    30\% uncertainty.}\label{fig:correlations_density_temp}
\end{figure*}

\end{appendix}
\end{document}